\documentclass[11pt]{article}
\pdfoutput=1
\usepackage[utf8]{inputenc}
\usepackage[T1]{fontenc}
\usepackage{lmodern, textcomp}
\usepackage[american]{babel}
\usepackage{csquotes}
\usepackage{comment}

\usepackage[nottoc]{tocbibind}

\usepackage{cite}
\usepackage{amsmath,amsfonts,amssymb}
\usepackage[small,bf,hang]{caption}
\usepackage{slashed}
\input epsf.sty
\usepackage{epsfig}
\usepackage[titletoc,toc]{appendix}
\usepackage{xcolor}
\usepackage{siunitx}

\usepackage[pdftex, breaklinks=true, linktocpage,
	colorlinks=true, urlcolor=blue, linkcolor=blue, citecolor=red
]{hyperref}

\usepackage{graphicx}
\graphicspath{ {./figs/} }

\def\hybrid{
        \topmargin -20pt
        \oddsidemargin 0pt
        \headheight 0pt \headsep 0pt
        \textwidth 6.55in 
        \textheight 9.5in 
        \marginparwidth .875in
        \parskip 5pt plus 1pt \jot = 1.5ex}

\hybrid

\renewcommand{\theequation}{\thesection.\arabic{equation}} \csname
@addtoreset\endcsname{equation}{section}

\newcommand{\p}{\partial}

\newcommand{\pt}{\partial_t}

\renewcommand{\theequation}{\thesection.\arabic{equation}}

\def\moth{\mathsurround=0pt}
\newdimen\zo \zo=0pt

\def\tick{\leaders\hrule height 0.5ex depth 0pt \hskip 0.5pt}
\def\upboxfill{$\moth \setbox\zo\hbox{\tick}%
  \hskip 3pt\hbox to 0pt{$\tick$\hss}\hrulefill \hbox to 7.5pt{$\tick$\hss}$}

\def\dtick{\leaders\hrule height .34pt depth 0.5ex \hskip 0.5pt}
\def\downboxfill{$\moth \setbox\zo\hbox{\dtick}%
  \hskip 2pt\hbox to 0pt{$\dtick$\hss}\hrulefill \hbox to 2pt{$\dtick$\hss}$}

\def\bec{\begin{center}}
\def\ec{\end{center}}

\def\b{\beta}

\def\d{\delta} 

\def\e{\epsilon}

\def\t{\tau}

\def\x{\xi}

\def\be{\begin{equation}}
\def\ee{\end{equation}}
\def\bea{\begin{eqnarray}}
\def\eea{\end{eqnarray}}
\def\ba{\begin{array}}
\def\ea{\end{array}}

\renewcommand{\vec}[1]{\boldsymbol{#1}}
\renewcommand{\Re}{\operatorname{Re}}
\renewcommand{\Im}{\operatorname{Im}}

\begin{document}

\begin{titlepage}

\rightline{\tt MIT-CTP/5347}
\hfill \today
\begin{center}
\vskip 0.5cm

{\Large \bf {Initial value problem in string-inspired}}

{\Large \bf { nonlocal field theory}}

\vskip 0.5cm

\vskip 2.0cm
{\large {Harold Erbin$^{1,2,3}$, Atakan Hilmi Fırat$^1$, and Barton Zwiebach$^1$}}

	\vskip 0.8cm

{\em  \hskip -.1truecm
$^1$
Center for Theoretical Physics \\
Massachusetts Institute of Technology\\
Cambridge MA 02139, USA
\\
\medskip
$^2$
NSF AI Institute for Artificial Intelligence and Fundamental Interactions
\\
\medskip
$^3$
Université Paris Saclay, CEA, LIST\\
Gif-sur-Yvette, F-91191, France
\\
\medskip
\tt erbin@mit.edu, firat@mit.edu, zwiebach@mit.edu \vskip 5pt }

\vskip 3.0cm
{\bf Abstract}

\end{center}
\vskip 0.5cm
\noindent
\begin{narrower}
\baselineskip15pt
We consider a nonlocal scalar field theory inspired by the tachyon action in open string field theory.  The Lorentz-covariant
action is characterized by a parameter $\xi^2$ that quantifies the amount of nonlocality.
Restricting to purely time-dependent configurations, we show that a field redefinition perturbative in $\xi^2$ reduces the action to a local
two-derivative theory with a $\xi^2$-dependent potential.
This picture is supported by evidence that the redefinition maps the wildly oscillating rolling tachyon solutions of the nonlocal theory to conventional rolling in the new scalar potential.  For general field configurations we exhibit an
obstruction to a local Lorentz-covariant formulation, but we can still achieve
a formulation local in time, as well as a light-cone formulation.
These constructions provide an initial value formulation and a Hamiltonian.  Their causality is consistent with a lack of  superluminal behavior in the nonlocal theory.
\end{narrower}
\end{titlepage}

\tableofcontents

\section{Introduction and summary}

A central feature of string theory is the nonlocality of its interactions,
displayed manifestly in all the current formulations
of Lorentz invariant string field theories.  Indeed, in coordinate space, the interactions
feature exponentials of Laplacian operators.  In momentum space, this nonlocality
makes it clear that when working in Euclidean space,  there are no ultraviolet
divergences in string theory~(see~\cite{Pius:2016jsl,deLacroix:2017lif}).
The nonlocality of string theory has been the subject of intense discussion
and much speculation.  It has been suggested that it could play a role in resolving
the difficult issues in black hole evaporation~\cite{Susskind:1994sm,Lowe:1994ah,
Polchinski:1995ta,Hata:1995di,Giddings:2006vu,Lowe:2014vfa,Dodelson:2017hyu,Naseer:2020lwr}.  In Lorentz-covariant formulations,
nonlocality naturally exists both along spatial coordinates and along the time
coordinate.  While spatial nonlocality is an intriguing feature,  nonlocality
along the time direction is considered problematic.
The theory could have trouble with unitarity and may have ghost states.

The possible complications of nonlocality along time
appear even at the classical level.  A theory with nonlocal time dynamics
lacks an initial value problem, and lacks a Hamiltonian formulation~(see, however,
the alternatives considered in~\cite{llosa,Gomis:2003xv,
Tomboulis:2015gfa}).  These
are serious complications, as it becomes unclear in what sense the theory is
predictive---it could take an infinite amount of initial data to evolve
a configuration.   Moreover, the theory could exhibit acausal behavior and theories whose equations of motion have a {\em finite} number
$N >2$ of time derivatives are known to have Ostrogradski's instability, a classical
version of the troubles associated with ghost states.

There are two points, however, that are often emphasized and suggest that
there the time-nonlocality of string theory need not be problematic.   One is that the
Ostrogradski analysis of instabilities does {\em not} apply to theories with infinite number
of derivatives---for all we know theories with infinite number of time derivatives could be fully consistent.  Second, string theory has a light-cone string field theory formulation in which the dynamics along the light-cone time $x^+$ is conventional and {\em local}.   This string field theory has an initial value problem and a Hamiltonian formulation, while
 still exhibiting nonlocality along the spatial directions, including particularly intricate nonlocalities along the light-cone direction $x^-$.  These nonlocalities make it sometimes difficult to use light-cone string field theory for computations that
involve zero-momentum fields.

Most physicists believe that covariant string field theory is consistent, even though it does not make the existence of an initial value problem manifest. For that purpose,
the physically equivalent light-cone formulation is available.
In this viewpoint,
one views both formulations as consistent and equivalent.
Alternatively,
as argued by Eliezer and Woodard over thirty years ago~\cite{Eliezer:1989cr}, it is conceivable that the light-cone formulation has filtered out solutions of the covariant theory associated with higher time derivatives.  In this second viewpoint, light-cone string field theory is physically Lorentz invariant and equivalent to the covariant theory {\em only} perturbatively.   While we do not try to resolve these conflicting viewpoints (for some discussion of the issues,~see\cite{Erler:2004hv}),
we note that no problems with causality have been identified in string field theory.  Moreover, building on results obtained in a number of papers, Erler and Matsunaga~\cite{Erler:2020beb} have largely  established the relation between covariant and light-cone open string field theories.  The subject is being steadily demystified.

In this paper, we largely focus on open string field theory truncated to the
tachyon field.  This is a nonlocal theory, with a unit-free, real
nonlocality parameter $\xi$ that controls the derivative expansion of the theory.  With unit-free fields and derivatives, the Lagrangian takes the form
\be
L \ =  \  \tfrac{1}{2}   \, \phi \bigl(\partial^2 +  1\bigr) \phi
+ \tfrac{1}{3} \bigl( e^{\xi^2 \partial^2} \phi \bigl)^3  \,.
\ee
Here $\partial^2 = - \partial_t^2 + \nabla^2$.  We analyze this field theory in the spirit of effective field theory,
working in a series expansion in $\xi^2$, or equivalently, an expansion in the number of derivatives~\cite{Criado:2018sdb}.  We use field redefinitions to see exactly to what degree one can eliminate
higher derivatives from the theory.  This approach was partly
 inspired by the analysis of Hohm and Zwiebach~\cite{Hohm:2019jgu} who found a canonical
 presentation for the most general duality covariant $\alpha'$-corrected action
 for cosmological solutions--that is, time dependent solutions.  After a set of field redefinitions, the action for the metric, $b$-field, and the dilaton is particularly simple.
 The metric and $b$-field appear only within the generalized metric, and all the nonlocality of the $\alpha'$ expansion can be effectively eliminated to find an action with only first-order time derivatives acting on the generalized metric.  One wonders what is the most general class of nonlocal theories for which such a transformation is possible. The present paper presents a detailed analysis of this question for nonlocal scalar field theory of the type inspired by string theory.  Our methods, however, can be easily extended  for scalar theories with multiple fields, arbitrary potentials, and arbitrary nonlocal interactions.  A small parameter must be identified
 to carry out the recursive elimination of higher derivatives.

 This paper can also be viewed as an alternative analysis of the physics
 of the closely related p-adic string models discussed in the early 2000's
 in order to understand the rolling of the open string tachyon in the process
 of D-brane decay~\cite{Moeller:2002vx, Fujita:2003ex,Moeller:2003gg,Yang:2002nm,Minahan:2001pd}.
 The time-dependence of the rolling field exhibits wild,
 ever growing oscillations, instead of a steady rolling towards the tachyon vacuum.
 It was also noted in~\cite{Moeller:2002vx} that arbitrary initial conditions
 are in fact inconsistent in the infinite derivative theory.  An analysis of the
 initial value problem both for p-adic strings and for the tachyon model
 was done in~\cite{Barnaby:2007ve,Gomis:2003xv}.  For the tachyon model,
 these authors argued that the physical phase space
  at the unstable vacuum is two dimensional---this is a picture
  consistent with our results here.  For the p-adic model, the unstable vacuum
  phase space is infinite dimensional and the canonical rolling solutions of~\cite{Moeller:2002vx} are selected due to an implicit initial condition.

  Our work is a detailed analysis of nonlocal field theory in the framework
  of effective field theory.   We first analyze the solely time-dependent theory
  and then the general spacetime-dependent theory.  While there seems to be an intuition that the nonlocality can be removed perturbatively and thus no violation of
  causality will be observed within the resulting formulation,  we have
  seen no detailed analysis of this procedure, and our work shows a number
  of interesting complications and subtleties.
   As a matter of definition, we have a {\em higher derivative} when
   two or more derivatives act on a single field.   A  higher-derivative {\em term}
   is one in which there is at least one higher derivative, after trying to eliminate it
   by making use of   integration by parts, while discarding total derivatives.
   Thus, for example, $\partial^2 \phi$
   is a higher derivative, but $\phi \partial^2 \phi \simeq - \partial \phi \partial \phi$ is
   not a higher-derivative term because it can be expressed with single derivatives
   after integration by parts (we use $\simeq$
   to sometimes emphasize equality up to total derivatives).
   A term with $k$ fields and more than $k$ derivatives is clearly a higher-derivative
   term.
      In a theory without higher-derivative terms we have no nonlocality.
   Eliminating all higher-derivative terms in a nonlocal theory amounts to eliminating
   the nonlocality of the theory.

 For the time dependent theory we find:

 \begin{enumerate}

 \item
 All higher-derivative terms can be removed by field redefinitions.
 Even more is true:  all terms of the form $\phi^k (\p_t \phi)^n$ with integers
 $n>2$ and $k \geq 0$, can also be eliminated.  Note that
 a term for the form $\phi^k \p_t \phi$ is a total derivative.
 The result is that the only appearance of time derivatives in the redefined
 theory is in the canonical kinetic term, proportional to~$(\p_t\phi)^2$.

 \item  Accompanying the kinetic term, there is a potential $\tilde V (\phi; \xi^2)$ that we
 determine perturbatively in $\xi^2$.  This is a scalar field theory in
 canonical form.
   It would be very illuminating to have
 a closed-form expression for the potential $\tilde V(\phi; \xi^2)$.
 This potential, we show,
 has a minimum at the same depth
 as the original potential of the theory.

 \item
 We show  that the potential
 $\tilde V (\phi; \xi^2)$ has a countable set of ambiguities that can be used to
 partially fix its $\xi^2$ expansion.
 This kind of ambiguity originates from a countable set of quasi-symmetries;
 field transformations of the Lagrangian
  that to leading order only change its non-derivative
 part.  Such transformations are especially powerful
 in theories with infinite number of derivatives.

 \end{enumerate}

 To test the redefined theory we examined rolling tachyons in the nonlocal model and considered
 the effect of the field redefinition to a canonical form.  Such rolling process was
 considered in~\cite{Gomis:2003xv}.  A few aspects of our analysis
 are novel:

 \begin{enumerate}

 \item  The rolling tachyon solution in the nonlocal theory, which also shows wild
 oscillations, can be written as a series
    of the form  $\sum_{n=1}^\infty b_n e^{nt}$, with the $b_n$ calculable coefficients.
    Along the lines of Fujita and Hata~\cite{Fujita:2003ex},  we achieved some analytic control over the
    the large $n$ behavior of the $b_n$ coefficients that make a convincing case
    that the series is convergent for any $\xi^2 > 0$, and for all times.
    This solution is most useful for large $\xi^2$.

 \item    We developed an alternative perturbative analysis of the rolling problem,
 using the exact analytic time-dependent rolling of the $\xi=0$ theory as the
 zeroth-order approximation~\cite{Zwiebach:2000dk}.
 Notably, at each order in an expansion in $\xi^2$
 the time dependence of the correction can be given in closed form.
  This solution is most useful for small $\xi$.

 \item   Equipped with the field redefinition that maps the nonlocal time dependent
 theory to a canonical theory, we insert the wildly oscillatory solution of the nonlocal theory
 into the redefinition and find evidence that it is
   mapped to a smooth, conventional
 rolling solution.  As expected classically, the tachyon rolls down, runs over the tachyon vacuum,
 reaches the turning point, and turns around.  Our numerical work, however,
 is not powerful enough to confirm that the solution  returns to the unstable vacuum.

 \item  Ever since the surprising oscillations of rolling tachyons
 in p-adic models and open-string field theory were found~\cite{Moeller:2002vx},
  their relation to the decay of branes~\cite{Sen:2002nu}  and the
 state of `tachyon matter' \cite{Sen:2002in} has been  somewhat opaque.
 Certainly the oscillations persist when writing exact open string field theory
 rolling solutions~\cite{Schnabl:2007az,Kiermaier:2007ba}, including a remarkably
 simple version obtained in~\cite{Kiermaier:2010cf}.   Moreover, it has been shown,
 first on-shell~\cite{Ellwood:2008jh,Kishimoto:2008zj} and
 then off-shell~\cite{Kiermaier:2008qu,Kudrna:2012re}, that the rolling solution belongs to the
 class of solutions for which the associated boundary state is that
 from conformal field theory.  This supports the idea that the wild oscillations
 build the stress-energy tensor of tachyon matter, a
 state with finite energy density but zero pressure.   It would be useful, however, to see more
 explicitly how a wildly oscillating tachyon ends up giving a smoothly decaying pressure,
 especially because at the level
 of field theory, the tachyon models fail dramatically to
 reproduce the behavior of the pressure.
 For other viewpoints and discussion, see~\cite{Ellwood:2007xr,Hellerman:2008wp,Erler:2019xof}.
Our results are puzzling in that the mapping to a canonical
 scalar field theory suggests that the
 rolling solution represents the process that begins and ends at the
 same configuration: possibly a brane at $t=-\infty$ decaying and
 then reconstructing itself for $t= + \infty$.  This would be the bosonic
 string analog of an $S$-brane---a spacelike brane~\cite{Gutperle:2002ai}.
 It is not completely clear that the nonlocal model solution also begins and
 ends at the same configuration; this would require better control over the map
 to the canonical theory.

 \end{enumerate}

 Encouraged by the removal of all higher derivatives from the
 solely time dependent theory,
 we then work with general configurations
 and try to eliminate perturbatively in $\xi^2$ all nonlocality,  along {\em both time and space}.
 This turns out to be impossible; some nonlocality is irreducible---infinite classes
 of higher-derivative terms cannot be redefined away.
 Our results show that:

 \begin{enumerate}
 \item  Removal of {\em all} higher-derivative terms
  is possible up to ${\cal O} (\xi^6)$,
 where we have terms with six derivatives and terms with
 less than six derivatives, arising from previous redefinitions.
 At this order, the kinetic term receives its first correction, proportional to
 the four-derivative term
 $(\partial \phi)^4 =  (\partial_\mu \phi \partial^\mu \phi)^2$.  This term is
 consistent with an initial value problem because at most one time derivative acts
 on any field---it is not a higher-derivative term.

  \item  At order ${\cal O} (\xi^8)$, corresponding to eight derivatives or less,
  we encounter the six-derivative structure $(\partial \phi)^2 \partial^2 (\partial \phi)^2$ that cannot be redefined away while keeping manifest Lorentz covariance.
  Faced with this, we settle for
  the removal of terms with higher-order {\em time} derivatives, in order to have an initial value problem.  This can be done, but requires breaking
 manifest Lorentz covariance.  The result
 is a theory where fields are acted by at most one time derivative but any number
 of spatial derivatives---the spatial nonlocality is not removable.  We give an
 algorithm to carry out the procedure to arbitrary order.

  \item   A light-cone formulation of the scalar field theory is also possible.
  Such a formulation uses light-cone coordinates $x^+ = \tau$, for
  time and $x^-$ for space, along with transverse spatial coordinates~${\bf x}_T$.
  In a light-cone formulation time derivatives only appear in the kinetic term and
  in fact just one time derivative appears.  Starting with the covariant analysis,
  the first obstacle is the $(\partial \phi)^4$ term mentioned above, being an
  interaction with $\partial_\tau$ derivatives.  A field redefinition is used to
  eliminate these derivatives, at the cost of introducing $x^-$ nonlocalities.
  Again, we explain how this procedure
  can be carried out to arbitrary order.

 \end{enumerate}

 A difficult but important question is the relation between the original
 nonlocal theory and the versions discussed above that have no higher-order
 time derivatives and thus have a well-defined initial value problem.
 Do the field redefinitions, constructed perturbatively in the nonlocality
 parameter, define an invertible map from the original theory to the
 local theory?  If so, the construction we have given shows how to make
 the absence of causality violations manifest, at least classically.
 If the map is not truly invertible, there could exist difficulties
 with the nonlocal theory that are not visible in the redefined formulation.

 Ideally, one would want to assess causality directly from the nonlocal
 quantum theory.  A simple test of micro-causality requires showing that commutators
 of field operators vanish for spacelike separations.  But given the absence
 of a standard Hamiltonian formalism for the nonlocal theory, it is not clear how to define
 field operators.   The Bogoliubov
 causality condition~\cite{bogoliubov:tp}, which applies in the path-integral
 formulation,  may be a viable tool for this problem~\cite{Tomboulis:2017rvd}.
 In this paper, we assess causality of the
 nonlocal \emph{classical} theory directly by testing for
 superluminality in dispersion relations~\cite{Leander:1996:RelationWavefrontSpeed,Aharonov:1969:SuperluminalBehaviorCausality, Shore:2003:CausalitySuperluminalLight, Adams:2006:CausalityAnalyticityIR, Shore:2007:SuperluminalityUVCompletion}.   We find no evidence of acausal
 behavior in this situation.

 This paper is organized as follows.  Section~\ref{themodandthe} introduces
 the nonlocal model and the nonlocality parameter, discussing how it
 arises in string field theory.  The analysis of solely time dependent
 backgrounds and the redefinition into a theory local in time
  is given in Section~\ref{redthepurtimdepthe}.  Section~\ref{sec:rolling} revisits
  the rolling tachyon calculation in the nonlocal theory and discusses  how it
  is mapped to conventional rolling in the local version of the theory.
  The discussion of field redefinitions for general spacetime dependent
  configurations is given in Section~\ref{sec:redef-cov}.  We see why it is
  necessary to break manifest Lorentz covariance and how to find a light-cone
  formulation of the theory.  Our analysis of causality via dispersion relations is given in Section~\ref{causfromsup}.

\section{The model and the nonlocality parameter}\label{themodandthe}

The scalar field theory model we wish to study is motivated by
bosonic open string field theory (OSFT) truncated to the tachyon field $\phi(x)$.
Scaling  $\phi$ to have no units, the tachyon potential
$V(\phi)$, up to an overall multiplicative constant
that carries the appropriate units, is given by~\cite{Sen:1999nx}
\be
V(\phi) \sim  -\tfrac{1}{2} \phi^2 - \tfrac{1}{3} \gamma^3  \phi^3 \, , \ \ \gamma = \tfrac{3\sqrt{3}}{4} \simeq 1.299\,.
\ee
The constant $\gamma$ is determined by the geometry of the Witten vertex.
Moreover,  $1/\gamma \simeq 0.7698$ is the parameter that controls the level expansion in cubic OSFT. Different choices for the cubic open string vertex, such as the hyperbolic vertex~\cite{Cho:2019anu,Firat:2021ukc}, would produce different values for $\gamma$,
along with higher-power nonlocal interactions for the open string tachyon $\phi$. In this paper, we are going to exclusively focus on the cubic interaction provided by the cubic OSFT truncated to the tachyon field.

The Lagrangian $L$ for this field, again up to an overall multiplicative constant, is given by
\be
L  \sim   \tfrac{1}{2} \phi \bigl( \alpha' \partial^2 + 1\bigr) \phi + \tfrac{1}{3}  \bigl( e^{\log \gamma (\alpha' \partial^2 +1)} \phi \bigl)^3\,.
\ee
We use a metric diag$(-1, 1, \ldots , 1)$ and thus  $\partial^2 = - \partial_t^2 + \nabla^2$.  The kinetic term above
fixes the mass via $-\alpha' p^2 +1=0$, giving $p^2 = 1/\alpha'$.  This is a tachyon, since the
familiar mass-shell condition is $p^2 = -m^2$.   It is also clear that for constant $\phi$ we find $L = -V$,
as expected.  The nonlocality appears in the cubic interaction; each field is acted by an exponential that includes
the Lorentz-covariant Laplacian~$\partial^2$.  The exponential vanishes for on-shell tachyons, but controls the off-shell
behavior of the theory.  The off-shell behavior
encodes important physics, such as
 the depth of the tachyon potential
and the dynamics of rolling tachyons.

To isolate a unit-free nonlocality parameter we use the magnitude of the scalar field mass to introduce unit-free
coordinates $\tilde x$ and unit-free derivatives $\tilde \partial$.  For the above tachyon, the magnitude of the mass-squared is $1/\alpha'$, so we set
\be
\sqrt{\alpha'} \partial = \tilde \partial \,.
\ee
With unit-free derivatives the Lagrangian becomes
\be
L  \sim   \tfrac{1}{2} \phi \bigl( \tilde\partial^2 + 1\bigr) \phi + \tfrac{1}{3}  \bigl( e^{\xi^2
(\tilde\partial^2 +1)} \phi \bigl)^3\,, \ \ \ \xi^2 = \log \gamma \,,
\ee
where we have the {\em nonlocality parameter} $\xi^2$, defined to be the constant that multiplies
$\tilde\partial^2$ in the exponential.  This
 parameter controls the amount of delocalization of
each field at the interaction.  For $\xi=0$ we have a completely local theory.  The delocalization
scale equals $\xi$ multiplied by the Compton wavelength of the particle represented by field.
For the cubic OSFT tachyon, the nonlocality parameter is
\be
\xi^2 = \log \tfrac{3\sqrt{3}}{4}  \simeq 0.26162\quad \implies \quad  \xi  \simeq 0.5115\,.
\ee

With this example in mind, we can introduce the more general model we will consider now.
 The Lagrangian,
again up to a constant that carries the units, is given by
\be
\label{jnbb}
L \sim  \tfrac{1}{2} \phi \bigl( \partial^2 + m^2\bigr) \phi + \tfrac{1}{3} g \bigl( e^{a^2 (\partial^2 + m^2)} \phi \bigl)^3\,.
\ee
Here $a$ and $m$ are real constants, so that $a^2 >0$ and $m^2 >0$, and $g$ is a coupling constant.
The constant $a$ has units of inverse mass, so that $a^2 m^2$ is unit-free.
The Lagrangian above represents a tachyon field ($p^2 = m^2$).
The model is consistent with string field theory delocalization~\cite{Pius:2016jsl}.
Indeed, in momentum space the exponential factor acting
on each field participating in the interaction is
\be
e^{-a^2 (p^2 + m^2 )}\  = \  e^{a^2 (p^0)^2 }  e^{- a^2 {\bf p}\cdot {\bf p}}  e^{-a^2 m^2 }\,.
\ee
This suppresses field configurations with large spatial momentum
$|{\bf p}|$, while amplifies those with large
energy $|p^0|$.

While (\ref{jnbb}) is the general model, a series of scalings of fields and coordinates can be used
to put the theory in canonical form.  Factoring out the constant part of the exponential factors, we see that
\be
\label{clbb}
L \ \sim \  \tfrac{1}{2} \phi \bigl( \partial^2 + m^2\bigr) \phi + \tfrac{1}{3} \tilde g \bigl( e^{a^2 \partial^2} \phi \bigl)^3\,, \ \ \ \tilde g = g e^{ 3a^2 m^2}\,.
\ee
Letting $\phi \to \phi/\tilde g$ we obtain
\be
\label{clbbs}
\tilde g^2 L \ \sim \  \tfrac{1}{2} \phi \bigl( \partial^2 +  m^2\bigr) \phi + \tfrac{1}{3} \bigl( e^{a^2 \partial^2} \phi \bigl)^3\,.
\ee
This is a convenient form for analysis, just depending on $m$ and $a$.
It is possible to simplify further the theory using the mass $m$ to introduce unit-free derivatives $\tilde \partial$
as $\partial = m \tilde \partial$.  This gives
\be
\label{clbbsinterm}
\tilde g^2 L \ \sim \  \tfrac{1}{2}   m^2 \, \phi \bigl( \tilde\partial^2 + 1\bigr) \phi + \tfrac{1}{3} \bigl( e^{a^2 m^2 \tilde\partial^2} \phi \bigl)^3\,.
\ee
Now, letting $\phi \to m^2 \phi$ we obtain
\be
\label{clbbsansinterm}
\tfrac{\tilde g^2}{m^6} \,  L \ \sim \  \tfrac{1}{2}   \, \phi \bigl( \tilde\partial^2 + 1\bigr) \phi + \tfrac{1}{3} \bigl( e^{\xi^2 \tilde\partial^2} \phi \bigl)^3\,, \ \ \ \xi^2 = a^2 m^2 \,.
\ee
The theory only depends on the nonlocality parameter $\xi$ now.   At this points, for most intents and purposes,
we can forget about the constants multiplying $L$ and delete all tildes, to find our final, simplest form
of the nonlocal theory:
\be
\label{clbbsnsdlinterm}
\boxed{\phantom{\Biggl(} \
L \ =  \  \tfrac{1}{2}   \, \phi \bigl(\partial^2 +  1\bigr) \phi
+ \tfrac{1}{3} \bigl( e^{\xi^2 \partial^2} \phi \bigl)^3 \,. \ }
\ee
Here both the field $\phi$ and the derivatives are unit-free.
We will call this the nonlocal tachyon theory.
 If we are interested in a nonlocal
theory of an (ordinary) massive scalar, then we must change the $+1$ in the above
kinetic term for $-1$.  Operationally this is achieved by letting $\phi\to -\phi$,
$\partial^2 \to - \partial^2$, $\xi^2 \to - \xi^2$, and then changing also the sign
of the Lagrangian.  These transformations can be applied to the various forms
$L$ takes after field redefinitions.

The p-adic string models~\cite{Brekke:1987ptq} are closely related to the field theory above:
\be
L_{\rm p-adic} = -\tfrac{1}{2}  \, \phi  \, p^{-\tfrac{1}{2} \partial^2 } \phi  + \tfrac{1}{p+1}\,   \phi^{p+1}\,,
\ee
where $p$ is a prime number. By redefining the field we can move the nonlocal factor to the interaction, where it then takes the form
\be
L_{\rm p-adic} = -\tfrac{1}{2}  \, \phi \phi  + \tfrac{1}{p+1}\,  \bigl(  \, p^{\tfrac{1}{4} \partial^2 } \phi \bigr)^{p+1}.
\ee
We identify the nonlocality parameter $\xi_p$ parameter of the p-adic model:
\be
\xi_p^2 = \tfrac{1}{4} \log p \ \  \to \ \   \xi_p = \tfrac{1}{2} \sqrt{\log p}  \,.
\ee
For $p=2$ we get $\xi_2 = \tfrac{1}{2} \sqrt{\log 2} \simeq 0.416277$.  The higher the value of $p$
the larger the value of $\xi_p$ and the more nonlocal the theory is.
Models sharing the same kinetic term as the p-adic string but involving more general interactions have been introduced to describe the 1PI effective action of neural networks~\cite{Halverson:2020trp,Erbin:2021kqf}. In this case, the nonlocality parameter is proportional to the inverse of the standard deviation of the weight distribution.

\section{Redefining the purely time-dependent theory} \label{redthepurtimdepthe}

In this section, we will focus on the time-dependent dynamics of the theory, assuming the field
has no space dependence.  For this, we start with
 the form of the Lagrangian in~(\ref{clbbsnsdlinterm}).  With spatial derivatives vanishing
when acting on $\phi$, the Lagrangian becomes
\be
L \ =  \  -\tfrac{1}{2}   \, \phi \bigl(\partial_t^2 - 1\bigr) \phi
+ \tfrac{1}{3} \bigl( e^{-\xi^2 \partial_t^2} \phi \bigl)^3\,.
\ee
Recall that in this presentation, fields, coordinates, and $\xi$ are all unit-free.
The potential $V(\phi)$ for this theory is, as usual, minus $L$ evaluated for
constant fields:
\be
V(\phi) = -\tfrac{1}{2} \phi^2  - \tfrac{1}{3} \phi^3 \,.
\ee
The goal of this section is to show that after expanding this
 theory in powers
of the nonlocality parameter $\xi^2$, we can use  a sequence of field redefinitions
to cast the theory in a form that only contains the conventional kinetic term above and
a $\xi^2$-dependent {\em potential}.

The expansion of the Lagrangian
 goes as follows:
\be
L =  L_0 + \xi^2 L_2  + \xi^4 L_4  + \xi^6 L_6 + \mathcal{O} (\xi^8)  \,,
\ee
with the first few terms given by
\be
\begin{split}
L_0 \ = &  - \tfrac{1}{2}  \phi (\partial_t^2 -1)  \phi + \tfrac{1}{3} \, \phi^3 =
\tfrac{1}{2} (\partial_t \phi)^2 + \tfrac{1}{2} \phi^2 +   \tfrac{1}{3} \, \phi^3 \,, \\
L_2 \ = &  -  \phi^2 \partial_t^2 \phi\,,  \\
L_4 \ = &  \phantom{-}  \tfrac{1}{2} \phi^2  \partial_t^4 \phi +  \phi (\partial_t^2 \phi)^2 \,.
\end{split}
\ee
Let us show the first step in the procedure to eliminate all the time derivatives
from the interactions.  For this we imagine letting
\be
\phi \to \phi + \delta \phi\, ,  \  \hbox{with} \ \ \delta \phi \sim {\cal O} (\xi^2) \,.
\ee
Then we have
\be
L[\phi + \delta \phi] = \ L_0 - \delta \phi \, ( \partial_t^2 \phi + V' (\phi)) -
\xi^2 \phi^2 \partial_t^2 \phi  + {\cal O} (\xi^4) \,.
\ee
Here prime on the potential denotes the derivative with respect to $\phi$. Note that because $\delta \phi$ is itself of order $\xi^2$ we only need the linear
variation of $L$ if we are ignoring terms of order $\xi^4$ and higher.
Choosing
 $\delta \phi = -\xi^2 \phi^2$ the terms with derivatives cancel
and we get
\be
L[\phi + \delta \phi] = \ L_0 +  \xi^2 \phi^2 V'(\phi)  + {\cal O} (\xi^4) \,.
\ee
The new potential $\tilde{V}$, correct to order $\xi^2$, is therefore
\be \label{eq:tildepotential}
\tilde{V}(\phi;\xi^2) \equiv
\tilde{V}_0 (\phi) + \xi^2 \tilde{V}_2 (\phi) + {\cal O} (\xi^4)
\  = \  V(\phi) - \xi^2 \phi^2 V'(\phi) + {\cal O} (\xi^4)\,,
\ee
so that
\be
\tilde{V}_0(\phi) = V(\phi) =  -\tfrac{1}{2} \phi^2  - \tfrac{1}{3} \phi^3 \quad \text{and} \quad \tilde{V}_2(\phi) = -\phi^2 V'(\phi) = \phi^3 + \phi^4.
\ee
More explicitly,
\be
\tilde{V}(\phi;\xi^2) =  -\tfrac{1}{2} \phi^2 + \bigl(  - \tfrac{1}{3} + \xi^2 \bigr)  \phi^3 + \xi^2  \phi^4 +  {\cal O} (\xi^4) \,.
\ee
Next, we will show that field redefinitions allow to turn the whole set of higher-derivative interactions into a potential order-by-order in $\xi^2$.

\subsection{Field redefinitions}
In order to show that the purely time-dependent nonlocal theory is equivalent to a standard two-derivative theory with a $\xi$-dependent potential,  we use an inductive argument.
We assume that after field redefinitions and integration by parts the
Lagrangian, now denoted as $\tilde{L}$ to differentiate from $L$ above, has been put in the form
\be
\label{lag-recprod}
\tilde{L} = \  L_0 \,  - \, \xi^2  \tilde V_2   -  \cdots   - \xi^{2k -2}  \tilde V_{2k- 2}\,
+  \xi^{2k} L'_{2k}  +   {\cal O} ( \xi^{2k+2}) \,.
\ee
Here $\tilde V_2, \cdots , \tilde V_{2k-2}$ are potentials, thus have no fields with time
derivatives.  We use $L'_{2k}$ to denote the new
term at order $\xi^{2k}$; it differs from $L_{2k}$ by the terms contributed by the field redefinitions
needed so far to obtain the potentials.
These terms do not all have $2k$ derivatives
because the field redefinition effectively replaced
$ \partial_t^2 \phi \to  - V'(\phi) = \phi +  \phi^2$ at previous orders and reduced the number of derivatives by two each time it is used.
We now want to show that we can turn each term with derivatives in $L'_{2k}$ into a term without derivatives.  Having shown that derivatives can be
removed from $L_2'=L_2$,
 this proves, by induction,
 that the redefinition of the
purely time-dependent nonlocal theory results
in a potential.

Under any variation $\phi \to \phi + \delta \phi$ we have
\be
\tilde{L}[\phi] \ \to \ \tilde{L}[\phi + \delta \phi]   =  \tilde{L}[\phi] + \Delta \tilde{L}\,.
\ee
Here, $\delta \phi$  will have a series
expansion in the parameter $\xi$.
For our Lagrangian in~(\ref{lag-recprod}) we have
\be
\label{varlagrangian}
\begin{split}
\Delta \tilde{L} =& \  - \delta \phi \Bigl[ \partial_t^2\phi
+ V'(\phi)
\Bigr]  \\
&  - \tfrac{1}{2}  \delta \phi ( \partial_t^2 - 1 ) \delta\phi  + \phi (\delta \phi)^2 + \tfrac{1}{3}
(\delta \phi)^3  \\
&  \ - \xi^2 \Delta \tilde V_2   -  \ \cdots  - \xi^{2k-2} \Delta \tilde V_{2k-2} + \xi^{2k} \Delta L'_{2k}
+  {\cal O} (\xi^{2k+2})\,,
\end{split}
\ee
where  $\Delta \tilde V_{2i}$ represents the exact variation of $\tilde V_{2i}$.
The goal now is to produce a field redefinition
\be
\phi \to \phi + \xi^{2k} \delta_{2k} \phi\,,
\ee
such that we turn the $L'_{2k}$ Lagrangian into a potential $\tilde V_{2k}$ that doesn't contain any derivatives.
Note from (\ref{varlagrangian}) that to order $\xi^{2k}$ only the first line contributes.
Terms with $(\delta \phi)^2$ are of order $\xi^{4k}$, $(\delta \phi)^3$ are of order $\xi^{6k}$ and terms that have a power
of $\xi^2$  or higher in their prefactor also lead to terms of order higher than $\xi^{2k}$.
Therefore,
\be
\label{varlagrangian-jn}
\begin{split}
\Delta \tilde{L} =& \  - \xi^{2k} (\delta_{2k} \phi) \Bigl[  \partial_t^2\phi
+ V'(\phi)
\Bigr]
+  {\cal O} (\xi^{2k+2})\,.
\end{split}
\ee
As a result
\be
\label{cldlcsvg}
\begin{split}
\tilde{L}   + \Delta \tilde{L}  =
&\ \  L_0 \  - \xi^2  \tilde V_2   -  \cdots   - \xi^{2k -2}  \tilde V_{2k- 2}\\
&  +  \xi^{2k}\Bigl(  \underbrace{L'_{2k} -  (\delta_{2k} \phi)
\bigl[ ( \partial_t^2\phi   + V'(\phi)
\bigr] }_{F_{2k}}\Bigr)
+   {\cal O} (\xi^{2k+2}),
\end{split}
\ee
where $F_{2k}$ is defined as shown above.

We wish to show that $\delta_{2k} \phi$ can be chosen to eliminate derivatives in $L'_{2k}$.
To see what we must do, consider a term $X$ in $L'_{2k}$ that has been put in the following form, possibly after some integration-by-parts,
\be
X = (\partial_t^2 \phi) X_0 [\phi, \partial_t\phi ] \, \in L'_{2k}\,,
\ee
where $X_0[ \phi, \partial_t \phi]$, denotes a function that depends on $\phi$ and
derivatives of $\phi$, possibly of many orders.
Denoting other terms in $L'_{2k}$ by dots,  we  have
\be
F_{2k} =  \cdots +  (\partial_t^2 \phi) X_0 [\phi , \partial_t\phi] -  (\delta_{2k} \phi) \bigl[
\partial_t^2 \phi  + V'(\phi)
\bigr]\,.
\ee
We now choose
\be \label{eq:2kfieldredef}
\delta_{2k} \phi =  X_0[\phi, \partial_t\phi]    + \delta'_{2k} \phi\,,
\ee
where $\delta'_{2k} \phi$  denotes additional redefinitions that may be needed.
With this choice,
\be
F_{2k} = \cdots- V'(\phi) X_0[\phi, \partial_t\phi]  -  (\delta'_{2k} \phi)
\bigl[ \partial_t^2\phi  + V'(\phi) \bigr]\,.
\ee
The rule is then clear, when eliminating a term of the form $(\partial_t^2 \phi) X_0$
with the field redefinition~\eqref{eq:2kfieldredef} we induced
a term of the form $-V'(\phi) X_0$:
\be \label{eq:secondderrule}
(\partial_t^2 \phi) X_0[\phi, \partial_t\phi] \ \to \  - V'(\phi) \,
X_0[\phi,\partial_t\phi]  \, .
\ee
We use the symbol $\to$ to denote terms that are equivalent using field redefinitions.
Note that the number of derivatives in this term decreases by two with this replacement.
Of course, the term on the right may need further elimination if $X_0$
still contains time derivatives.  If that is the case,  we must show that we can write
\be \label{eq:Form}
-V(\phi)   X_0[\phi , \partial_t\phi] \simeq  (\partial_t^2 \phi) X_1 [\phi, \partial_t\phi] \,,
\ee
for some $X_1[\phi, \partial_t\phi]$, possibly after some integration-by-parts.
This term can now be removed by choosing
\be
\delta_{2k} \phi \, = \,  X_0[\phi, \partial_t\phi]  + X_1[\phi, \partial_t\phi] +  \delta''_{2k} \phi\,,
\ee
and the procedure continues if $X_1$ still contains
derivatives.  We can recursively eliminate all higher derivatives by this procedure as long as we establish that writing higher-derivative terms of the
form $(\partial_t^2 \phi)  X [\phi, \partial_t\phi]$ is always possible. We will show that this is indeed the case in general in the following subsection.

\subsection{The recursive argument} \label{therecarg}

We now show how to recursively eliminate all derivatives
from a term.
For this consider a general term  $T \in L'_{2k}$.  If the term has explicit
factors of $\partial_t^2 \phi$ these can be eliminated as explained above
by field redefinitions that effectively replace each factor of $\partial_t^2 \phi$
by $-V'(\phi)$.  Therefore, we can assume that the general term $T$ can be
written without any second derivatives of $\phi$:
\be
T =  (\partial_t^{k_1} \phi ) \ (\partial_t^{k_2} \phi) \cdots (\partial_t^{k_\ell} \phi) \  (\partial_t \phi)^r \ \phi^s\,.
\ee
Here $r, s\geq 0$ are integers.  Moreover, all $k_i$'s are integers larger than 2.
We order the $k$'s as follows
\be
3 \leq k_1 \leq k_2 \leq \cdots \leq k_\ell  \,.
\ee
We say that the term $T$ is a term of {\em index} $\ell$.  This means that there are $\ell$ factors
with three or more derivatives on a field.
We also call the $k_1$ the {\em lowest order} of $T$.

We are going to show that a series of steps can turn $T$ into a set of
terms of index $\ell -1$, thus reducing the index by one unit.
To do this we first show how
to reduce the lowest order recursively.
Consider $T$ and
integrate by parts a time derivative acting on the first term:
\be
\begin{split}
T \simeq & \   -(\partial_t^{k_1-1}\phi )  \partial_t\bigl[ \ (\partial_t^{k_2} \phi) \cdots (\partial_t^{k_\ell} \phi) \  (\partial_t \phi)^r \ \phi^s \bigr] \\
= & \ -(\partial_t^{k_1-1}\phi )  \partial_t\bigl[ \ (\partial_t^{k_2} \phi) \cdots (\partial_t^{k_\ell} \phi)\bigr] \  (\partial_t \phi)^r \ \phi^s \\
& \  + r \, (\partial_t^{k_1-1}\phi )   \ (\partial_t^{k_2} \phi) \cdots (\partial_t^{k_\ell} \phi) \
(\partial_t \phi)^{r-1}  ( \partial_t^2 \phi) \ \phi^s\\
& \  + s\, (\partial_t^{k_1-1}\phi )   \ (\partial_t^{k_2} \phi) \cdots (\partial_t^{k_\ell} \phi) \
(\partial_t \phi)^{r+1}  \phi^{s-1}.
\end{split}
\ee
On the last right-hand side we have three expressions on three lines.
On the first line we have a collection of terms
obtained by acting with the derivative on the bracket $[ \cdots ]$.
All the terms of the first line
have index $\ell$ but lowest order reduced by one unit.   The same is true for the term
on the third line.  On the second line we have a $(\partial^2_t \phi)$
which can be replaced by $-V'(\phi)$,
yielding a term with index $\ell$ and lowest order reduced by one unit.
This shows we can lower the lowest order recursively.
Assume we have lowered the lowest order down to three.  We now find
that an additional step results on the lowering of the index.
Indeed, consider a general term $T'$ of index $\ell$ whose lowest order is three and integrate by parts as follows
\be
T'
=  (\partial_t^{3} \phi) \ (\partial_t^{k_2} \phi) \cdots (\partial_t^{k_\ell} \phi) \  (\partial_t \phi)^r \ \phi^s
\simeq  -(\partial_t^2\phi )  \partial_t \Bigl[  (\partial_t^{k_2} \phi) \cdots (\partial_t^{k_\ell} \phi) \  (\partial_t \phi)^r \ \phi^s \Bigr] \,.
\ee
It is clear that upon the replacement $\partial_t^2 \phi \to -V'(\phi)$ outside the square brackets, and subsequent action of the derivative on the square bracket along with replacing resulting $\p_t^2 \phi$,
all that
is left are terms of index $\ell-1$.   This shows that the index can be reduced
recursively down to zero.

Having shown that $T$ can be transformed by field redefinitions into terms
of index zero, the general term we must consider now contains only powers
of first derivatives of the field and powers of the field:
$(\pt \phi)^{2p} \phi^q$,
with $p,q$ nonnegative integers.
Note that the number of derivatives is always even
because all terms in the Lagrangian had even number of derivatives and we reduce
the number of derivatives by two with each field redefinition.
We  now have
\be
\begin{split}
(\pt \phi)^{2p} \phi^q = &\  \  \pt \phi ( \pt \phi)^{2p-1}  \phi^q
\simeq  \ -  \phi   \pt  \bigl[  (\pt\phi)^{2p-1}  \phi^q \bigr]\,,
\end{split}
\ee
where we separated out one of the first derivatives and then integrated by parts.
Evaluating it we get
\be
\begin{split}
(\pt \phi)^{2p} \phi^q = &\
-  (2p-1) \, \phi   (\pt^2 \phi) (\pt\phi)^{2p-2}  \phi^q  \ - q \phi \, (\pt \phi)^{2p} \phi^{q-1}  \\
= & \ -  (2p-1) \,   (\pt^2 \phi) (\pt\phi)^{2p-2}  \phi^{q+1}  \ - q \, (\pt \phi)^{2p} \phi^{q}\,.
\end{split}
\ee
Moving the second term to the left hand side we find that
\be \label{eq:IntLikeThing}
(\pt \phi)^{2p} \phi^q  = - {2p-1 \over 1 + q} (\pt^2 \phi)  ( \pt \phi)^{2p-2} \phi^{q+1}.
\ee
Letting $\partial_t^2 \phi \to -V'(\phi) = \phi+ \phi^2$,
we can do the replacement:
\be \label{eq:firstderrepl}
\boxed{\phantom{\Biggl(} \ \
(\pt \phi)^{2p} \phi^q   \ \,  \to \, \
- {2p-1 \over 1 + q}   ( \pt \phi)^{2p-2} (\phi^{q+2} + \phi^{q+3} )\,.\ \ }
\ee
This reduces the number of first derivatives by two.
Applied recursively, this converts the term with a power of first derivatives into
a set of terms without derivatives.  This is what we wanted to show.  For arbitrary $T$
the index is lowered to zero, and then, using~(\ref{eq:firstderrepl})
recursively the result is a set of terms without derivatives---that is, a contribution
to the potential. By induction and the argument from previous subsection, we then conclude that eliminating all derivatives, except those in the kinetic term,
is possible to all orders in $\xi^2$. This is what we wanted to establish.

We can summarize the procedure to remove derivatives algorithmically.
The steps go as follows at a given order of $\xi^2$:

\begin{enumerate}
\item Remove all $\pt^2 \phi$ factors by letting $\pt^2 \phi \to -V'(\phi)$.

\item In each term containing at least one derivative of order $\geq 3$,
integrate by parts the lowest higher-order derivative. The result is a collection of monomials.

\item Repeat steps 1 and 2 with all monomials, until the collection of monomials contains no derivatives of order $\geq 3$ and thus, in fact, no derivatives of order greater than one.

\item In each monomial containing first-order derivatives,
use~(\ref{eq:firstderrepl}) recursively until all derivatives
are eliminated.
\end{enumerate}

After employing this algorithm to $\mathcal{O}(\xi^8)$, the potential $\tilde V (\phi; \xi^2)$ we arrive to
is the following:
\be
\begin{split}
\label{eq:PotentialResult}
\tilde{V}(\phi;\xi^2) =   -\tfrac{1}{2} \phi ^2  &  + \left[-\tfrac{1}{3}+\xi^2-\tfrac{3}{2} \xi^4
+2 \xi^6-\tfrac{9}{4}  \xi^8 +\cdots \right] \phi ^3 \\
& + \left[\xi^2 -\tfrac{19}{3} \xi^4+\tfrac{419}{18} \xi^6
-\tfrac{4595}{72}  \xi^8 +\cdots\right] \phi ^4\\
& +\left[-\tfrac{16}{3} \xi^4+\tfrac{517}{9} \xi^6
-\tfrac{12331}{36} \xi^8 +\cdots \right] \phi ^5 \\
& + \left[\tfrac{118}{3} \xi^6-\tfrac{9194}{15}  \xi^8
+\cdots \right]
\phi ^6 \\
& +\left[-\tfrac{15812}{45} \xi^8 +\cdots \right] \phi ^7 + \mathcal{O} (\phi^8) \ .
\end{split}
\ee
The dots in square brackets indicate
$\mathcal O(\xi^{10})$ contributions.
Note the
 simple pattern:  the $\phi^3$
coefficient begins at ${\cal O} (\xi^0)$,  the $\phi^4$
coefficient begins at ${\cal O} (\xi^2)$;  the $\phi^5$ coefficient at ${\cal O}(\xi^4)$, and so on.  We get two extra powers of $\xi$ on the leading term
 for each extra power of $\phi$.  This can be seen
from the removal of the original higher-derivative cubic terms,
recalling that each power of $\xi^2$ comes with two time derivatives  and
$\partial_t^2 \phi \to \phi + \phi^2$, which implies that
the elimination of two derivatives acting on a
field adds one or at most two non-derivative fields.   The two-derivative
term $ \xi^2(\partial_t^2 \phi) \phi^2 \to \xi^2 (\phi+ \phi^2) \phi^2$
generates cubic and quartic potentials, but no higher.   The four-derivative term $\xi^4 (\partial_t^2 \phi) (\partial_t^2 \phi) \phi \to \xi^4 (\phi + \phi^2)^2 \phi$ generates cubic, quartic and quintic potentials, but no higher.  To get the lowest $\xi$ power in the coefficient
of a given power of $\phi$, the $\partial_t^2 \phi \to \phi^2$ part of the replacement is the relevant one.

The field redefinition $\d \phi$ required to get this theory  with potential $\tilde V (\phi; \xi^2)$ is given by:
\be
\begin{split}
\label{eq:FieldRedefBeforeTot}
\d \phi =   -\xi^2 \phi ^2  & + \xi^4 \Bigl[\tfrac{3}{2}  \phi ^2
+\tfrac{13}{3} \phi^3
+(\pt \phi)^2+2 \phi  (\pt^2 \phi) \Bigr] \\
& - \xi^6 \Bigl[ 2 \phi^2 +\tfrac{178}{9} \phi ^3+\tfrac{91}{3} \phi ^4
+\tfrac{4}{3}  (\pt \phi)^2 +\tfrac{46}{3} \phi  \left( \pt \phi \right)^2 +\tfrac{8}{3}  \phi (\pt^2 \phi)\\
& \hspace{1cm}
+18 \phi ^2  (\pt^2 \phi)+\tfrac{4}{3} (\pt^2 \phi)^2 +\tfrac{4}{3}  \left( \pt \phi \right)(\pt^3 \phi )+\tfrac{4}{3} \phi   (\pt^4 \phi)\Bigr] \\
& + \xi^8 \Bigl[ \tfrac{9}{4}  \phi ^2+\tfrac{526}{9}  \phi ^3
+\tfrac{2264}{9}  \phi^4  +\tfrac{1338}{5} \phi ^5
+\tfrac{5}{2} (\pt \phi)^2 +\tfrac{1037}{6}\phi  (\pt \phi)^2
+\tfrac{2605}{6} \phi ^2 (\pt \phi)^2 \\
& \hspace{1cm}
+\tfrac{7}{2}  \phi  (\pt^2 \phi)
+\tfrac{637}{6}  \phi ^2 (\pt^2 \phi)
+\tfrac{671}{3} \phi ^3  (\pt^2 \phi)
+4 (\pt \phi)^2(\pt^2 \phi)
+ (\pt^2 \phi)^2 \\
& \hspace{1cm}
+\tfrac{32}{3} \phi   (\pt^2 \phi)^2
+\tfrac{1}{2}   (\pt \phi)(\pt^3 \phi)
-\tfrac{53}{3} \phi (\pt \phi)(\pt^3 \phi)
+\tfrac{5}{6}  (\pt^3 \phi)^2
+\tfrac{3}{2}  \phi (\pt^4 \phi) \\
& \hspace{1cm}
+\tfrac{35}{2} \phi ^2  (\pt^4 \phi)
+2    (\pt^2 \phi) (\pt^4 \phi)
+ (\pt \phi)(\pt^5 \phi)
+\tfrac{2}{3} \phi   (\pt^6 \phi) \Bigr] + \mathcal{O} (\xi^{10}). \\
\end{split}
\ee
As it turns out, the redefinitions we have considered do not
fix the redefined potential uniquely. We will examine this ambiguity next.

\subsection{Quasi-symmetries and the nonuniqueness of the scalar potential}
\label{nonuniscapot}

Consider a conventional scalar field theory with the canonical 
kinetic term and a potential polynomial in the field, 
restricted to configurations that are only time-dependent.
A redefinition, polynomial in the field and involving 
a bounded number of derivatives, changing the potential,
 would also change the derivative terms in the Lagrangian.  
 There are, however, field transformations
that in fact change the potential while leaving
the kinetic term unchanged at {\em linearized order}. 
We will call these
transformations quasi-symmetries.  In a nonlocal theory,
these quasi-symmetries can be used in a novel way.  The full
nonlinear variation leads to further derivative terms that 
can be eliminated recursively, as we discussed.  The result
is a theory with a different potential but the same derivative
structure---just the canonical kinetic term. 
We will use below quasi-symmetries
to choose a natural form for the potential.

Consider a scalar field theory, still only time-dependent, with a scalar potential.
The Lagrangian~is
\be
L =  -\tfrac{1}{2} \phi \partial_t^2 \phi - V(\phi)\,.
\ee
Now consider the  variation
\be
\label{cl-first-var}
\hat\delta_1 \phi =  \dot \phi ^2  + V(\phi) \,,
\ee
where the dot indicates time derivative.
 The change of the Lagrangian to the first order is
\be
\begin{split}
L + \hat\delta_1 L = & \
L - \delta \phi ( \ddot \phi + V'(\phi))  + {\cal O} ((\delta\phi)^2) \\
= & \  L - (\dot \phi ^2  + V ) ( \ddot \phi + V')   + {\cal O} ((\delta\phi)^2)\,, \\
= & \  L - \dot \phi ^2 \ddot \phi   - V  \ddot \phi -
\dot \phi ^2 V'  -V V'   + {\cal O} ((\delta\phi)^2)\, .
\end{split}
\ee
We claim that the three terms following $L$ are a total time derivative
and therefore we can ignore their variation in the action.  Indeed,
\be
\dot \phi ^2 \ddot \phi  = \tfrac{1}{3} \partial_t  \dot\phi^3  \,, \ \ \
V  \ddot \phi +
\dot \phi ^2 V'  = \partial_t ( V \dot \phi) \,.
\ee
It follows that to first order the field redefinition just changes the potential by
the addition of $V V'$
\be
L + \hat\delta_1 L =  L  - V(\phi) V'(\phi)  + {\cal O} ((\delta\phi)^2) \, ,
\ee
ignoring higher orders. For the theory we are considering,
$V = -\tfrac{1}{2} \phi^2 - \tfrac{1}{3} \phi^3$
and therefore the variation of the potential is
\be
\label{newpot1}
- V V' = \, - \, \bigl( \tfrac{1}{2} \phi^3  + \tfrac{5}{6} \phi^4 + \tfrac{1}{3} \phi^5 \bigr)\,.
\ee
In the context we are working, it is natural to make this redefinition
go accompanied with a power of $\xi$ and an arbitrary constant $c_1$.  We
set
\be
\delta \phi =  \xi^4  c_1 \, \bigl( \dot \phi ^2  + V(\phi) \bigr) \,.
\ee
This variation to leading order affects just the ${\cal O} (\xi^4)$ terms,
adding $(-\xi^4 c_1 V V')$ to the potential.  The choice of power of
$\xi$ is consistent with the pattern noted on the potential $\tilde V$: this term
contributes to the leading coefficient of $\phi^5$ and to subleading coefficients
of $\phi^4$ and $\phi^3$.  A lower power of $\xi$ would have spoiled the pattern;
a higher power of $\xi$ would not.    Notice that with this choice,
the variation of the  ${\cal O} (\xi^2)$
terms in the action starts contributing at order $\xi^6$, and variations proportional
to $(\delta\phi)^2$ are of  ${\cal O}(\xi^8)$.  Thus, this redefinition changes the
potential at order $\xi^4$ and adds other terms, generally with derivatives,
at higher orders, where they will themselves be transformed into higher
order contributions to the potential in the recursive procedure.

Another viewpoint on this ambiguity of the potential emerges when we consider
total derivatives in the action.  These are irrelevant, of course, but, as it turns out,
our algorithm applied to a total derivative term
can produce a potential and an associated field redefinition.
So starting from two Lagrangians differing by a total derivative one can arrive two different potentials.  Since we have started with two physically equivalent theories and
field redefinitions don't change the physics, these two different potentials must describe the same theory.

In order to see this in practice,  consider adding the following total derivative term to the Lagrangian
\be
A_{3} = -\tfrac{1}{3} \pt [ (\pt \phi)^3 ] =  -(\pt \phi)^2 (\pt^2 \phi).
\ee
We can now eliminate the second-order derivative using
 equation~\eqref{eq:secondderrule} and get
\be
A_{3} =  -(\pt \phi)^2 (\pt^2 \phi) \to  - (\pt \phi)^2 (\phi + \phi^2).
\ee
Carrying on, we eliminate first-order derivatives using equation~\eqref{eq:firstderrepl} and finally see
\be \label{eq:TotDerExample}
A_{3} \to  \tfrac{1}{2} (\phi^3 + \phi^4 )  + \tfrac{1}{3} (\phi^4 + \phi^5) =
\tfrac{1}{2} \phi^3 + \tfrac{5}{6} \phi^4 +  \tfrac{1}{3}\phi^5 .
\ee
These are, as anticipated, exactly the same terms we obtained in~(\ref{newpot1}) up to a sign which can be accounted by redefining $-A_3$ instead.

The field variation in~(\ref{cl-first-var}) is the first of a series of transformations
that change the potential.  To see the pattern, let us consider the
next quasi-symmetry.  It happens to be
\be
\label{cl-sec-var}
\hat\delta_2 \phi =  \dot\phi^4  + 3 \dot\phi^2  V + \tfrac{3}{2} V^2 \,.
\ee
Following the previous computation, this variation just changes the potential
to leading order because
\be
\hat\delta_2 L = -(\dot\phi^4  + 3 \dot\phi^2  V + \tfrac{3}{2} V^2 ) (\ddot \phi + V') =
-\tfrac{3}{2} V^2 V'  +  \hbox{total-derivatives} \,.
\ee
This is checked by expansion, with terms combining in pairs to create total
derivatives.   The general quasi-symmetry can be obtained after a bit of work and
reads as follows
\be\label{ser9-terms}
\begin{split}
\hat\delta_n \phi =& \ \  \dot\phi^{2n} \,  + (2n-1) \dot\phi^{2n-2} V  \\
& \hskip 25pt+ (2n-1) (2n-3) \dot\phi^{2n-4}\ \tfrac{1}{2}V^2 \\
&\hskip 25pt + (2n-1) (2n-3)(2n-5) \dot\phi^{2n-6} \ \tfrac{1}{3!}V^3\\
& \hskip 25pt \ \ \vdots  \ \ \ \ \ \ \ \ \vdots \\
&  \hskip 25pt+ (2n-1)!! \ \tfrac{1}{n!} V^n \,.
\end{split}
\ee
The closed form expression is
\be
\boxed{\phantom{\Biggl(} \ \
\hat\delta_n \phi =  \sum_{p=0}^n   \frac{(2n-1)!!}{(2n-1-2p)!!} \ (\dot \phi)^{2(n-p)} \  \frac{V^p}{p!}\,, \ \ }
\ee
with the convention that $(-1)!! = 1$.  The associated contribution
to the potential is then
read from the variation of the Lagrangian:
\be \label{eq:TotalDerCont}
L + \hat\delta_n L =  L -  \tfrac{ (2n-1)!!}{n!} V^n V'  + {\cal O} \left((\delta \phi)^2\right) \,.
\ee

Associated with the classes of scalar potential ambiguities above
we have total derivative terms that, using the algorithm, produce the
same potentials.   Consider the set of total derivatives
\be
A_{p} \equiv -\tfrac{1}{p} \pt [ (\pt \phi)^p ] =  -(\pt \phi)^{p-1} (\pt^2 \phi)\,, \ \ \
p= 3, 5, \ldots \ \ .
\ee
The potential that arises from the quasi-symmetry $\hat\delta_n \phi$ also arises, up to a multiplicative constant,  from the total derivative $A_{2n+1}$.
Attempts to find other classes
of ambiguities did not work. Henceforth, we will assume
these are the only possible ambiguities for the potential.

\subsection{Choices of potentials} \label{chopot}

We now use the ambiguities above
to construct  families of equivalent
potentials where we can make choices for a simpler potential.
For this purpose, and {\em before} performing any field
redefinitions,  we shift the original Lagrangian by adding a series
of total derivatives:
\be \label{eq:TotDerShift}
L \to  L + \sum_{p =3, 5, \dots} \xi^{3p-5} f_p (\xi^2) A_p\,.
\ee
The multiplicative factor $\xi^{3p-5}$ in front of $A_p$
is inserted to preserve the $\xi$ regularities noted in $\tilde V (\phi; \xi^2)$, as discussed
below equation~(\ref{eq:PotentialResult}).
The $f_p (\xi^2)$ are functions assumed to be of the form
\be
f_p (\xi^2) = c_{p,0} + \xi^2 c_{p,2} +  \xi^4 c_{p,4} + \mathcal{O}(\xi^6),
\ee
where $c_{p,i}$ are real constants whose values are unconstrained.
Applying now the algorithm to the shifted Lagrangian~\eqref{eq:TotDerShift},
we find, to $\mathcal{O}(\xi^8)$,
\begin{align*}
\tilde{V}(\phi;\xi^2)
= -\tfrac{1}{2} \, \phi^2
&- \Bigl[
\tfrac{1}{3}
- \xi^2
+ \tfrac{3}{2} \xi^4 (1 + c_{3,0})
- 2 \xi^6
\bigl( 1 + \tfrac{4}{3} \, c_{3,2} \bigr)
+ \tfrac{9}{4} \xi^8
\bigl( 1 + \tfrac{2}{3} \, c_{3,4} \bigr) + \cdots
\Bigr] \phi^3
\\ & -
\Bigl[
- \xi^2
+ \xi^4 \bigl(
\tfrac{19}{3}
+ \tfrac{5}{2} \, c_{3,0}
\bigr)
- \xi^6
\bigl(
\tfrac{419}{18}
+ \tfrac{15}{2} \, c_{3,0}
+ \tfrac{5}{2} \, c_{3,2}
\bigr)
\\ & \hspace{2cm}
+ \xi^8
\bigl(
\tfrac{\num{4595}}{72}
+ \tfrac{45}{4} \, c_{3,0}
+ \tfrac{45}{8} \, c_{3,0}^2
+ \tfrac{15}{2} \, c_{3,2}
+ \tfrac{5}{2} \, c_{3,4}
\bigr) + \cdots
\Bigr] \phi^4
\\ &
-\Bigl[
\xi^4 \bigl( \tfrac{16}{3}
+ c_{3,0} \bigr)
- \xi^6
\bigl(
\tfrac{517}{9}
+ 15 c_{3,0}
+ c_{3,2}
\bigr)
\\ & \hspace{2cm}
+ \xi^8
\bigl(
\tfrac{\num{12 331}}{36}
+ 75 c_{3,0}
+ \tfrac{63}{4} \, c_{3,0}^2
+ 15 \, c_{3,2}
+ c_{3,4}
\bigr) +\cdots
\Bigr] \phi^5
\\ &
- \Bigl[
-\xi^6
\bigl(
\tfrac{118}{3}
+ 7 c_{3,0}
\bigr)
+ \xi^8
\bigl(
\tfrac{\num{9194}}{15}
+ \tfrac{364}{3} \, c_{3,0}
+ 14 \, c_{3,0}^2
+ 7 \, c_{3,2}
+ c_{3,4}
\bigr) + \cdots
\Bigr] \phi^6
\\ &
- \Bigl[
\xi^8
\bigl(
\tfrac{\num{15 812}}{45}
+ \tfrac{164}{3} \, c_{3,0}
+ 4 \, c_{3,0}^2
\bigr) + \cdots
\Bigr] \phi^7 + \mathcal{O} (\phi^8).
\stepcounter{equation}
\tag{\theequation}
\label{eq:Potential}
\end{align*}
As before, the dots in square brackets indicate $\mathcal{O}(\xi^{10})$.
Setting all the constants $c_{p,q}$ to zero reduces this potential back to~\eqref{eq:PotentialResult}.
The field redefinitions in this case are (keeping terms up to $\mathcal{O}(\xi^6)$ for brevity)
\be
\begin{split}
\label{eq:FieldRedef}
\d \phi =   -\xi^2 \phi ^2
&+ \xi^4 \Bigl[
\tfrac{3}{2}(1+c_{3,0})\phi ^2 +\left(\tfrac{13}{3}+c_{3,0}\right) \phi^3 + (1-3 c_{3,0}) (\pt \phi)^2+2 \phi  (\pt^2 \phi)  \Bigr]\\
&- \xi^6 \Bigl[ \left(2 + \tfrac{3}{2} c_{3,2} \right) \phi^2+\left(\tfrac{178}{9}+6c_{3,0}+c_{3,2}\right) \phi ^3
+ \left(\tfrac{91}{3} +5 c_{3,0} \right)\phi ^4
\\& \hspace{1cm}
+\left( \tfrac{4}{3} - 3 c_{3,2} \right) (\pt \phi)^2   + \left( \tfrac{46}{3} - 6 c_{3,0} \right)\phi  \left( \pt \phi \right)^2 + \tfrac{8}{3} \phi (\pt^2 \phi)  +18 \phi ^2  (\pt^2 \phi)
\\& \hspace{1cm}
+\tfrac{4}{3} (\pt^2 \phi)^2  +\tfrac{4}{3} \left( \pt \phi \right)  (\pt^3 \phi ) +\tfrac{4}{3} \phi   (\pt^4 \phi) \Bigr] + \mathcal{O}(\xi^8).
\end{split}
\ee
The $c_{p,q}$ constants define equivalence classes of potentials for the redefined theory. Choosing some values for the constants is choosing a representative
for the potential.

A natural representative is obtained by demanding
that the coefficients of odd powers of $\phi$ be polynomials in $\xi^2$, rather than
never-ending power series.
This condition can be satisfied because the term $A_{2n+1}$ contributes to the potential
powers of $\phi$ that begin with $\phi^{2n+1}$, as one can see from~\eqref{eq:TotalDerCont} easily, where $n \in \mathbb{Z}_{\geq 0}$.
However, this term would be always multiplied with some power of $\xi^2$ and this power would be always greater or equal to $3n-1$ by~\eqref{eq:TotDerShift}. In other words, the total derivative $A_{2n+1}$ would contribute terms of the form $\sim \xi^{2m} \phi^{2n+1}$ for $m \geq 3n-1$ to the potential. Since each of these terms comes with a constant multiplying them (see~\eqref{eq:TotDerShift}), one can set terms of this form in the redefined potential to zero by adjusting the constants. Obviously, this will turn the coefficients of odd powers of $\phi$ to polynomials in $\xi^2$. Note that the coefficients of even powers of $\phi$ would not be constrained in this procedure, they are still
never-ending power series in $\xi^2$.

The first few $c_{p,q}$ that specify this choice are given by
\begin{equation} \label{eq:CanonicalGauge}
\begin{gathered}
c_{3,0}
= -1,
\qquad
c_{3,2}
= -\tfrac{4}{3},
\qquad
c_{3,4}
= -\tfrac{3}{2},
\end{gathered}
\end{equation}
as one can see easily from~\eqref{eq:Potential}. Further specifying $c_{p,q}$ relevant to the order $\mathcal{O}(\xi^{14})$, we find the potential in this choice of representation is given by
\begin{align*}
\tilde{V}(\phi;\xi^2)
= -\tfrac{1}{2} \, \phi^2
&+ \Bigl[
-\tfrac{1}{3}
+ \xi^2
\bigg] \phi^3 \\
& + \Bigl[
\xi^2
- \tfrac{23}{6} \, \xi^4
+ \tfrac{112}{9} \, \xi^6
- \tfrac{400}{9} \, \xi^8
+ \tfrac{\num{5056}}{45} \, \xi^{10}
- \tfrac{\num{30 848}}{135} \, \xi^{12}
+ \tfrac{\num{372 224}}{945} \, \xi^{14}
+ \cdots
\Bigr] \phi^4
\\ &
+ \Bigl[
-\tfrac{13}{3} \, \xi^4
+ \tfrac{370}{9} \, \xi^6
- \tfrac{\num{2356}}{9} \, \xi^8
\Bigr] \phi^5
\\ &
+ \Bigl[
\tfrac{97}{3} \, \xi^6
- \tfrac{\num{7444}}{15} \, \xi^8
+ \tfrac{\num{40 016}}{27} \, \xi^{10}
- \tfrac{\num{16 951 588}}{\num{2025}} \, \xi^{12}
+ \tfrac{\num{365 040 328}}{\num{4725}} \, \xi^{14}
+ \cdots
\Bigr] \phi^6
\\ &
+ \Bigl[
-\tfrac{\num{13 532}}{45} \, \xi^8
+ \tfrac{\num{1 645 424}}{15} \, \xi^{10}
- \tfrac{\num{246 594 764}}{\num{6075}} \, \xi^{12}
+ \tfrac{\num{18 403 444 376}}{\num{42 525}} \, \xi^{14}
\Bigr]  \phi^7
\\ &
+ \Bigl[
\tfrac{\num{1 057 238}}{405} \, \xi^{10}
- \tfrac{\num{528 895 198}}{\num{8505}} \, \xi^{12}
+ \tfrac{\num{293 278 365 536}}{\num{297 675}} \, \xi^{14}
+ \cdots
\Bigr]  \phi^8
\\ &
+ \Bigl[
-\tfrac{\num{17 612 426}}{567} \, \xi^{12}
+ \tfrac{\num{1 376 189 404}}{\num{1323}} \, \xi^{14}
+ \cdots
\Bigr]  \phi^9
\\ &
+ \Bigl[
\tfrac{\num{17 745 598 574}}{\num{42 525}} \, \xi^{14}
+ \cdots
\Bigr]  \phi^{10}
+ \mathcal{O} (\phi^{11}).
\stepcounter{equation}
\tag{\theequation}
\label{eq:CanonicalPotential}
\end{align*}
As we have argued, the coefficients of odd powers of $\phi$ are going to be some polynomial given by the choice of $c_{p,q}$, and we already see this to be the case for $\phi^3$, $\phi^5$ and $\phi^7$, consistent with our analysis.

In this case field redefinitions are (reporting up to $\mathcal{O}(\xi^6)$ for brevity)
\be
\begin{split}
\label{eq:FieldRedefAfterTot}
\d \phi =   -\xi^2 \phi ^2
&+ \xi^4 \Bigl[ \, \tfrac{10}{3} \phi^3+ 4(\pt \phi)^2+2\phi  (\pt^2 \phi) \Bigr] \\
&- \xi^6 \Bigl[\tfrac{112}{9}\phi ^3 +\tfrac{76 }{3}\phi ^4 + \tfrac{16}{3} (\pt \phi)^2 +\tfrac{64}{3} \phi  \left( \pt \phi \right)^2
\\& \hspace{1cm}
+\tfrac{8}{3} \phi (\pt^2 \phi) +18 \phi ^2  (\pt^2 \phi) +\tfrac{4}{3} (\pt^2 \phi)^2 +\tfrac{4}{3}  \left( \pt \phi \right) (\pt^3 \phi )+\tfrac{4}{3} \phi   (\pt^4 \phi) \Bigr]  + \mathcal{O}(\xi^8).
\end{split}
\ee

We offer now some very preliminary observations
on convergence, using data up to ${\cal O} (\xi^{22})$.\footnote{We thank Ted Erler for raising this point.}
Consider first the coefficients of even powers of $\phi$.
These are infinite series in $\xi^2$ of the form
$\sum_n a_{2n} \xi^{2n}$.  We find $\log |a_{2n}| <  2n$ and possibly
$|a_{2n} |\lesssim (2n)^\beta$, for $\beta$ a positive number.  This behavior
is consistent with a radius of convergence for $\xi^2$ that could be
as large as one.  With the coefficients of $\phi^n$ convergent, one can ask
if the potential itself  $\sum_n c_n (\xi^2) \phi^n$ has a region of convergence in
$\phi$,  for various fixed values of $\xi^2$.   Here, at least for $\xi \lesssim 0.5$,
we find $\log |c_n| < n$, consistent with a finite radius of convergence.
A proper assessment, however,
 would require more data and possibly, taking into account the freedom in choosing the potential.  We leave a complete analysis for the future.

A key property of the redefined potentials $\tilde V(\phi; \xi^2)$ is that the depth
at the critical point is independent of $\xi^2$.    The critical point depends on
$\xi^2$ and on the coefficients $c_{p,q}$ used to construct equivalence classes,
but the value of the critical point does not depend on either.  The depth of the
tachyon potential has a physical interpretation in string field theory:  it gives
the tension of the unstable D-brane.

For zero nonlocality $\xi^2$, the redefined potential is equal to the
original potential $V$ since no redefinitions are needed in the first place:
\be
\tilde V (\phi; \xi^2=0) =  V(\phi) = - \tfrac{1}{2} \phi^2 - \tfrac{1}{3} \phi^3.
\ee
The (stable) minimum of the potential $V$ is at $\phi = -1$ with $V(\phi=-1) = -1/6$.  We claim that for the critical point $\phi_*$ of $\tilde V (\phi; \xi^2)$ one finds $\tilde V ( \phi_*; \xi^2) = -1/6$.

The explanation of this result is simple.  Consider the time-dependent
nonlocal theory we started with written as follows:
\be
L \ =  \  - K (\phi,
\partial \phi ; \xi^2)  - V(\phi) \,,
\ee
where $K$ denotes all terms containing time derivatives of the fields and
$V$ is the potential above.  The field redefinitions that bring this Lagrangian
to canonical kinetic term plus potential form, can be separated as follows:
\be
\delta \phi =   g (\phi; \xi^2) +  h( \phi,
\partial \phi ; \xi^2).
\ee
Here $g$ contains no derivatives of the fields while $h$ contains all terms
with derivatives.  Letting $\phi\to \phi + \delta \phi$ in $L$ is supposed
to give us the canonical answer.  But it is now clear that any variation of a
field in $K$ still gives a terms with derivatives as well as any variation of $V$ by $h$, and only $g$-type variations
of fields in $V$ would contribute the new potential.  So, in fact, we have
\be
\tilde V (\phi; \xi^2) = V ( \phi + g (\phi;\xi^2)) \,.
\ee
This shows that $\tilde V$ is just a redefinition of $V$, explaining why the critical
values of $\tilde V$ and $V$ must coincide in general.

We have tested our potentials $\tilde V$ and verified with {\em Mathematica} that, when computed
to order $\xi^{2p}$ with $p$ an integer, the value of $\tilde V$ at the critical point is indeed $-1/6$ with corrections of order $\xi^{2p+2}$, up to $p = 7$.  In fact, first few of this can be tested
directly using perturbation theory. Consider a potential $\tilde V$
written as in~\eqref{eq:tildepotential}:
\be
\tilde V (\phi; \xi^2) =  \tilde{V}_0 (\phi)  + \xi^2 \tilde{V}_2 (\phi)  + \xi^4 \tilde{V}_4 (\phi) + \mathcal{O} (\xi^6).
\ee
Let $\phi_0$ denote the critical point of $V (\phi) = \tilde V_0 (\phi)$.  This is also the critical
point of $\tilde V$ when $\xi^2=0$.  We wish to see how this critical point, $\phi_*(\xi^2)$,
moves as $\xi^2$ becomes nonzero, and what is the value of $\tilde V$
at such point.  A calculation shows that
\be
\tilde V( \phi_*(\xi^2) ; \xi^2)  = \tilde  V_0(\phi_0) + \xi^2  \tilde V_2 (\phi_0)
+ \xi^4  \Bigg[ \tilde V_4 (\phi_0) - {\tilde V_2' (\phi_0) \tilde V_2'(\phi_0)  \over 2 \tilde V_0''(\phi_0) } \Bigg] + {\cal O} (\xi^6) \,.
\ee
For the depth of the potential not to change from the value $V (\phi_0)= \tilde V_0 (\phi_0)$, the coefficients
of the nonzero powers of $\xi^2$ must vanish.   The simplest test is to see that
$\tilde V_2(\phi_0=-1)=0$.  From the potential in~(\ref{eq:CanonicalPotential})  we read
\be
\tilde V_0= -\tfrac{1}{2} \phi^2 - \tfrac{1}{3} \phi^3 \,,   \ \
\tilde V_2 =   \phi^3 + \phi^4 \,, \ \ \tilde V_4 =  -\tfrac{23}{6} \phi^4  -\tfrac{13}{3} \phi^5 \,.
\ee
Clearly $\tilde V_2 (\phi=-1)=0$.  One can also check that the coefficient of $\xi^4$
also vanishes for $\phi=-1$.

\section{Rolling Tachyons}
\label{sec:rolling}

This section discusses the
dynamics of the purely time-dependent tachyon
in the nonlocal theory.   We consider the situation where the scalar field
is at the unstable $\phi=0$ vacuum at $t =-\infty$ and it rolls towards
the minimum at $\phi= -1$.  Following\cite{Moeller:2002vx} we first solve
for the scalar field solution in the nonlocal theory, working perturbatively in
$e^{t}$, which is small for large negative $t$. We note that this solution has appeared in the literature before in~\cite{Gomis:2003xv}.
Alternatively, using an exact rolling solution for the local limit  $\xi^2=0$ as a starting point,
we can find exact rolling solutions for the nonlocal theory
in an expansion in powers of $\xi^2$.  Finally,  we take
the resulting wildly oscillatory nonlocal theory
rolling solution and apply the field redefinition obtained in
section~\ref{redthepurtimdepthe}  that maps the theory to a standard kinetic term
and a potential $\tilde V(\phi; \xi^2)$.  We find evidence that the mapped solution
describes conventional rolling in $\tilde V(\phi; \xi^2)$.  This supports
the consistency of the picture we have developed.

\subsection{Rolling tachyon nonperturbatively in $\xi^2$}

Consider again the nonlocal Lagrangian for the solely time-dependent field:
\be
L =  \tfrac{1}{2}  \phi ( -\partial_t^2 + 1)
\phi  + \tfrac{1}{3} \bigl( e^{-\xi^2 \partial_t^2} \phi \bigr)^3.
\ee
The potential
\be \label{eq:CubicPotential}
{V}(\phi)  = -\tfrac{1}{2}  \phi^2  - \tfrac{1}{3} \phi^3\, ,
\ee
has an unstable vacuum at $\phi=0$,  and a stable vacuum at $\phi=-1$.
For rolling that begins at $\phi=0$,
the turning point is at $\phi= -3/2$.
The equation of motion following from $L$ is
\be \label{eq:EOMvm}
 ( \partial_t^2 - 1) \phi  =   e^{-\xi^2 \partial_t^2}\bigl( e^{-\xi^2 \partial_t^2} \phi \bigr)^2.
\ee
The left hand side vanishes for the rolling ansatz  $\phi = - e^{t}$, where the
tachyon starts at the unstable vacuum for $t =-\infty$
and rolls towards the minimum at $\phi = -1$.  Any other coefficient in this
ansatz can be absorbed by a shift of $t$. This is therefore the starting point for a series solution with coefficients $b_n$
for all $n> 0$, and
with $b_1= -1$:
\be \label{eq:NL}
\phi = \sum_{n=1}^\infty b_n e^{n t}  =  - e^{t} + b_2 e^{2t} + b_3 e^{3t} + \mathcal{O} (e^{4t}).
\ee
After substitution into~(\ref{eq:EOMvm}) we get a recursive solution for the coefficients:
\be \label{eq:Recursion}
b_n =  {1\over n^2 -1} \sum_{p=1}^{n-1}  b_p b_{n-p}  e^{-2 \xi^2 (n^2 - np + p^2) }\,.
\ee
The first few coefficients are found to be
\be
\begin{split}
b_1=  & \  -1 \,, \\
b_2 = & \ \  \tfrac{1}{3}  e^{-6\xi^2} \,, \\
b_3 = &  \ - \tfrac{1}{12}  e^{-20 \xi^2} \,, \\
b_4 = & \ \tfrac{1}{15} \bigl( \tfrac{1}{6} e^{-46 \xi^2} + \tfrac{1}{9} e^{-36
\xi^2} \bigr) \,.
\end{split}
\ee
With the coefficients determined, the rolling solution is that in~(\ref{eq:NL}).
Just as in p-adic string theory and open string field theory in the level
expansion~\cite{Moeller:2002vx}, the solution above exhibit wildly oscillatory
behavior. The field overshoots the turning point and the oscillation amplitude grows in
time.  We can see above that the sign of the first few $b_n$ coefficients alternates
with $n$.  A little thought shows this property holds in general on
account of~(\ref{eq:NL}).  Each $b_n$ is a sum of terms, all
with the same sign.

\begin{figure}[t!]
	\includegraphics[scale=0.6]{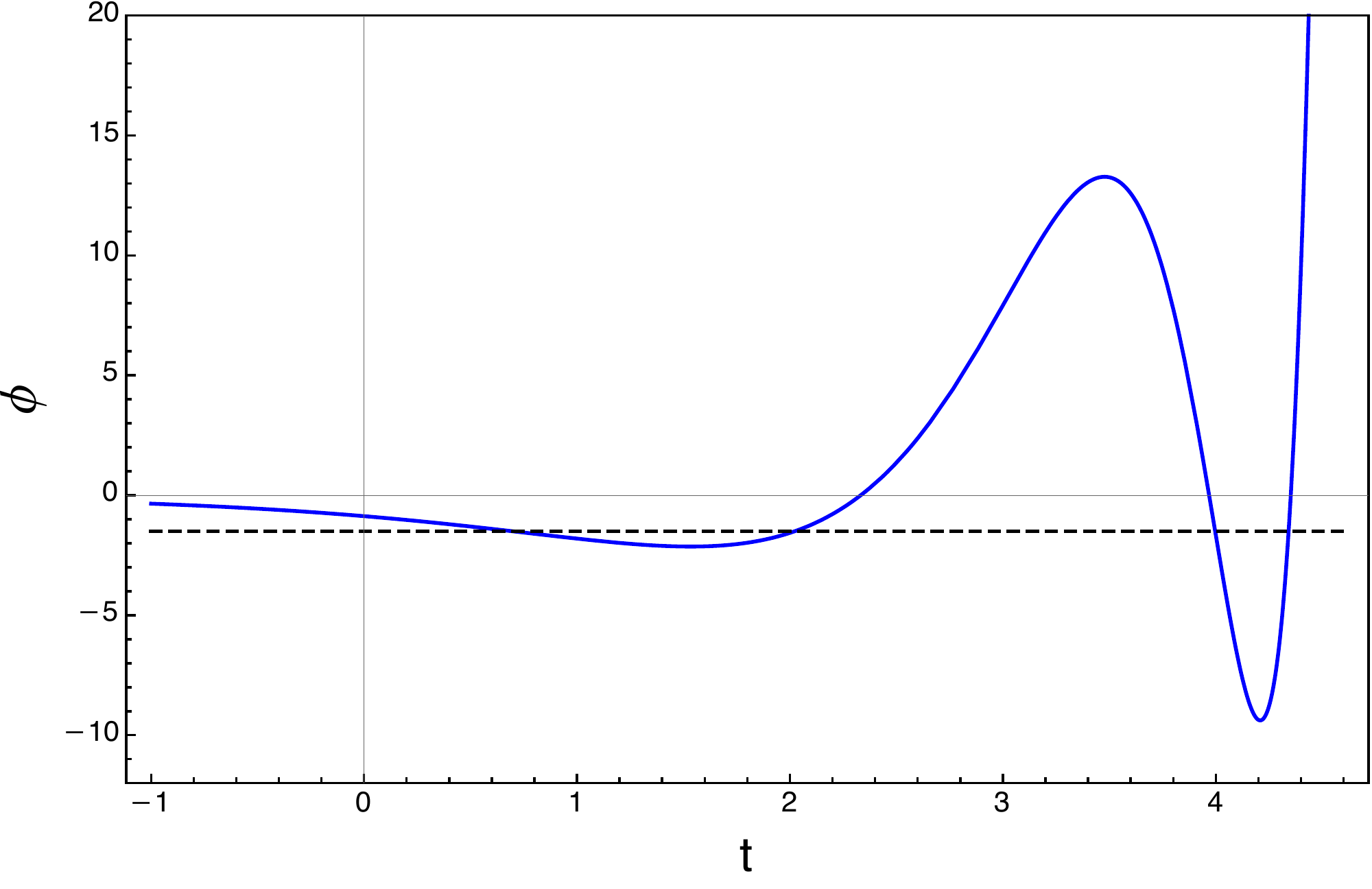}
	\centering
	\caption{The rolling solution in nonlocal theory with
	$\xi=0.39$, keeping terms up to
	$n=30$.
	The turning point at $\phi = -3/2$ is marked with black dashed line. Similar to the rolling solutions in~\cite{Moeller:2002vx}, this also overshoots the turning points and oscillations get larger with time.}
	\label{fig:NL_Rolling}
\end{figure}

A rolling solution with $\xi = 0.39$ is shown in figure~\ref{fig:NL_Rolling}.
The solution overshoots the turning points of the potential~\eqref{eq:CubicPotential} and the resulting oscillations gets larger with time, similar to those observed in~\cite{Moeller:2002vx}. This oscillatory
behavior becomes more prominent with increasing $\xi$, as this increases
the degree of nonlocality in the theory.

To understand the convergence of the series~\eqref{eq:NL}
describing the rolling,
we now attempt to find the behavior of the coefficients $b_n$ for $n \to \infty$.
A similar analysis for a related solution was
done by Fujita and Hata~\cite{Fujita:2003ex}.
Our analysis below is not rigorous, but the result is supported by
numerical work.  The upshot is that for any time $t$ the series solution above
converges whenever $\xi^2 > 0$.  For $\xi=0$, the series solution converges only until the field reaches the turning point.
This will be studied in the next subsection.

\begin{figure}[t!]
	\includegraphics[scale=0.4]{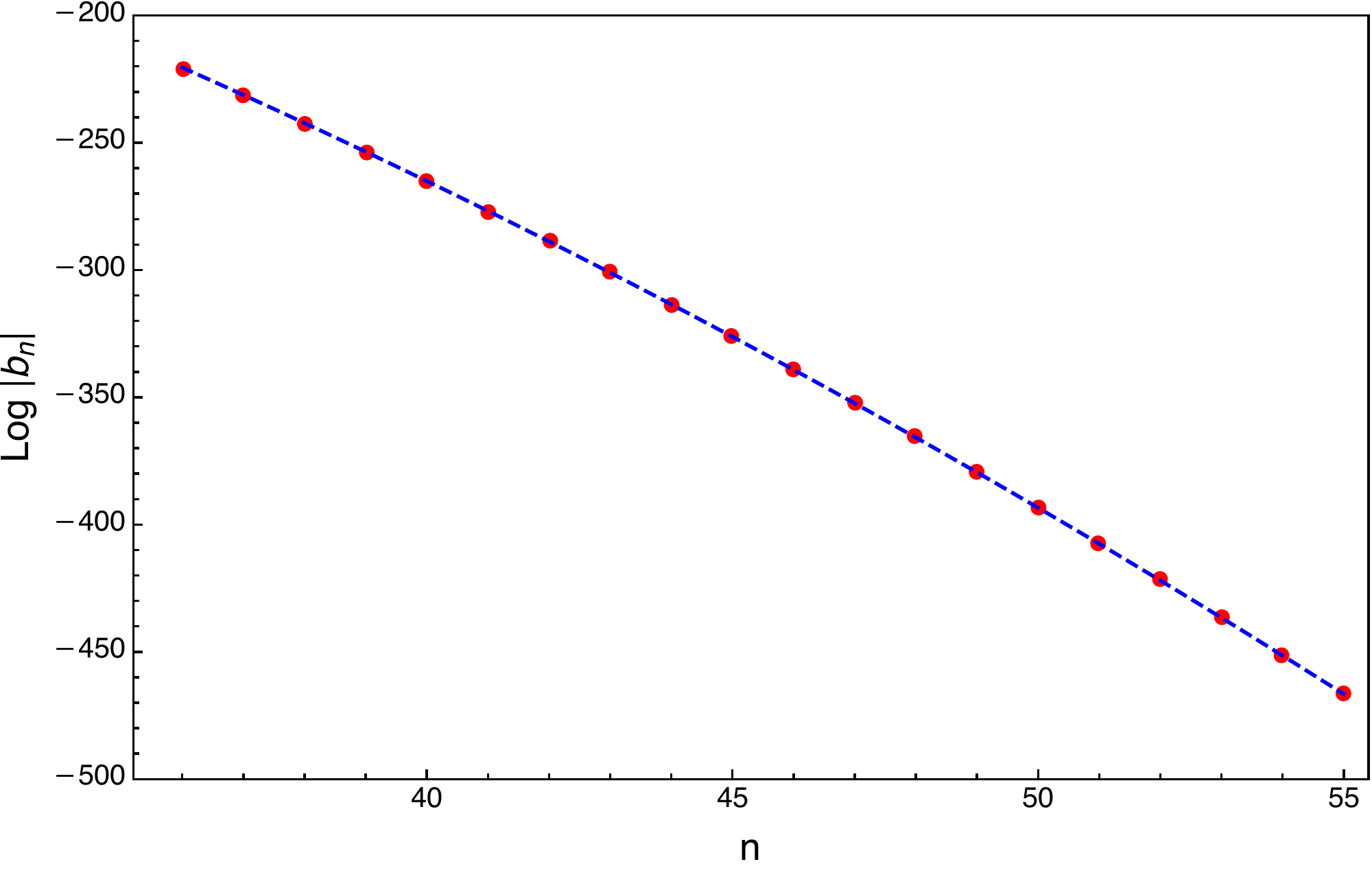}
	\includegraphics[scale=0.4]{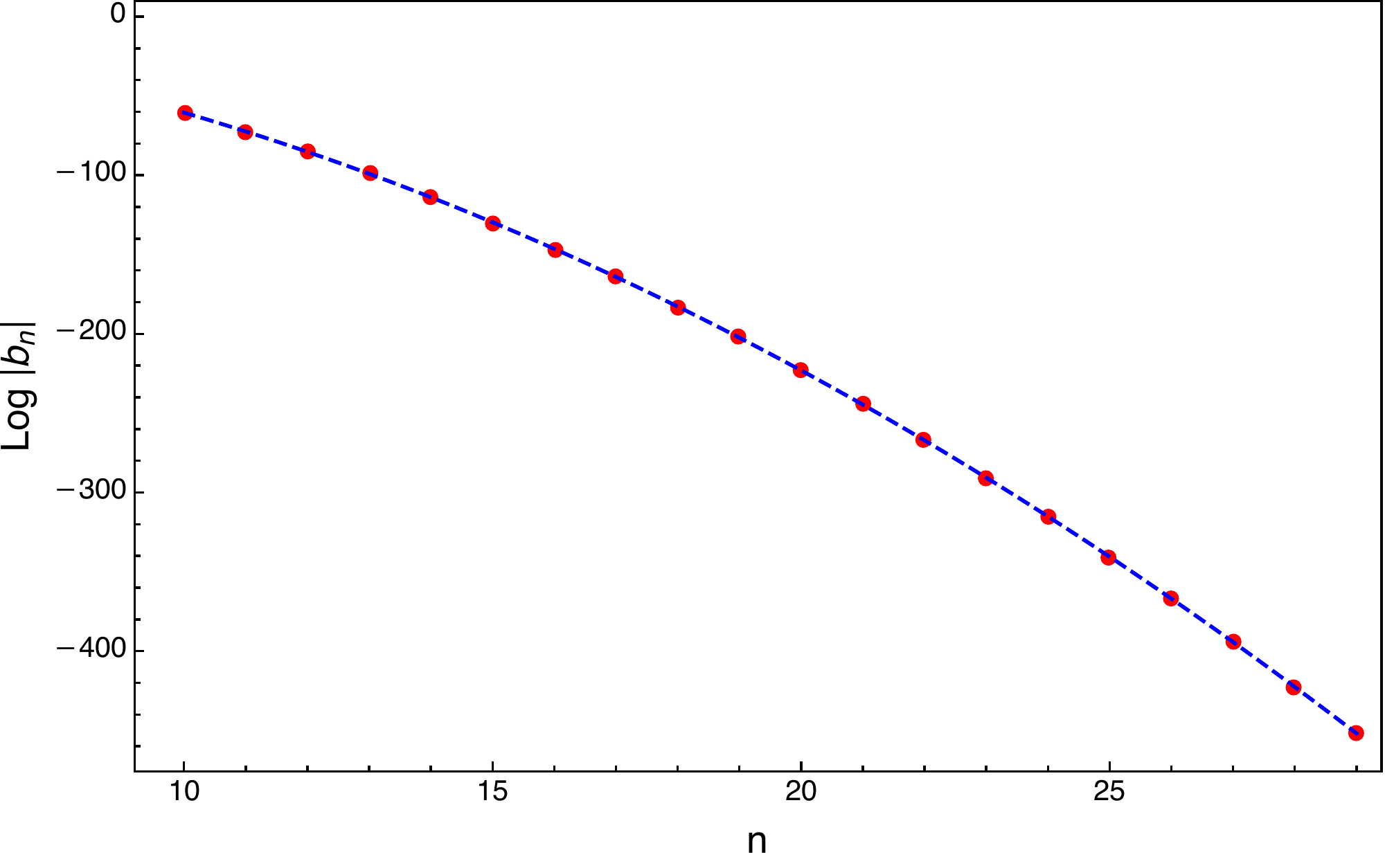}
	\centering
	\caption{The dots are $\log |b_n| $ for $36 \leq n \leq  55$ for $\xi =0.2$ (left)
	and  $10 \leq n \leq  29$ for $\xi =0.4$ (right),  obtained numerically from~\eqref{eq:NL}.  The curves give $\log |b_n| $ from the asymptotic result~\eqref{eq:AsymptSol},
	 with $\beta = 2.08$ (left) and $\beta = 1.99$ (right).
	 The match of the logarithms is solid, giving evidence for~\eqref{eq:AsymptSol}.
	 We found similar fits for other values of $\xi$ using $\beta \approx 2$.}
	\label{fig:large_n}
\end{figure}

By the recursion~\eqref{eq:Recursion} it is clear that the coefficients $b_n$ are given by sums of the terms of the form $C_i e^{-A_i \xi^2}$, with $A_i$ and $C_i$ are some
$\xi^2$-independent constants.  The constants $A_i$ are all positive; this
follows because the exponent in~(\ref{eq:NL}) contains the factor
$n^2 - np +p^2$, which is positive in the range $1 \leq p \leq n-1$.
We now assume that $b_n$ is dominated by a single term of the form
$C_i e^{-A_i \xi^2}$,  the one with the lowest value of $A_i$.  Unless the
$C_i$ vary wildly, this is the least suppressed term for any
nonzero value of $\xi^2$.
Looking at the recursion~\eqref{eq:Recursion}, the sum is
modulated by the exponential factor, which is largest at the minimum
of $n^2 - np + p^2$.  This minimum, with value $3n^2/4$, occurs for $p= n/2$ (this is the exact value
for the integer $p$ for
even $n$, and the approximate value for odd $n$).  If this term in the
sum dominates,
we have the approximate relation valid for very large $n$:
\be
\label{jnbddflk}
b_n  \simeq   \frac{1}{n^2} \Bigl( b_{\tfrac{n}{2}}\Bigr)^2 \,  e^{- \tfrac{3}{2} n^2 \xi^2} \,.
\ee
Additionally, as stated above, we assume that for large $n$:
\be \label{eq:Asymptoticbn}
b_n\,  \simeq   \, C(n) e^{-\alpha(n) \xi^2} \,,
\ee
with constants $C(n)$ and $\alpha(n) >0$ to be determined.
Inserting this ansatz into~(\ref{jnbddflk}) we find the conditions:
\be
\begin{split}
	\alpha(n) = 2 \alpha\left( \tfrac{n}{2}\right) + \tfrac{3}{2} n^2 \quad \text{and} \quad C(n) = \tfrac{1}{n^2} C^2\left(\tfrac{n}{2}\right).
\end{split}
\ee
By inspection, these are satisfied by
\be
\alpha(n) = 3n^2 \quad \text{and } \quad C(n) = 16 n^2 e^{-\beta n},
\ee
where $\beta \in \mathbb{R}$ is an undetermined constant.
 From this,  $b_n$ at large $n$ is given by
\be \label{eq:AsymptSol}
b_n \, \simeq \,  (-1)^n 16n^2 e^{-\beta n} \,  e^{-3 n^2 \xi^2},
\ee
where have also included the correct sign for the coefficient for $b_n$.
We have determined $\beta \simeq 2$ by fitting--for large $n$--the
above expression to numerically calculated coefficients.
A couple of fits are shown in
figure~\ref{fig:large_n}. In fact, the exact value of $\beta$ as well
as the prefactor $16n^2$ do not affect the following discussion of convergence.

Consider now the series expansion~\eqref{eq:NL} for the rolling tachyon and imagine
evaluating the sum for some fixed value $t_0$ of the time.
To ascertain convergence,  we use the ratio test.  We consider the absolute value of the ratio of consecutive
terms in the expansion, as $n \to \infty$
\be
\left| {b_{n+1} e^{(n+1) t_0} \over b_{n}  e^{n t_0} } \right | =
\left| {b_{n+1} \over b_{n}} \right | e^{t_0}
 \simeq  {(n+1)^2 \over n^2} e^{-\beta} e^{t_0}
 e^{- 3 (2n +1) \xi^2} \,.
\ee
Due to the last exponent, it is clear that for {\em any} value of $t_0$ the
ratio goes to zero as $n \to \infty$.
So the ratio test shows the
convergence of the series~\eqref{eq:NL} for the nonlocal theory ($\xi^2 > 0$) for
any value of $t \in \mathbb{R}$.  In practical terms, as the value of time increases we have
to include a larger number of terms in the series in order to see the convergence.

\subsection{Rolling tachyon perturbatively in $\xi^2$}

The previous subsection discussed the rolling tachyon
 nonperturbatively in $\xi^2$.  The resulting
series expansion has coefficients that involve exponentials of $\xi^2$, and the series
itself is convergent for all values of time.  The shortcoming of this solution is
that we have not been able to sum the series to arrive a form valid in all times.
In this section we work perturbatively
in $\xi^2$ in order to achieve this.  Thus our first step in the analysis would be finding the rolling solution for the local
$\xi^2=0$ theory.  As we will see, an exact analytic solution is possible.
Moreover, working perturbatively in $\xi^2$ allows to find analytic expressions
at each order.  This discussion gives further insight into the nature of rolling
solutions, and is particularly helpful for small nonlocality.

For the local theory ($\xi^2=0$)
the rolling solution obtained from the recursion~\eqref{eq:Recursion} is given by:
\be \label{eq:LocalSeries}
\phi (t) = -e^t + \tfrac{1}{3} e^{2t} - \tfrac{1}{12} e^{3t} +
\tfrac{1}{54} e^{4t} - \tfrac{5}{1296} e^{5t} + \mathcal{O} (\e^{6t}).
\ee
It turns out one can sum this series and get a closed form expression:
\be \label{eq:ExactSol0}
\phi_0(t) = - {e^t \over (1+ \tfrac{1}{6} e^t)^2}  \,.
\ee
This solves the equation of motion~\eqref{eq:EOMvm} when $\xi^2=0$,
 with $\phi_0 \to -e^t$ as $t \to -\infty$.  The solution shows the scalar rolling down reaching
the turning point
$\phi = -3/2$ at $e^t = 6$ and then going back to $\phi=0$ at $t = + \infty$.
This result also shows that the series~\eqref{eq:LocalSeries}, arising from expansion of~(\ref{eq:ExactSol0}),
converges only for $e^t < 6$.
Observe that the radius of convergence for the series~\eqref{eq:LocalSeries} turns out to be the
time where the tachyon reaches the turning point.
The general solution~(\ref{eq:ExactSol0}) is valid for all times and for $e^t > 6$, so it
can be expanded in powers of $e^{-t}$.  In fact, for large $t$ we have
$\phi \simeq - 36 e^{-t}$.

Note that we have already argued that the nonlocal version~\eqref{eq:NL} of the series solution~(\ref{eq:LocalSeries}) converges for {\em all} times, due to the exponential damping associated with $\xi^2$.
But additionally, it is also clear from the structure of the $b_n$ coefficients that
\be
|b_n(\xi^2=0)| \ > \ |b_n(\xi^2  \neq 0)|.
\ee
It follows that we can conclude \emph{rigorously} that the series~\eqref{eq:NL} converges for $e^t < 6$
for any  $\xi^2 \neq 0$.

The exact solvability of the local limit of the cubic potential
was noticed long ago in the context of lump
solutions~\cite{Zwiebach:2000dk}.
With  $\xi^2=0$, the equation of motion~\eqref{eq:EOMvm}
can be solved by energy conservation.  Since the total energy is zero when the
tachyon is at the unstable critical point, and it does not change in time,
we have that
\be
0 = {1 \over 2} \left({d \phi \over d t}\right)^2 + {V}(\phi)  \implies
{d\phi\over dt}  = - \sqrt{-2V(\phi)}\,,
\ee
where the sign is chosen so that the field rolls towards more negative values towards to the stable vacua $\phi =1$.
This can be easily integrated, which gives
\be
dt = -  {d\phi\over \phi \sqrt{1 + \tfrac{2}{3} \phi} } \quad \to \quad
t - t_0  = 2 \tanh^{-1} \sqrt{1+\tfrac{2}{3}\phi \,}\,.
\ee
Solving for $\phi$ in terms of $t$ one quickly gets
\be
\phi (t) =  \, - \frac{ 6 e^{t - t_0} } {\bigl( 1 + e^{t-t_0} \bigr)^2} \,.
\ee
The condition $\phi \to - e^t$ as $t \to -\infty$ fixes $e^{-t_0} = {1\over 6}$,
and then the solution above coincides with~\eqref{eq:ExactSol0}.

To analyze the nonlocal theory perturbatively in $\xi^2$
we begin by expanding the equation of motion~\eqref{eq:EOMvm}  in powers of $\xi^2$:
\be
\begin{split}
(\partial_t^2 -1) \phi = & \
\phi^2 - \xi^2 \left[ 2 (\partial_t \phi)^2 + 4
\phi (\partial_t^2 \phi) \right] \\
& \ \ \ \, + \xi^4 \left[4 \phi (\partial_t^4  \phi) + 6 (\partial_t \phi)^2 + 8 (\partial_t \phi)(\partial_t^3 \phi) \right] + \mathcal{O}(\xi^6).
\end{split}
\ee
Let the rolling solution $\phi$ for this equation take the form:
\be
\phi = \phi_0 + \xi^2 \phi_2 + \xi^4 \phi_4 + \mathcal{O}(\xi^6)\,,
\ee
with $\phi_0, \phi_2,$ and $\phi_4$ functions of time to be determined.
Inserting this series to the expanded equation of motion above, the first
three equations that follow are
\be
\begin{split}
	\mathcal{O}(\xi^0): \quad (\partial_t^2 -1) \phi_0 &= \phi_0^2, \\
	\mathcal{O}(\xi^2): \quad (\partial_t^2 -1) \phi_2 &= 2 \phi_0 \phi_2 -  2  (\partial_t \phi_0)^2 - 4 \phi_0 (\partial_t^2 \phi_0), \\
	\mathcal{O}(\xi^4): \quad (\partial_t^2 -1) \phi_4 &=  2 \phi_0 \phi_4 + \phi_2^2 - 4 (\partial_t \phi_0) (\partial_t \phi_2) - 4 \phi_0 (\partial_t^2 \phi_2) - 4 \phi_2 (\partial_t^2 \phi_0)\\
	& \hspace{2.5 cm} + 4 \phi_0 (\partial_t^4 \phi_0) + 6 (\partial_t \phi_0)^2 + 8 (\partial_t \phi_0)(\partial_t^3 \phi_0)\,.
\end{split}
\ee
The first one is the equation of motion for the local theory
whose rolling solution is already given in~\eqref{eq:ExactSol0}.
We have second-order, non-homogeneous linear ordinary differential equations (ODE) at each order and these can be solved recursively.
Since we know $\phi_0$,  we now solve for $\phi_2$,  and with $\phi_0$ and
$\phi_2$ we can solve for $\phi_4$, and so on.

\begin{figure}[t!]
	\includegraphics[scale=0.4]{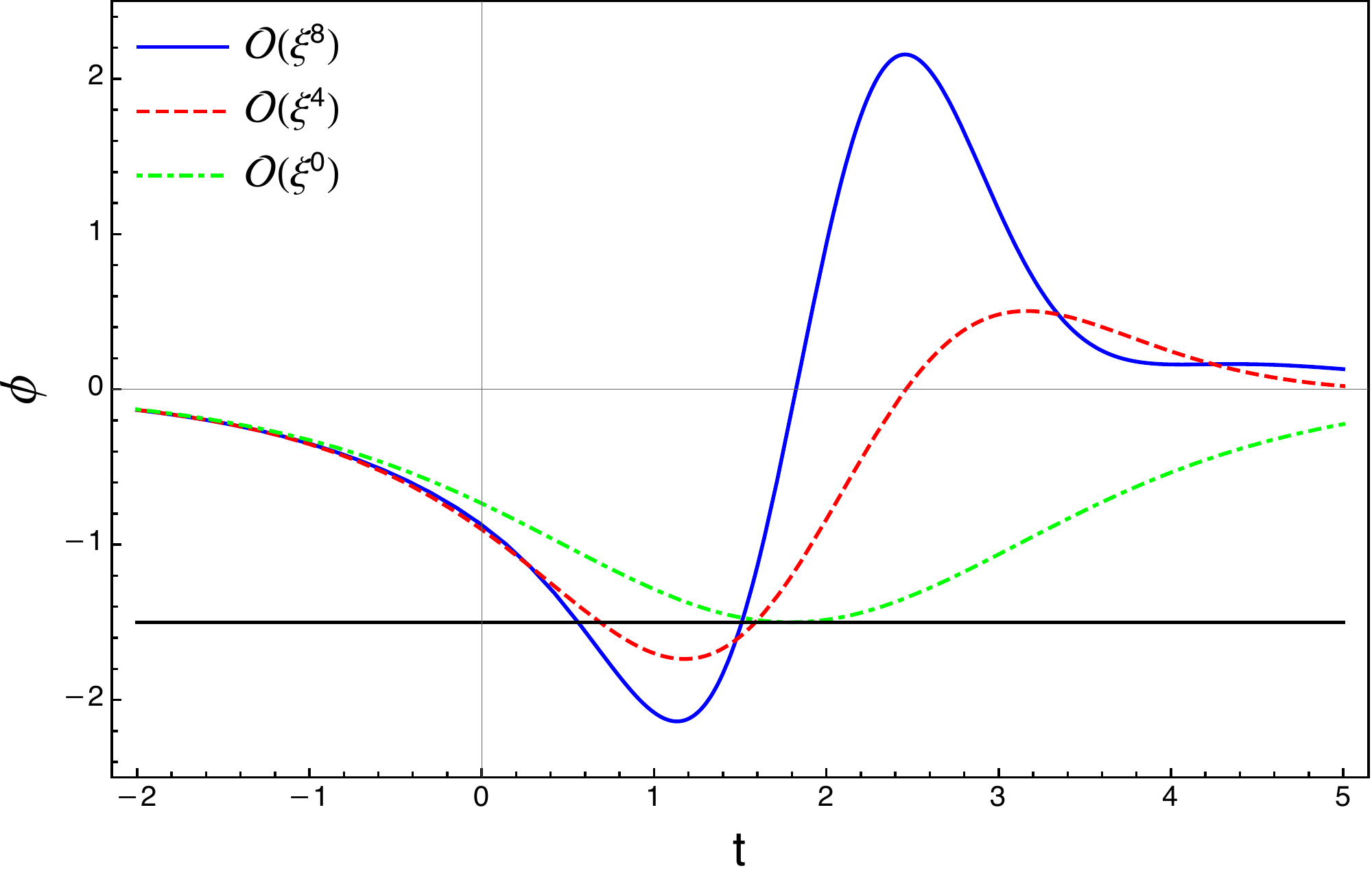}
	\includegraphics[scale=0.4]{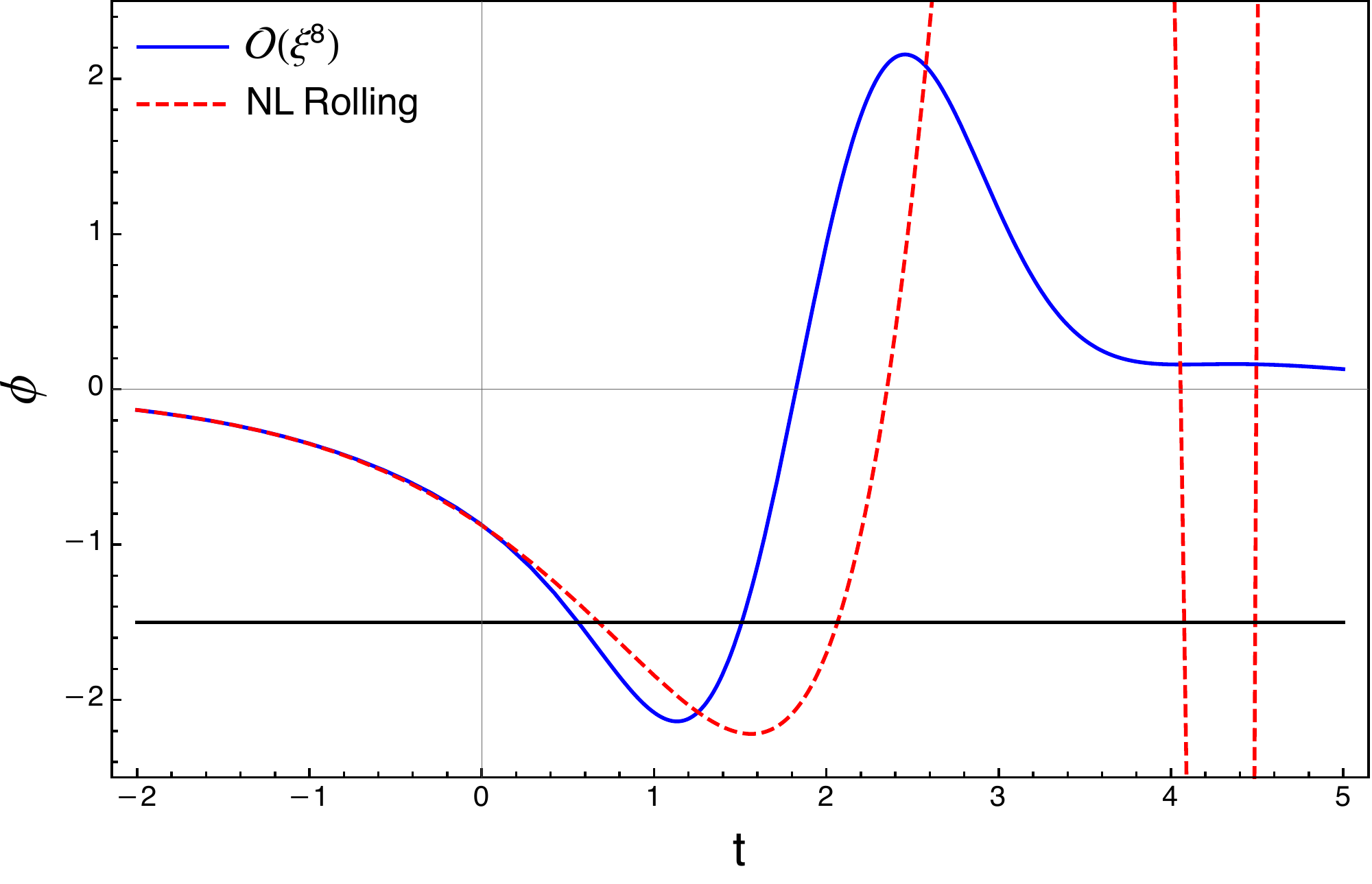}
	\centering
	\caption{In both plots $\xi=0.4$.
	Left:  The rolling solution~\eqref{eq:TruncatedRolling} to orders $\mathcal{O}(\x^4i)$ and $\mathcal{O}(\xi^8)$, shown together with the rolling in the local theory.
	Right: The rolling solution to order $\mathcal{O}(\xi^8)$ compared to the nonlocal
	theory rolling
	based on~\eqref{eq:NL}, including $b_n$ up to $n \leq 14$.
	The turning point at $\phi=-3/2$ is marked with straight black lines.
	}
	\label{fig:truncated_rolling}
\end{figure}

Let us solve for $\phi_2$.
Inserting the exact solution $\phi_0$ in~\eqref{eq:ExactSol0} and solving the resulting ODE with Mathematica's DSolve, we obtain the solution:
\be
\phi_2(t) = \frac{432 e^{2 t} \left(e^t-6\right)}{\left(6+e^t\right)^4}.
\ee
The solution from DSolve has two constants of integration.  To fix them
 we imposed two conditions:  (i)  $\phi_2 \to 0$ as $t \to -\infty$, and (ii)
  $\phi_2$, expanded in powers of $e^{t}$ must contain no $e^t$ term.
These are imposed because we want to start the rolling at $\phi = 0$,
    and the $-e^t$ term that drives the rolling
    is already provided by $\phi_0$ in~\eqref{eq:ExactSol0}.
    Carrying the same procedure for the next order  we find
\be \label{eq:TruncatedRolling}
\phi(t) = - {36 e^t \over (6+ e^t)^2} + \frac{432 e^{2 t} \left(e^t-6\right)}{\left(6+e^t\right)^4} \xi^2 -\frac{864 e^{2 t} \left(2 e^{3 t}-129 e^{2 t}+576 e^t-324\right)}{\left(6+e^t\right)^6} \xi^4 + \mathcal{O}(\xi^6).
\ee
This procedure can be repeated recursively to arbitrary order
 to obtain the \textit{exact} rolling solutions in the $\xi^2$-truncated theories.
For higher orders we observed that it was easier to solve the required ODE by guessing an ansatz based on the pattern in~\eqref{eq:TruncatedRolling}, rather than using DSolve.
In figure~\ref{fig:truncated_rolling} to the left, we show the rolling
solution~\eqref{eq:TruncatedRolling}, extended to ${\cal O}(\xi^8)$
and then truncated at various orders; to the right, we show the ${\cal O}(\xi^8)$
solution together with the series solution in $e^{nt}$.
It seems clear that as one includes
higher orders in $\xi^2$, the behavior of the rolling solution approaches
that of the nonlocal theory.  Indeed, we have checked that the solution~\eqref{eq:TruncatedRolling} matches with the rolling solution~\eqref{eq:NL} perturbatively up to the order $\mathcal{O}(\xi^8,e^{14 t})$.

\subsection{Rolling solution after field redefinition}

Now that we have control over the rolling solution of the nonlocal theory,
we can test our work in redefining the theory into a local one for the purely time-dependent
fields.  The expectation
 is that the wild oscillations of the nonlocal theory tachyon
should turn
into smooth rolling in the potential $V$ for the new redefined field.
We can test this concretely, since we have already determined the field redefinition.
This is a consistency check for
 the field redefinition, and for the claimed equivalence of the nonlocal theory
to the local $\xi^2$-dependent theory.

 To deal with the field redefinition, we have to introduce a bit of notation that we
 did not use in the previous section.   We will still call $\phi$ the original field of the
 nonlocal theory, but we will call $\phi'$ the field of the redefined theory.
 Our work before was
 based on replacements of the form
 \be
 \phi  \ \to \phi + \delta \phi (\phi, \partial \phi) \,,
 \ee
 implemented directly on the Lagrangian as in $L[\phi] \to L [\phi + \delta \phi ]$,
 a replacement that yields an equivalent Lagrangian.
But now the field in the final Lagrangian must be
 called $\phi'$, this actually means that we are setting
 \be
 \phi = \phi' + \delta \phi (\phi', \partial \phi')  \,,
 \ee
so in order to demonstrate the taming of the wild $\phi$-field oscillations,
 we must calculate $\phi'$
 in terms of $\phi$ and its derivatives,  and then use
  the rolling solution for $\phi$ in~\eqref{eq:TruncatedRolling}
  to calculate the rolling solution $\phi'$ in the redefined theory.
  Both the field redefinition and this rolling solution are perturbative
  in~$\xi^2$.

To calculate $\phi'$ we have to perturbatively invert the field redefinition~(\ref{eq:FieldRedef})
we found before.
 In the $\phi, \phi'$
notation we are now using, this reads to $\mathcal{O}(\xi^4)$,   
\be \label{eq:FieldRedefPassive}
\begin{split}
	\phi &=   \phi' -\xi^2 {\phi'}^2  + \xi^4 \Bigl[(1+c_{3,0})\tfrac{3}{2} {\phi'}^2
	+\left(
	\tfrac{13}{3} +c_{3,0}\right) {\phi'}^3
	\\ & \hspace{5cm}
	+(1-3 c_{3,0}) (\partial_t \phi')^2+2 \phi'  (\partial_t^2 \phi')  \bigg] + \mathcal{O}(\xi^6),
\end{split}
\ee
The inversion can be done perturbative by writing the ansatz
\be
\phi' = \phi_0' + \xi^2 \phi_2' + \xi^4 \phi_4' + \mathcal{O}(\xi^6),
\ee
and inserting this expression into~\eqref{eq:FieldRedefPassive}
to solve recursively for $\phi_0', \phi_2', \phi_4', \ldots$.
The end result can be found after some algebra and it is
\be \label{eq:FieldRedefActive}
\begin{split}
\phi' &=\phi+\xi^2 \phi^2 +  \xi^4 \Bigl[  \tfrac{3}{2} (-c_{3,0}-1) \phi^2  + \left(-c_{3,0} - \tfrac{7}{3} \right)\phi^3 \\ & \hspace{4 cm}- \left(1 - 3 c_{3,0} \right) (\partial_t\phi)^2 -2 \phi (\partial_t^2 \phi)  \Bigr] + \mathcal{O}(\xi^6).
\end{split}
\ee

Setting $\phi$ equal to  the rolling solution~\eqref{eq:TruncatedRolling},
we can evaluate the right-hand side of the equation above and find that
\be \label{eq:RedefinedRolling}
\begin{split}
\phi'(t) = &- {36 e^t \over (6+ e^t)^2} + \frac{432 e^{2 t} \left(e^t-3\right)}{\left(6+e^t\right)^4} \xi^2
\\ &-\frac{216 e^{2 t} }
{\left(6+e^t\right)^6}  \left(8 e^{3 t} + \left(-9 c_{3,0} -345\right) e^{2 t}+36\left(19+3c_{3,0}\right) e^t+324\left(-c_{3,0}-1\right)\right)\, \xi^4 + \mathcal{O}(\xi^6).
\end{split}
\ee
We have carried this procedure
up to $\mathcal{O}(\xi^8)$ and
checked that the resulting $\phi'$ satisfies the equation for
conventional rolling in the potential $\tilde V(\phi'; \xi^2)$ in~\eqref{eq:Potential}, truncated to $\mathcal{O}(\xi^8)$.

\begin{figure}[t!]
	\includegraphics[scale=0.4]{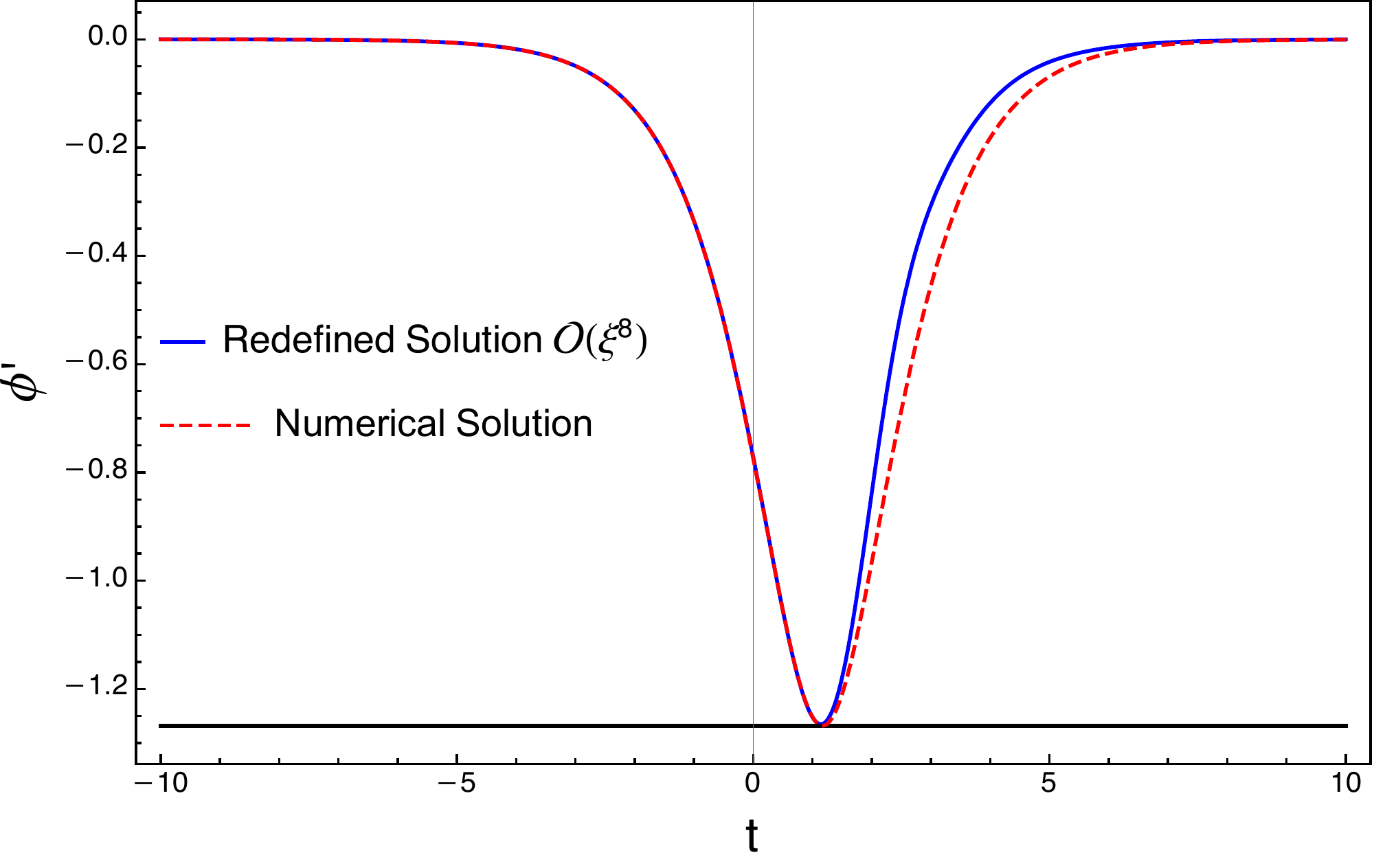}
	\includegraphics[scale=0.4]{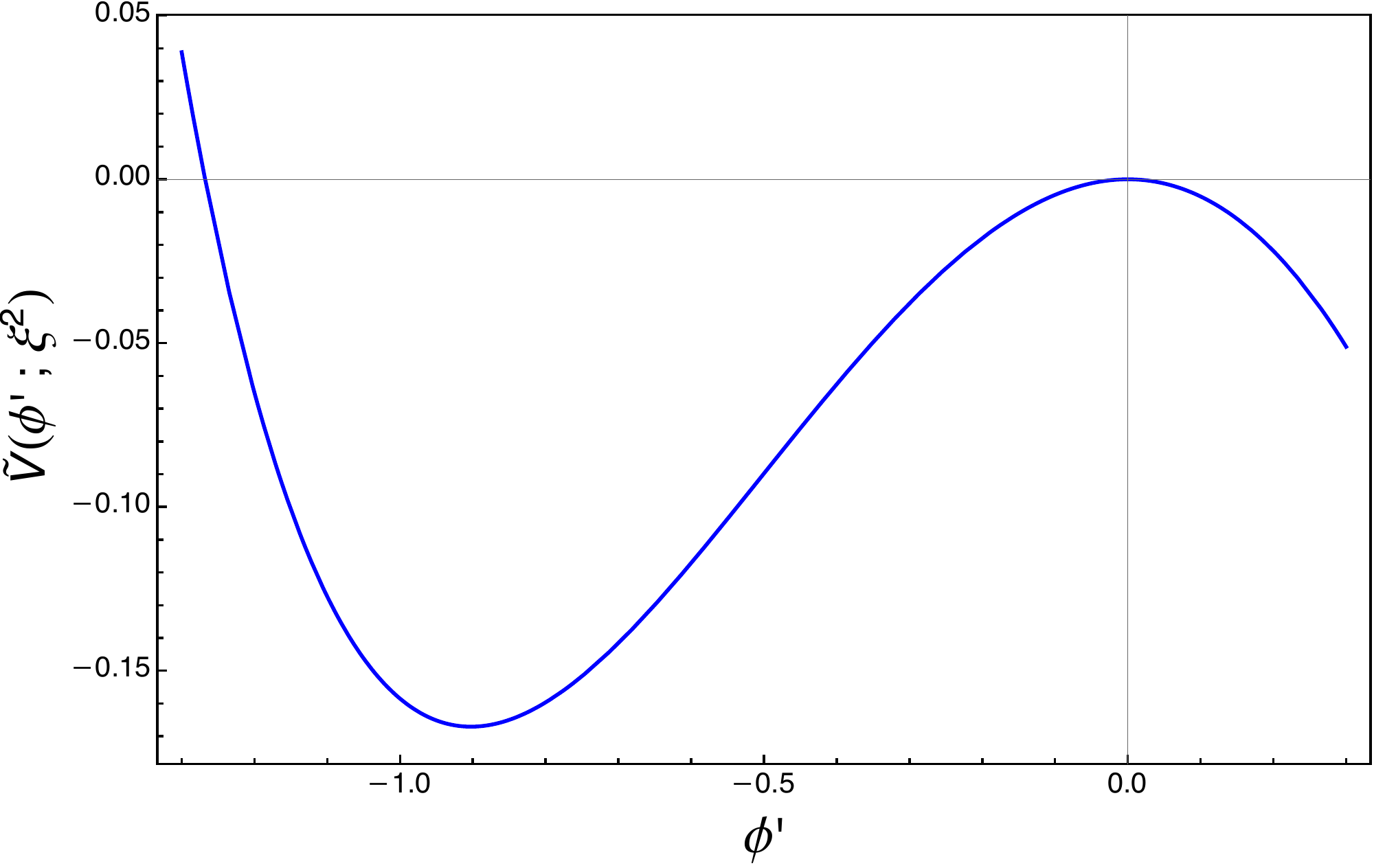}
	\centering
	\caption{Plots calculated to ${\cal O} (\xi^8)$ and setting $\xi=0.3$.
		Left: The redefined solution $\phi'(t)$
		compared to the rolling solution
		in $\tilde V(\phi'; \xi^2)$.
		The black line denotes the value $\phi' \simeq -1.27$ of the
		turning point in $\tilde V(\phi'; \xi^2)$. Neither solution overshoots this turning point.
		Right:  The potential $\tilde V(\phi'; \xi^2)$.}
	\label{fig:redefined_rolling}
\end{figure}
\begin{figure}[t!]
	\includegraphics[scale=0.6]{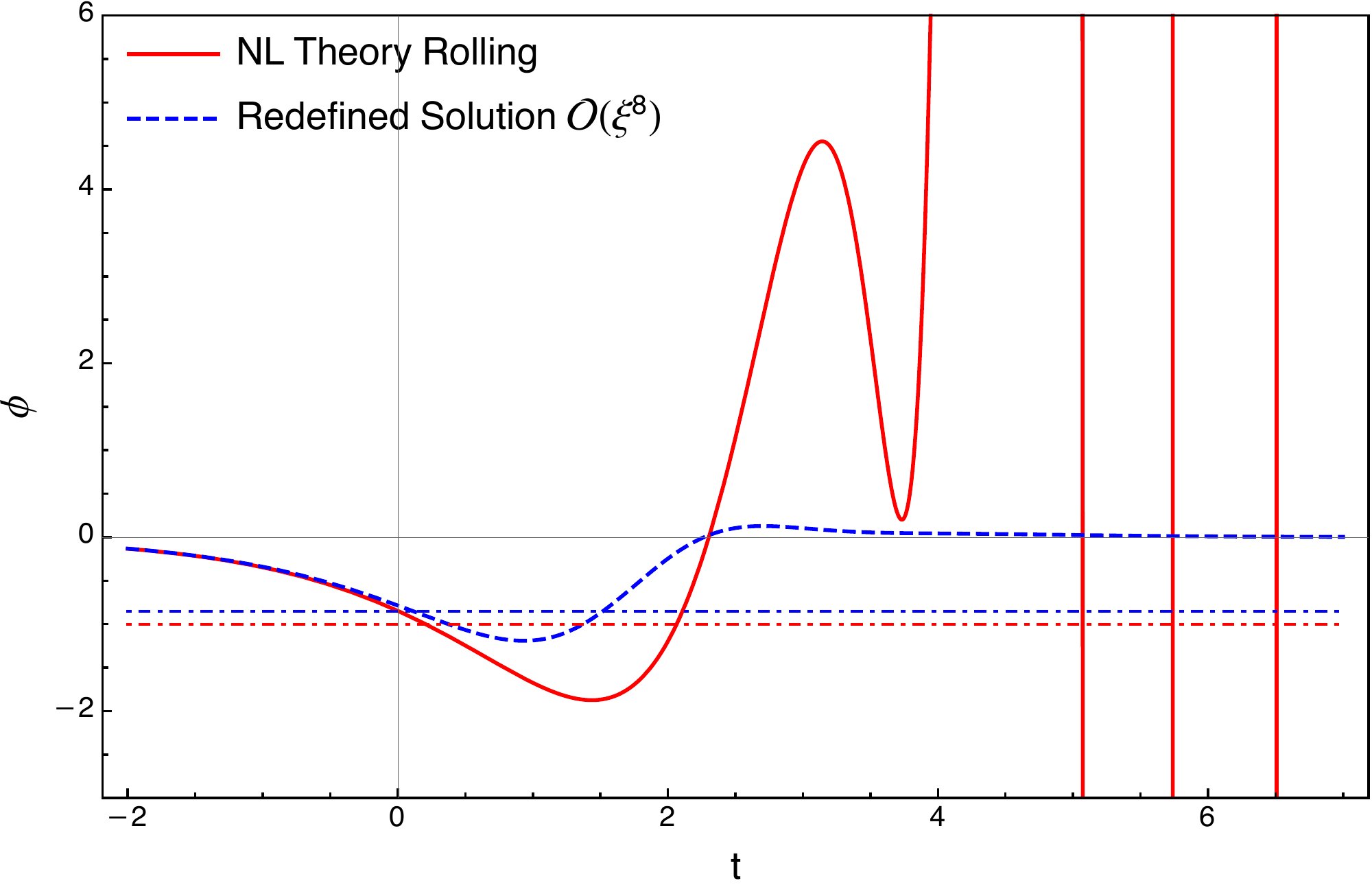}
	\centering
	\caption{Comparison between the solutions~\eqref{eq:NL} (solid red) and~\eqref{eq:RedefinedRolling} (dashed blue) when $\xi=0.35$.
	The color-coded dot-dash horizontal lines mark the stable vacua of their respective potentials. Notice their first pass through the minimum occur roughly at the same time.}
	\label{fig:redefined_comparison}
\end{figure}

Assume the constants $c_{i,j}$ take the values in~\eqref{eq:CanonicalGauge}, fixing
the ambiguity of the potential.  Working to ${\cal O}(\xi^8)$ and taking $\xi=0.3$,
figure~\ref{fig:redefined_rolling}, left, shows the numerical rolling solution
for the potential $\tilde V(\phi', \xi^2)$ as well as redefined solution $\phi'(t)$
above.   As expected, but nonetheless still quite striking, the overshooting
of the turning point has disappeared for the field variable $\phi'$. The redefined and numerical solutions match before the turning point, but they differ slightly afterwards.
 We have observed that including more orders in $\xi$ improves the matching between these solutions.  The potential $\tilde V(\phi; \xi^2)$ itself is shown to the right.
The comparison of the redefined rolling solution~\eqref{eq:RedefinedRolling} with the rolling solution~\eqref{eq:NL} of the original nonlocal theory for $\xi =0.35$ is shown in figure~\ref{fig:redefined_comparison}. We checked that similar behavior holds  for potentials related by quasi-symmetries.   The lack of late-time oscillations in the redefined solution is
an automatic feature of the expansion.  One must go to much higher orders in $\xi^2$
to truly test the disappearance of oscillations. 

\section{Noncovariant redefinitions of the general nonlocal theory}
\label{sec:redef-cov}

In this section, we extend the previous analysis of the solely time-dependent
theory to the case of the completely general spacetime-dependent Lorentz-covariant Lagrangian:
\begin{equation}
\label{eq:lagrangian-cov}
L
= \tfrac{1}{2} \, \phi \partial^2 \phi
+ \tfrac{1}{2} \, \phi^2
+ \tfrac{1}{3} \, \big( e^{\xi^2 \partial^2} \phi \big)^3.
\end{equation}
We ask if it is possible to remove higher-order derivatives perturbatively in $\xi^2$ using a field redefinition while keeping manifest Lorentz covariance.
Up to and including ${\cal O}(\xi^6)$ terms we see that this is possible.
At ${\cal O}(\xi^8)$, however,  we encounter an obstruction: we cannot
eliminate certain higher-derivative terms covariantly. So instead, we settle for removing just higher-order {\em time} derivatives, since this suffices for setting up a well-posed initial value problem. We show that this can be achieved at ${\cal O}(\xi^8)$, and then provide a general argument that this can be achieved for all orders.

Defining an \emph{obstruction} as a higher-derivative term   
 that cannot be removed from the theory covariantly,
we will adapt the following strategy for field redefinitions.  At any fixed order in $\xi^2$,
after dealing with the lower-order terms we must
\begin{enumerate}
	\item Remove all higher-order derivative terms that can be
	dealt with covariantly
	\item Remove terms of the form $\phi^n (\partial \phi)^2$ for $n\geq 1$ in order to have a standard kinetic term.
	\item Remove higher-order time derivatives from obstruction terms by breaking the manifest Lorentz covariance.
	\item Repeat the steps above if additional higher-derivative or $\phi^n (\p \phi)^2$ terms are introduced by the previous step.
\end{enumerate}
Since the field redefinition $\delta\phi$ multiplies $\partial^2 \phi$ (or $\partial_t^2 \phi$), we need to apply successive redefinitions to decrease the total derivative order in steps of two. The end result would be a collection of terms in which there are only first order time derivatives, coming in arbitrary numbers.
Note that not all terms with first-order derivatives can be removed since equation~\eqref{eq:IntLikeThing}, which allowed us to do this in the time-dependent case, does not have a generalization to the general spacetime-dependent case.
An immediate consequence is that the potential obtained in this section will be different from the one obtained in the purely time-dependent theory.

Alternatively, it is possible to write the Lagrangian~\eqref{eq:lagrangian-cov} in the light-cone frame and give the theory an initial value
formulation by eliminating \textit{all} derivatives with respect to   
light-cone time $\tau \equiv x^+ \equiv (x^0+x^1)/2$, except the one that appears in the kinetic term. We will see this requires the familiar light-cone nonlocalities---inverse $x^- \equiv (x^0-x^1)/2$ derivatives.

In subsection~\ref{eoirdj}
we explicitly implement the algorithm above to ${\cal O}(\xi^8)$ to display
the first obstruction to a Lorentz covariant field redefinition, and to show
how one implements the elimination of higher-order time derivatives in this case.
Then, in subsection~\ref{04einf}, we provide a general algorithm to carry out
the procedure to all orders.
A brief description of Hamiltonian formulation of the theory is
given in subsection~\ref{hamform-4}.  The light-cone formulation of the theory
is discussed in subsection~\ref{lcjnlvlvg}.

\subsection{Redefinitions up to ${\cal O}(\xi^8)$}\label{eoirdj}

We begin by expanding the Lagrangian~\eqref{eq:lagrangian-cov} in powers of $\xi^2$.
To  $\mathcal{O}(\xi^8)$, we find
\begin{equation}
\begin{aligned}
L = \tfrac{1}{2} \, \phi \partial^2 \phi
&
+ \tfrac{1}{2} \, \phi^2
+ \tfrac{1}{3} \, \phi^3
+ \xi^2 \, \phi^2 \partial^2 \phi
+ \xi^4 \left(
\tfrac{1}{2} \, \phi^2 \partial^4 \phi
+ \phi (\partial^2 \phi)^2
\right)
\\ &
+ \tfrac{1}{3}\,  \xi^6 \left(
\tfrac{1}{2} \, \phi^2 \partial^6 \phi
+ 3 \phi \partial^2 \phi \partial^4 \phi
+ (\partial^2 \phi)^3
\right)
\\ &
+ \tfrac{1}{3} \xi^8\left(
\tfrac{1}{8} \, \phi^2 \partial^8 \phi
+ \phi \partial^2 \phi \partial^6 \phi
+ \tfrac{3}{4} \, \phi (\partial^4 \phi)^2
+ \tfrac{3}{2} \, (\partial^2 \phi)^2 \partial^4 \phi
\right)
+ \mathcal O(\xi^{10}).
\end{aligned}
\end{equation}
Using the field redefinition
\begin{equation}
\phi
\longrightarrow \phi + \xi^2 \d_2\phi + \xi^4 \d_4\phi + \xi^6 \d_6 \phi + \xi^8 \d_8 \phi + \mathcal{O}(\xi^{10}) \,,
\end{equation}
the field-redefined Lagrangian $(L \to \tilde L)$ takes the following form:
\begin{equation}
	\tilde L
		\equiv \sum_{n = 0}^{\infty} \xi^{2n} \tilde L_{2n}\,,
\end{equation}
where
\begin{equation}
\tilde L_0
=\ \tfrac{1}{2} \, \phi \partial^2 \phi
+ \tfrac{1}{2} \, \phi^2
+ \tfrac{1}{3} \, \phi^3 \,,
\end{equation}
\begin{equation}
	\tilde L_2
	=
	\phi^2 \partial^2 \phi
	+ \d_2 \phi \partial^2 \phi
	+ (\phi^2+ \phi) \d_2 \phi \,,
\end{equation}
\begin{equation}
\begin{aligned}
	\tilde L_4
	=
	\tfrac{1}{2} \, \phi^2 \partial^4 \phi
	&
	+ \phi (\partial^2 \phi)^2
	+ \delta_4\phi \, \partial^2 \phi
	+ (\phi^2 + \phi) \, \delta_4 \phi
	\\ &
	+ 2 \phi \d_2 \phi \partial^2 \phi
	+ \phi^2 \partial^2 \d_2 \phi
	+\tfrac{1}{2} \d_2 \phi \partial^2 \d_2 \phi
	+ \d_2 \phi (\tfrac{1}{2} + \phi) \d_2 \phi \,,
\end{aligned}
\end{equation}
\begin{equation}
\begin{aligned}
\tilde L_6
=
\tfrac{1}{6} \phi^2 \partial^6 \phi
&
+ \phi (\partial^2 \phi) (\partial^4 \phi)
+\tfrac{1}{3} (\partial^2 \phi)^3
+\d_6 \phi \partial^2 \phi
+(\phi^2 + \phi) \partial \phi_6
+ \d_4 \phi \partial^2 \d_2 \phi
\\ &
\nonumber
+(2 \phi + 1) \d_2 \phi \d_4 \phi
+ \tfrac{1}{3} (\d_2 \phi)^3
+\phi^2 \partial^2 \d_4 \phi
+2 \phi \d_4 \phi \partial^2 \phi
+(\d_2 \phi)^2 \partial^2 \phi
\\ &
\nonumber
+2 \phi \d_2 \phi \partial^2 \d_2 \phi
+ \phi \d_2 \phi \partial^4 \phi
+\tfrac{1}{2} \phi^2 \partial^4 \d_2 \phi
+ \d_2 \phi (\partial^2 \phi)^2
+ 2 \phi (\partial^2 \phi) (\partial^2 \d_2 \phi) \,,
\end{aligned}
\end{equation}

\begin{equation}
\begin{aligned}
\tilde L_8
=
\tfrac{1}{24} \phi^2 \partial^8 \phi
&
+ \tfrac{1}{3} \phi (\partial^2 \phi) (\partial^6 \phi)
+\tfrac{1}{4} \phi (\partial^4 \phi)^2
+\tfrac{1}{2} (\partial^2 \phi)^2 (\partial^4 \phi)
+\tfrac{1}{6} \phi^2 \partial^6 \d_2 \phi
+\tfrac{1}{3} \phi \d_2 \phi \partial^6 \phi
\\ &
+\phi (\partial^2 \phi) \partial^4 \d_2 \phi
+ \phi (\partial^2 \d_2 \phi) \partial^4 \phi
+ \d_2 \phi (\partial^2 \phi) (\partial^4 \phi)
+ (\partial^2 \phi)^2 \partial^2 \d_2 \phi
+ \phi \d_2 \phi \partial^4 \d_2 \phi
\\ &
+ \tfrac{1}{2} (\d_2 \phi)^2 \partial^4 \phi
+\phi (\partial^2 \d_2 \phi)^2
+2 \d_2 \phi (\partial^2 \d_2 \phi) \partial^2 \phi
+ (\d_2 \phi)^2 \partial^2 \d_2 \phi
+ 2 \phi \d_4 \phi \partial^2 \d_2 \phi
\\ &
+ 2 \phi \d_2 \phi \partial^2 \d_4 \phi
+2 \d_2 \phi \d_4 \phi \partial^2 \phi
+(\d_2 \phi)^2 \d_4 \phi
+\d_2 \phi \partial^2 \d_6 \phi
+(1+2\phi) \d_2 \phi \d_6 \phi
\\ &
+\tfrac{1}{2} \phi^2 \partial^4 \d_4 \phi
+\phi \d_4 \phi \partial^4 \phi
+ 2 \phi (\partial^2 \phi) \partial^2 \d_4 \phi
+ \d_4 \phi (\partial^2 \phi)^2
+\tfrac{1}{2} (\d_4 \phi)^2
+\tfrac{1}{2} \d_4 \phi \partial^2 \d_4 \phi
\\ &
+\phi (\d_4 \phi)^2
+ \phi^2 \partial^2 \d_6 \phi
+2\phi \d_6 \phi \partial^2 \phi
+\d_8 \phi \partial^2 \phi
+(\phi^2 + \phi) \d_8 \phi \,.
\end{aligned}
\end{equation}

Starting at ${\cal O}(\xi^2)$, choosing $\d_2 \phi = -\phi^2$ eliminates the
term $\phi^2 \partial^2 \phi$ and leaves us with no derivatives:
\be
\tilde L_2 = -\phi^3 - \phi^4\,.
\ee
Using the chosen $\d_2 \phi$ and integrating-by-parts, we see $\tilde L_4$ takes the form
\be
\tilde L_4 \simeq \partial^2 \phi \bigg[
\d_4 \phi
+\tfrac{1}{2} \, \partial^2 \phi^2
+ \phi \partial^2 \phi
-2\phi^3 \bigg]
+ (\phi^2 + \phi) \, \d_4 \phi
+ \tfrac{1}{2} \phi^4
+\phi^5
+ 2 \phi^2 (\partial \phi)^2.
\ee
We now choose
\be
\d_4 \phi = \d_4 \phi' -\tfrac{1}{2} \, \partial^2 \phi^2 - \phi \partial^2 \phi +2\phi^3 ,
\ee
including a $\d_4 \phi' $ for further redefinitions. We then find
\be
\begin{split}
	\tilde L_4 &\simeq
	\partial^2 \phi \, \d_4 \phi'
	+ (\phi^2 + \phi) \, \bigg[\d_4 \phi' -\tfrac{1}{2} \, \partial^2 \phi^2 - \phi \partial^2 \phi +2\phi^3\bigg]
	+ \tfrac{1}{2} \phi^4
	+\phi^5
	+ 2 \phi^2 (\partial \phi)^2 \\
	& \simeq
	\partial^2 \phi \bigg[\d_4 \phi' - \tfrac{1}{2} \phi^2 - (\phi^2 + \phi) \phi \bigg]
	+(\phi^2 + \phi) \d_4 \phi'
	+\tfrac{5}{2} \phi^4
	+3 \phi^5
	+ 4 \phi^2 (\partial \phi)^2,
\end{split}
\ee
after arranging terms and integrating by parts. Further, we can pick
\be
\d_4 \phi' =\tfrac{1}{2} \phi^2 + (\phi^2+\phi)\phi = \phi^3 + \tfrac{3}{2} \phi^2 ,
\ee
and this eliminates the remaining higher-order derivative terms and yields
\be \label{eq:L4cov}
\begin{split}
	\tilde L_4
	\simeq
	(\phi^2 + \phi) \big( \phi^3 + \tfrac{3}{2} \phi^2 \big)
	+\tfrac{5}{2} \phi^4
	+3 \phi^5
	+ 4 \phi^2 (\partial \phi)^2
	= \tfrac{3}{2} \phi^3 + 5 \phi^4 + 4 \phi^5 + 4 \phi^2 (\partial \phi)^2.
\end{split}
\ee

We can also eliminate the term $4 \phi^2 (\partial \phi)^2$ covariantly. For generality, we will show how to eliminate the generic term $\phi^n (\partial \phi)^2$, where $n$ is a positive integer.  Integrating-by-parts, we see such term can be written as
\be
\begin{split} \label{eq:covfirstorder}
	\phi^n (\partial \phi)^2
	&\simeq - \phi \partial_\mu (\phi^n \partial^\mu \phi)
	= -n \phi^n (\partial \phi)^2 - \phi^{n+1} \partial^2 \phi \\
	&\implies \phi^n (\partial \phi)^2 \simeq -\tfrac{1}{n+1} \phi^{n+1} \partial^2 \phi.
\end{split}
\ee
Notice the similarity between this argument and that given in~\eqref{eq:IntLikeThing}. The above holds only when we have two derivatives, otherwise the contractions of Lorentz indices do not work out.  Suppose now  that the Lagrangian at order $\xi^{2k}$
contains  a term of the form $\b \phi^n (\partial \phi)^2$, with $\beta$ a constant. This term can be eliminated with a field redefinition  $\phi \to \phi + \xi^{2k} \d_{2k} \widetilde \phi$, as follows:
\be
\begin{split}
	T_{2k} & \equiv \d_{2k} \widetilde \phi (\partial^2 \phi + \phi + \phi^2) + \b \phi^n (\partial \phi)^2 \\
	&\simeq  \partial^2 \phi ( \d_{2k} \widetilde \phi -\tfrac{\b}{n+1} \phi^{n+1})+ \d_{2k} \widetilde \phi (\phi + \phi^2).
\end{split}
\ee
We then fix the redefinition and find:
\be
\d_{2k} \widetilde \phi = \tfrac{\b}{n+1} \phi^{n+1} \quad \to \quad
T_{2k} = \tfrac{\b}{n+1} \phi^{n+2} +\tfrac{\b}{n+1} \phi^{n+3}.
\ee
In particular, this shows the term $4 \phi^2 (\partial \phi)^2$ in~\eqref{eq:L4cov} can be redefined to
\be
4 \phi^2 (\partial \phi)^2 \to \tfrac{4}{3} \phi^{4} +\tfrac{4}{3} \phi^{5}  \ \
\hbox{with} \ \ \d_4 \widetilde \phi = \tfrac{4}{3} \phi^3\,.
\ee
Combining this with the rest at this order we see that
\be
\tilde L_4 \simeq  \tfrac{3}{2} \phi^3 + \tfrac{19}{3} \phi^4 + \tfrac{16}{3} \phi^5.
\ee
Including $\d_4 \widetilde \phi $ in $\d_4 \phi$, the total field redefinition at order $\mathcal O(\xi^4)$ to reach this form is
\be \label{eq:lc4fr}
\d_4 \phi  = -\tfrac{1}{2} \, \partial^2 \phi^2 - \phi \partial^2 \phi + \tfrac{13}{3} \phi^3  +\tfrac{3}{2} \phi^2.
\ee
As one can see, all derivatives have been eliminated to this order as well. The results
so far are the same as those for the purely-time-dependent case~\eqref{eq:PotentialResult} after replacing $\partial_t \to \partial$ and taking a
 minus sign into account, as we essentially implemented a similar procedure.

Repeating the analysis to ${\cal O} (\xi^6)$ we find
\be
\begin{aligned}
	\tilde L_6 &\simeq -\big(\tfrac{3}{2}\phi^3 + \tfrac{178}{9} \phi^4 + \tfrac{472}{9} \phi^5 + \tfrac{112}{3} \phi^6 \big) -\tfrac{8}{3} (\partial \phi)^4,
\end{aligned}
\ee
after fixing the field redefinition as follows
\be
\begin{aligned}
	\d_6 \phi &= -\tfrac{1}{6} \partial^4 \phi^2 -\phi \partial^4 \phi-\tfrac{1}{3} (\partial^2 \phi)
	^2 + 18\phi^2 \partial^2 \phi \\
	&\quad + \tfrac{8}{3} \phi \partial^2 \phi + \tfrac{70}{3} \phi (\partial \phi)^2 + \tfrac{7}{3} (\partial \phi)^2 - \tfrac{85}{3} \phi^4 -\tfrac{151}{9} \phi^3 - \tfrac{3}{2} \phi^2.
\end{aligned}
\ee
The $(\partial \phi)^4$ term in $\tilde L_6$ cannot be written as $(\partial \phi)^4 = \partial^2 \phi [\cdots]$ so it cannot get eliminated covariantly. But as explained before, this
is not a higher derivative term, and it is no obstacle for an initial value formulation.
The obstacle appears at next order, as we now show.

Repeating now the analysis to ${\cal O}(\xi^8)$, successive field redefinitions remove all higher-order derivatives except for the term of the form $(\partial \phi)^2 \partial^2 (\partial \phi)^2$, which is an obstruction:
\begin{equation}
\label{eq:cov-L8-covonly}
\begin{aligned}
\tilde L_8 &\simeq \tfrac{9}{8} \phi^3 + \tfrac{223}{6} \phi^4 + \tfrac{483}{2} \phi^5 + 485 \phi^6 + \tfrac{2695}{9} \phi^7\\
&\quad + \left( \tfrac{52}{3} + 96 \phi \right) (\partial \phi)^4
- \tfrac{4}{3} \, (\partial \phi)^2 \partial^2 (\partial \phi)^2  \,.    
\end{aligned}
\end{equation}
We omit the total redefinition to this order since it is rather long and unenlightening.
The term $(\partial \phi)^2 \partial^2 (\partial \phi)^2$ in $\tilde L_8$ cannot be removed covariantly:  no integration by parts allows it to be written in the
factorized form $\partial^2 \phi [ \cdots ]$ required for
covariant elimination.\footnote{We have not  
 attempted to prove this claim;  our trying convinced us we cannot factorize it.}
We will therefore break manifest Lorentz covariance
and focus on removing only higher-order \emph{time} derivatives.

Focusing on the obstruction term $ (\partial \phi)^2 \partial^2 (\partial \phi)^2$, we first evaluate the part of the term
multiplying $(\partial\phi)^2$ by breaking derivatives into temporal and spatial parts:
\begin{align*}
\partial^2 (\partial \phi)^2
&
= (- \partial_t^2 + \vec\nabla^2)
\big(- \dot \phi^2 + (\vec\nabla \phi)^2 \big)
\\ &
= \partial_t^2 (\dot \phi^2)
- \partial_t^2 (\vec\nabla \phi)^2
- \vec\nabla^2 (\dot \phi^2)
+ \vec\nabla^2 (\vec\nabla \phi)^2
\\ &
= 2 \dot \phi \phi^{(3)}
+ 2 \ddot \phi^2
- 2 \vec\nabla \phi \cdot \vec\nabla \ddot \phi
- 4 (\vec\nabla \dot \phi)^2
- 2 \dot \phi \vec\nabla^2 \dot \phi
+ 2 \vec\nabla \phi \cdot \vec\nabla (\vec\nabla^2 \phi)
+ 2 (\nabla_i \nabla_j \phi)^2
\\ &
= 2 \big[ \dot \phi \phi^{(3)}
+  \ddot \phi^2
-  \vec\nabla \phi \cdot \vec\nabla \ddot \phi
- 2 (\vec\nabla \dot \phi)^2
+  \partial^\mu \phi \partial_\mu (\vec\nabla^2 \phi)
+  (\nabla_i \nabla_j \phi)^2 \big].
\end{align*}
Here, we use $\phi^{(n)} \equiv \partial_t^n \phi$. Performing  
the  field redefinition $\phi \to \phi + \xi^8 \d_8 \hat \phi$ and letting
$\tilde T_8$ denote the part of $\tilde L_8$
that contains the field redefinition and the obstruction term, we have
\begin{align*}
\widetilde T_8
&
= - \ddot \phi \, \delta_8 \hat\phi
+ (\vec\nabla^2 \phi + \phi^2 +\phi) \delta_8 \hat\phi
\\ & \qquad
- \tfrac{8}{3} \, (\partial \phi)^2 \Big(
\dot \phi \phi^{(3)}
+ \ddot \phi^2
- \vec\nabla \phi \cdot \vec\nabla \ddot \phi
- 2 (\vec\nabla \dot \phi)^2
+ \partial^\mu \phi \partial_\mu (\vec\nabla^2 \phi)
+ (\nabla_i \nabla_j \phi)^2
\Big)
\\ &
\simeq \ddot \phi \, \Big(
- \d_8 \hat\phi
-\tfrac{8}{3} \, (\partial \phi)^2 \ddot \phi
-\tfrac{8}{3} \, \nabla^i \big( (\partial \phi)^2 \nabla_i \phi \big)
+\tfrac{8}{3} \, \partial_t \big( (\partial \phi)^2 \, \dot \phi \big)
\Big)
+ (\vec\nabla^2 \phi + \phi^2 + \phi) \d_8 \hat\phi
\\ & \qquad
- \tfrac{8}{3} \, (\partial \phi)^2 \Big(
- 2 (\vec\nabla \dot \phi)^2
+ \partial^\mu \phi \, \partial_\mu (\vec\nabla^2 \phi)
+ (\nabla_i \nabla_j \phi)^2
\Big).
\end{align*}
Now fixing
\begin{equation}
\delta_8 \hat\phi
= \delta_8 \hat\phi'
- \tfrac{8}{3} \, (\partial \phi)^2 \ddot \phi
- \tfrac{8}{3} \, \partial^\mu \big( (\partial \phi)^2 \partial_\mu \phi \big),
\end{equation}
we see that $\tilde L_8$ becomes, after some calculation, 
\begin{align*}
\widetilde T_8 &
\simeq \ddot \phi \, \Big(
- \delta_8 \hat\phi'
- \tfrac{8}{3} \, (\partial \phi)^2
(\vec\nabla^2 \phi + \phi^2 + \phi)
\Big)
+ (\vec\nabla^2 \phi + \phi^2 + \phi) \delta_8 \hat\phi'
\\ & \qquad
- \tfrac{8}{3} \, (\partial \phi)^2 \Big(
(\nabla_i \nabla_j \phi)^2
- \partial^\mu \phi \, \partial_\mu (\phi^2 + \phi)
- 2 (\vec\nabla \dot \phi)^2 \Big).
\end{align*}
We then take
\begin{equation}
\delta_8 \hat\phi'
= \delta_8 \hat\phi''- \tfrac{8}{3} \, (\partial \phi)^2 (\vec\nabla^2 \phi + \phi^2 + \phi),
\end{equation}
and after some simplification, we find
\begin{equation}
\begin{aligned}
\widetilde T_8 &
\simeq (\partial^2 \phi + \phi^2 + \phi) \delta_8 \hat\phi''
+ \tfrac{8}{3} (2 \phi + 1) (\partial \phi)^4
- \tfrac{8}{3} (\phi^2 + \phi)^2 (\partial \phi)^2
\\ & \qquad
- \tfrac{8}{3} \, (\partial \phi)^2 \Big(
(\vec\nabla^2 \phi)^2
+ (\nabla_i \nabla_j \phi)^2
- 2 (\vec\nabla \dot \phi)^2
+ 2 (\phi^2 + \phi) \vec\nabla^2 \phi \Big).
\end{aligned}
\end{equation}
Finally, we can eliminate the term $- \tfrac{8}{3}(\phi^2 + \phi)^2 (\partial \phi)^2 = - \tfrac{8}{3}(\phi^2 + 2 \phi^3 + \phi^4) (\partial \phi)^2$ by taking
\begin{equation}
\delta_8 \hat\phi'' = -\tfrac{8}{15} \phi^5 - \tfrac{4}{3} \phi^4 + \tfrac{8}{9} \phi^3\,.
\end{equation}
This yields
\begin{equation}
\begin{aligned}
\widetilde T_8 &
\simeq -\tfrac{8}{9} \phi^4 - \tfrac{20}{9} \phi^5 - \tfrac{28}{15} \phi^6 -\tfrac{8}{16} \phi^7
+ \tfrac{8}{3} (2 \phi + 1) (\partial \phi)^4
\\ & \qquad
-\tfrac{8}{3}\, (\partial \phi)^2 \Big(
(\vec\nabla^2 \phi)^2
+ (\nabla_i \nabla_j \phi)^2
- 2 (\vec\nabla \dot \phi)^2
+ 2 (\phi^2 + \phi) \vec\nabla^2 \phi \Big).
\end{aligned}
\end{equation}
Combining $\widetilde T_8$ with the rest of $\tilde L_8$, we finally get
\begin{equation}
\begin{aligned}
\tilde L_8 & \simeq \tfrac{9}{8} \phi^3 + \tfrac{653}{18} \phi^4 + \tfrac{4307}{18} \phi^5 + \tfrac{7247}{15} \phi^6 + \tfrac{13451}{45} \phi^7
+ \left( 20 + \tfrac{304}{3} \phi \right) (\partial \phi)^4
\\ & \qquad
-\tfrac{8}{3}\, (\partial \phi)^2 \Big(
(\vec\nabla^2 \phi)^2
+ (\nabla_i \nabla_j \phi)^2
- 2 (\vec\nabla \dot \phi)^2
+ 2 (\phi^2 + \phi) \vec\nabla^2 \phi \Big).
\end{aligned}
\end{equation}
We no longer have higher time derivatives, but higher spatial derivatives remain.

Summarizing, the total Lagrangian after field redefinition can be written as:
\begin{equation}
\label{eq:lagrangian-redef-from-cov}
\tilde L
\simeq - \tilde K(\phi, \dot\phi, \vec\nabla \phi; \xi^2)
- \tilde V(\phi; \xi^2),
\end{equation}
where
\begin{equation}
\begin{aligned}
\tilde K(\phi, \dot\phi, \vec\nabla \phi; \xi^2)
& = \tfrac{1}{2} \, (\partial \phi)^2
+ \left(
\tfrac{8}{3} \, \xi^6
- 20 \xi^8
\right) (\partial \phi)^4
- \tfrac{304}{3} \xi^8 \phi (\partial \phi)^4
\\ & \quad\
+ \tfrac{8}{3} \, \xi^8 (\partial \phi)^2 \Big(
(\vec\nabla^2 \phi)^2
+ (\nabla_i \nabla_j \phi)^2
- 2 (\vec\nabla \dot \phi)^2
+ 2 (\phi^2 + \phi) \vec\nabla^2 \phi
\Big) + \, {\cal O} (\xi^{10}),
\end{aligned}
\end{equation}
and
\begin{equation}
\begin{aligned}
\tilde V(\phi; \xi^2)
=
- \tfrac{1}{2} \, \phi^2
& - \tfrac{1}{3} \, e^{- 3 \xi^2} \, \phi^3
+\left[\xi^2 - \tfrac{19}{3} \xi^4 + \tfrac{178}{9} \xi^6 - \tfrac{653}{18} \xi^8 + \cdots \right] \, \phi^4
\\ &
+\left[-\tfrac{16}{3}\xi^4 + \tfrac{472}{9} \xi^6 - \tfrac{4307}{18} \xi^8 + \cdots\right] \, \phi^5
\\ &
+\left[\tfrac{112}{3} \xi^6 - \tfrac{7247}{15} \xi^8 + \cdots \right] \, \phi^6
+\left[-\tfrac{13451}{45} \xi^8 + \cdots \right] \, \phi^7 + \mathcal{O}(\phi^8).
\end{aligned}
\end{equation}
In the potential we have summed the series in front of the cubic term:
\begin{equation}
	e^{- 3 \xi^2}
		=
		1
		- 3 \xi^2
		+ \tfrac{9}{2} \xi^4
		- \tfrac{9}{2} \xi^6
		+ \tfrac{27}{8} \xi^8
		+ \cdots
\end{equation}
It is possible to prove this result as follows.  
First note that since the field redefinition $\delta \phi$  is at least quadratic in $\phi$, cubic terms in the Lagrangian can be generated only by variations linear in $\delta \phi$ of quadratic
terms in the Lagrangian.   In other words, cubic terms are generated from
$\delta \phi (\partial^2 \phi + \phi + \phi^2)$, and only when trying to remove
higher derivatives from the original cubic interactions of the nonlocal theory---quartic
and higher order terms in $\phi$ induced by the redefinition process
cannot generate cubic terms.
As a consequence, the effective rule for the generation of cubic terms from  the field redefinition is the replacement $\partial^2 \phi \to - \phi$ on cubic terms.
This means that $(e^{\xi^2 \partial^2} \phi)^3 \to (e^{- \xi^2} \phi)^3$.
This makes clear the on-shell three-point amplitude in the redefined theory
agrees to all orders
in $\xi^2$ with the one derived from the original theory.
We have also checked that the on-shell four-point amplitude with the redefined Lagrangian agrees with the one computed from the original Lagrangian up to and including $\mathcal O(\xi^4)$.

As a check of this potential, we have also verified that its value at the critical point,
computed to ${\cal O}(\xi^8)$, is indeed $-1/6+ {\cal O} (\xi^{10})$.
This is the same consistency check we used for the potential in the solely
time-dependent theory.

\subsection{General algorithm}\label{04einf}

In this subsection, we extend the algorithm provided in
section~\ref{therecarg} to the Lorentz covariant Lagrangian \eqref{eq:lagrangian-cov}.
As we have just seen, it is not possible in general to remove
all higher-order derivatives covariantly and at some point we simply need to settle for removing
higher-order {\em time} derivatives.  The purpose of this section is to show how
these derivatives can be removed recursively.  The end result would be a theory where
fields are only acted upon by a single time derivative or none, but an arbitrary
number of spatial derivatives---in other words spatial nonlocality would remain.

Consider a general term at some order in the $\xi^2$ expansion of the theory:
\begin{equation}
\label{gen-nonvcoform}
\begin{aligned}
T & = \big( W_1(\vec\nabla) \partial_t^{k_1} \phi \big) \,
\big( W_2(\vec\nabla) \partial_t^{k_2}  \phi \big)
\cdots
\big( W_\ell(\vec\nabla)\partial_t^{k_\ell} \phi \big) \,
\big( Y_1(\vec\nabla) \partial_t \phi \big)
\cdots
\big( Y_m(\vec\nabla) \partial_t\phi \big)
\\ & \hspace{2cm}
\times
\big( Z_1(\vec\nabla) \phi \big)
\cdots
\big( Z_n(\vec\nabla) \phi \big)\,,
\end{aligned}
\end{equation}
where the $W_1, \ldots, W_\ell, Y_1, \ldots, Y_m, Z_1, \ldots, Z_n$ are
monomials
built using spatial derivatives $\vec \nabla$.  The monomials
may have free indices; contractions, which we do not display, may occur
between different factors.  Some $W, Y,$ or $Z$'s may be just trivial---that is, equal to one.  Moreover, we take
\begin{equation}
\label{condefeoi}
3 \le k_1 \le \cdots \le k_\ell.
\end{equation}
An example of such a term is $(\nabla_i \partial_t^3 \phi) \, \partial_t \phi \, \nabla_i \phi$ for which $X_1 = Z_1 = \nabla_i$ and $Y_1 = 1$.
Following our earlier notation,  $\ell$ is called the index of the term $T$,
and $k_1$ is called the lowest order of the term $T$.

Let us first make a general point about integration by parts: the spatial derivatives
do not interfere with the manipulation of time derivatives and do not affect the
way we do redefinitions. Indeed,   suppose we have a term of the form
\be
T' \, = \, \big( W_1(\vec\nabla) \partial_t^2 \phi \big) \bigl[ \cdots \bigr]\,,
\ee
where the dots represent arbitrary additional terms in the form
of~(\ref{gen-nonvcoform}).  We now have, integrating by parts the spatial
derivatives in $W_1$,
\be
T'
\, \simeq \,  (-1)^{n_1} \partial_t^2\phi  \,
W_1(\vec\nabla) \bigl[ \cdots \bigr] \,.
\ee
Here $n_1$ is the number of derivatives in $W_1$.
With $\partial_t^2\phi$ appearing multiplicatively,
the effect of a field redefinition is implemented by the replacement $\partial_t^2\phi\to \vec\nabla^2 \phi + \phi+ \phi^2$, so we get
\be
T' \simeq  (-1)^{n_1} ( \, \vec\nabla^2 \phi + \phi+ \phi^2 )
W_1(\vec\nabla) \bigl[ \cdots \bigr]  \, = \,
[ W_1(\vec\nabla)  ( \, \vec\nabla^2 \phi + \phi+ \phi^2 ) ]
\bigl[ \cdots \bigr]\,,
\ee
where we again integrated by parts $W_1(\vec\nabla)$ resulting in the cancellation
of the sign factor.  The end result is that the replacement of $\partial_t^2 \phi$
in $T'$ could have been done from the get-go, ignoring the spatial derivatives
acting on the field.  This example also shows that $\partial_t^2$ operators
on fields can be eliminated directly, and this is why the constraint~(\ref{condefeoi}) 
involves $k$'s that are greater than or equal to three.

We can now proceed with a procedure analogous to that
 in section~\ref{therecarg}.   
Integrating by parts a single time derivative acting on the first term of $T$, we have:
\begin{equation}
\begin{aligned}
T
&
\simeq - (W_1(\vec\nabla)\partial_t^{k_1 - 1} \phi )\; \partial_t \Big[
\big( W_2(\vec\nabla) \partial_t^{k_2}\phi \big)
\cdots
\big( W_\ell(\vec\nabla)\partial_t^{k_\ell}  \phi \big)
\Big] \,
\big( Y_1(\vec\nabla) \partial_t\phi \big)
\cdots
\big( Y_m(\vec\nabla) \partial_t \phi \big)
\\ & \hspace{3cm}
\times
\big( Z_1(\vec\nabla) \phi \big)
\cdots
\big( Z_n(\vec\nabla) \phi \big)
\\ & \qquad
- (W_1(\vec\nabla)\partial_t^{k_1 - 1} \phi )\, \Big[
\big( W_2(\vec\nabla) \partial_t^{k_2}\phi \big)
\cdots
\big( W_\ell(\vec\nabla) \partial_t^{k_\ell}\phi \big) \,
\times
\big( Z_1(\vec\nabla) \phi \big)
\cdots
\big( Z_n(\vec\nabla) \phi \big)
\Big]
\\ & \hspace{3cm}
\times \Big[
\big( Y_1(\vec\nabla) \partial_t^2\phi \big)
\cdots
\big( Y_m(\vec\nabla) \partial_t\phi \big)
+ \cdots
+ \big( Y_1(\vec\nabla) \partial_t \phi \big)
\cdots
\big( Y_m(\vec\nabla) \partial_t^2\phi \big)
\Big]
\\ & \qquad
- (W_1(\vec\nabla)\partial_t^{k_1 - 1} \phi )\,\Big[
\big( W_2(\vec\nabla) \partial_t^{k_2} \phi \big)
\cdots
\big( W_\ell(\vec\nabla) \partial_t^{k_\ell}\phi \big) \,
\big( Y_1(\vec\nabla) \partial_t \phi \big)
\cdots
\big( Y_m(\vec\nabla) \partial_t \phi \big)
\Big]
\\ & \hspace{3cm}
\times
\partial_t \Big[
\big( Z_1(\vec\nabla) \phi \big)
\cdots
\big( Z_n(\vec\nabla) \phi \big)
\Big].
\end{aligned}
\end{equation}
After distributing the time derivative in the first and third terms, we obtain terms of the same form as $T$ but with the lowest order reduced by one unit.
All contributions from the second term contain factors of $\partial_t^2 \phi$
which can be removed by a field redefinition; here also the lowest order has been
reduced by one unit.
This shows that one can reduce the lowest order
recursively, until it becomes three.
Then, we can perform a last integration by part and obtain a term proportional to $\partial_t^2 \phi$ which can be eliminated by a field redefinition.
At this point the index has been reduced by one unit.  Reducing the index recursively
until it becomes zero means that we have shown that any general term can
be reduced to the form:
\begin{equation}
T'' = \ \big( Y_1(\vec\nabla) \partial_t\phi \big)
\cdots
\big( Y_m(\vec\nabla) \partial_t \phi \big) \,
\big( Z_1(\vec\nabla) \phi \big)
\cdots
\big( Z_n(\vec\nabla) \phi \big),
\end{equation}
which proves our claim that all higher-order time derivatives can be removed.

We conclude by explaining why it is not possible to remove first-order derivatives.
Considering the term $T''$ above, and integrating by parts the first time derivative:
\begin{equation}
\begin{aligned}
T''
&
\simeq Y_1(\vec\nabla) \phi \, \Big[
\big( Y_2(\vec\nabla) \partial_t^2\phi \big)
\cdots
\big( Y_m(\vec\nabla) \partial_t \phi \big)
+ \cdots
\Big] \,
\Big[
\big( Z_1(\vec\nabla) \phi \big)
\cdots
\big( Z_n(\vec\nabla) \phi \big)
\Big]
\\ & \qquad
+ Y_1(\vec\nabla) \phi \, \Big[
\big( Y_2(\vec\nabla) \partial_t\phi \big)
\cdots
\big( Y_m(\vec\nabla) \partial_t \phi \big)
\Big] \,
\Big[
\big(  Z_1(\vec\nabla) \partial_t\phi \big)
\cdots
\big( Z_n(\vec\nabla) \phi \big)
+ \cdots
\Big].
\end{aligned}
\end{equation}
All terms on the first line can be
written with fewer time derivatives using the field redefinition.
In order to write $T''$ with fewer time derivatives,
as in the strategy to obtain~\eqref{eq:IntLikeThing} in the time-dependent
case, it is crucial for the terms on the second line
to be proportional to $T''$ itself.
Here, this is not possible for unless all $Y_i, Z_i =1$, making it apparent
that in general first-order time derivatives cannot be removed.

\subsection{Hamiltonian for the redefined theory}  \label{hamform-4}  

Since the Lagrangian is of the form~\eqref{eq:lagrangian-redef-from-cov} after field redefinition, it is now a simple matter to write down a Hamiltonian for the nonlocal theory. To this end, first note that the canonical momenta $\Pi$ associated
with $\phi$ is 
given by a series in $\xi^2$:
\be \label{eq:CanonicalMomenta}
\Pi = \frac{\partial \tilde L}{\partial \dot{\phi}} = -\frac{\partial \tilde K}{\partial \dot{\phi}}
= \dot{\phi} \bigg[1 + \tfrac{32}{3} \xi^6  (\partial \phi)^2  + \mathcal{O} (\xi^8) \bigg]
= \dot{\phi} \bigg[1 - \tfrac{32}{3} \xi^6  \dot\phi^2 + \tfrac{32}{3} \xi^6  (\vec \nabla\phi)^2  + \mathcal{O} (\xi^8) \bigg].
\ee
We must  
invert this expression and determine $\dot{\phi}$ in terms of $\Pi$,  in order to write
the Hamiltonian. So let us make the ansatz
\be
\dot{\phi} = p_0 + \xi^2 p_2 + \xi^4 p_4 + \xi^6 p_6 + \mathcal{O} (\xi^8),
\ee
where $p_{2i}$ are some functions of $(\phi, \vec \nabla \phi, \Pi, \vec \nabla \Pi)$ and solve for $p_{2i}$ order-by-order in $\xi^2$ after inserting
this expansion in~\eqref{eq:CanonicalMomenta}.
We find
\be \label{eq:dotphi}
\dot{\phi} = \Pi  + \tfrac{32}{3} \xi^6 \Pi^3 -  \tfrac{32}{3} \xi^6 \Pi  (\vec \nabla \phi)^2 + \mathcal{O} (\xi^8).
\ee
This inversion was possible, even though the right-hand side of~\eqref{eq:CanonicalMomenta} is non-linear in $\dot{\phi}$, because we are working perturbatively in $\xi^2$.
After we insert the ansatz for $\dot{\phi}$, we are able to solve for $p_{2i}$ order-by-order. Note that the function $p_{2j}$ appears for the first time, linearly,
 at order $\xi^{2j}$ in the expansion of~\eqref{eq:CanonicalMomenta}.  It can
 therefore be solved for in terms of (known) lower $p_{2i}$'s. Thus, it is clear that this procedure can be extended to higher-orders in $\xi^2$ straightforwardly.

Now, substituting the equation~\eqref{eq:dotphi} for $\dot{\phi}$ in $\tilde K(\phi, \dot{\phi}, \vec\nabla \phi)$, we get the expression for which we replaced $\dot{\phi}$ with the canonical momenta $\Pi$ in $\tilde K$:
\be
\begin{split}
	\tilde K(\phi, \vec\nabla \phi,\Pi, \vec\nabla \Pi) &= \tfrac{1}{2} \bigl[- \Pi^2 + (\vec\nabla \phi)^2 \bigr]\\
	&\quad + \xi^6 \bigl[- \tfrac{32}{3} \Pi^4 + \tfrac{32}{3} \Pi^2 (\vec \nabla \phi)^2 + \tfrac{8}{3} (-\Pi^2 +\vec \nabla \phi)^2 \bigr] + \mathcal{O} (\xi^8),
\end{split}
\ee
which yields the following Hamiltonian after Legendre-transforming $L$ in~\eqref{eq:lagrangian-redef-from-cov}:
\be
\begin{split}
	H &= \dot{\phi} \Pi - \tilde L = \dot{\phi} \Pi + \tilde K + \tilde V  
	\\&= \bigl[\tfrac{1}{2} \Pi^2 + \tfrac{1}{2}(\vec\nabla \phi)^2 -\tfrac{1}{2} \phi^2 -\tfrac{1}{3}\phi^3  \bigr]
	\\ & \hspace{3cm}
	+ \xi^2 \bigl[\phi^3 + \phi^4 \bigr]
	- \xi^4 \bigl[\tfrac{3}{2} \phi^3 + \tfrac{19}{3} \phi^4 + \tfrac{16}{3} \phi^5 \bigr]
	\\ & \hspace{3cm}
	+ \xi^6 \bigl[\tfrac{8}{3} (-\Pi^2 +\vec \nabla \phi)^2  + \tfrac{3}{2}\phi^3 + \tfrac{178}{9} \phi^4 + \tfrac{472}{9} \phi^5 + \tfrac{112}{3} \phi^6  \bigr]
	+ \mathcal{O} (\xi^8).
\end{split}
\ee
It is clear that Hamiltonian can be found arbitrarily high orders in $\xi^2$.
The existence of a Hamiltonian makes that initial-value formulation for the nonlocal theory manifest and supports the claim that this theory is causal. Lastly, notice that Hamiltonian reduces to the one for the local cubic tachyonic theory when $\xi^2 =0$, as it should be.

\subsection{Light-cone formulation}
\label{lcjnlvlvg}

In this section, we 
consider the nonlocal theory in the light-cone frame and show that it becomes manifestly first-order in  light-cone time derivatives after a suitable field redefinition.
For $d>1$ spatial dimensions, light-cone coordinates 
are  defined by
\be
x^\pm \equiv \tfrac{1}{2} (x^0 \pm x^1)
\implies x^2 = -2 x^+ x^- + \vec{x}_T^2,
\ee
with $\vec{x}_T = (x_2, \cdots, x_d)$ collectively denotes the transverse directions. Here and henceforth $T$ subscript will denote the transverse directions to $x^{\pm}$. Similarly for derivatives and momentum we have
\be
\begin{split}
	\partial_\pm \equiv \tfrac{1}{2} (\tfrac{\partial}{\partial x^0} + \tfrac{\partial}{\partial x^1})
	&\implies \partial^2 = -2 \partial_- \partial_+ + \vec\nabla_T^2, \\
	p^\pm \equiv \tfrac{1}{2} (p^0 \pm p^1)
	&\implies p^2 = -2 p^+ p^- + \vec{p}_T^2.
\end{split}
\ee
We can also write
\be
p \cdot x = - p^+ x^- -p^- x^+ + \vec{p}_T \cdot \vec{x}_T.
\ee
In particular, Fourier transformation of the $x^-$ dependence
introduces $p^+$ dependence:
\be \label{eq:Fourierofphi}
\phi (x^-) = \int \frac{d p^+}{2 \pi} e^{-i p^+ x^-} \tilde{\phi}(p^+).
\ee
While light-cone field theories are often written in momentum space
and thus using $p^+$ rather than $x^-$,  we will work in coordinate space
throughout.  To translate, one can use $\partial_- = - i p^+$.

With $\tau \equiv x^+$,  the action is written as
$S = \int d\t d\vec{x}_T dx^-  \, L $,
with Lagrangian $L$ given by
\be
L\, = \, \tfrac{1}{2} \phi (-2 \partial_- \partial_\t + \vec\nabla_T^2 + 1) \phi + \tfrac{1}{3} (e^{-2\xi^2 \partial_- \partial_\t + \xi^2 \vec\nabla_T^2} \phi)^3 \, .
\ee

In the light-cone formulation of field theories, the light-cone time derivative $\partial_\tau$ is supposed to only appear in the standard kinetic term, and does so to first-order. This means we should be able to put the nonlocal theory in the form
\be
L\, = \, \tfrac{1}{2} \phi (-2 \partial_- \partial_\t + \vec\nabla_T^2 + 1) \phi + L_{int} (\phi, \vec\nabla_T \phi;  \partial_-) \, .
\ee
after performing appropriate field redefinitions. Here the interaction term, $L_{int}$, is expected to involve arbitrary powers of transverse derivatives $ \vec\nabla_T$ (i.e. being non-local in transverse directions), but no light-cone time derivative $\partial_\t$.
The price one has to pay to put the theory in this form
is to introduce \textit{nonlocality} in the $x^-$ direction at each order in $\xi^2$ so that $L_{int}$ involves the inverse of $\p_-$.

In order to show that it is possible to obtain the form described above, let us start with the covariant form of the action~\eqref{eq:lagrangian-cov}
and reduce it to the form above until we hit an obstruction for which we cannot eliminate light-cone time-derivatives while keeping covariance. As we have showed in the previous subsection, this will happen starting at the order $\mathcal O (\xi^6)$ for which we have
\be
\begin{aligned}
	\tilde L_6 &\simeq -\big(\tfrac{3}{2}\phi^3 + \tfrac{178}{9} \phi^4 + \tfrac{472}{9} \phi^5 + \tfrac{112}{3} \phi^6 \big) -\tfrac{8}{3} (\partial \phi)^4.
\end{aligned}
\ee
The term $(\partial \phi)^4$ cannot be eliminated covariantly and contains light-cone time derivatives.  To eliminate them we break manifest Lorentz covariance, which means specializing to the light-cone frame.

Let us first discuss the removal of terms.  Suppose we have a term of the form
$(\partial_\tau \phi) X [ \phi, \partial_\tau \phi]$ and consider a variation
$\delta \phi$ to remove the explicit $\tau$ derivative
\be
\begin{split}
T \ = &  \ \delta \phi  (\partial^2 \phi + \phi + \phi^2)  +
(\partial_\tau \phi) X [ \phi, \partial_\tau \phi]\\
= & \ \delta \phi (-2 \partial_-\partial_\t  \phi + \vec \nabla_T^2 \phi + \phi + \phi^2)
+\Bigl[  {1\over 2 \partial_-} (2 \partial_-\partial_\tau \phi)\Bigr]
 X [ \phi, \partial_\tau \phi]\\
 = & \ \delta \phi (-2  \partial_- \partial_\t\phi + \vec \nabla_T^2 \phi + \phi + \phi^2)
 - (2 \partial_-\partial_\tau \phi) {1\over 2 \partial_-} X [ \phi, \partial_\tau \phi]\,,
 \end{split}
\ee
where we introduced inverse $x^-$ derivatives and noticed that integrating by parts is allowed with inverse derivatives, as one can verify by either writing $1/\partial_-$ in Schwinger representation or by switching to the momentum basis.  We now choose
\be
\delta \phi = -  {1\over 2 \partial_-} X [ \phi, \partial_\tau \phi]\,,
\ee
This yields:
\be
T \ =  \ (\vec \nabla_T^2 \phi + \phi + \phi^2)
\Bigl[  -  {1\over 2 \partial_-} X [ \phi, \partial_\tau \phi] \Bigr]\,
=  \Bigl[   {1\over 2 \partial_-}  (\vec \nabla_T^2 \phi + \phi + \phi^2) \Bigr]
X [ \phi, \partial_\tau \phi] \,.
\ee
Summarizing the rule, we have
\be
\label{lc-rule}
(\partial_\tau \phi) X [ \phi, \partial_\tau \phi] \ \to \ \Bigl[   {1\over 2 \partial_-}  (\vec \nabla_T^2 \phi + \phi + \phi^2) \Bigr]
X [ \phi, \partial_\tau \phi] \ \ \ \hbox{with} \ \  \ \delta \phi = -  {1\over 2 \partial_-} X [ \phi, \partial_\tau \phi]\,.
\ee

Consider now the problematic term at $\xi^6$:
\be
- \tfrac{8}{3} (\partial \phi)^4 =
 - \tfrac{8}{3}(-2\partial_\t \phi \partial_- \phi + (\vec \nabla_T \phi)^2)^2 = -\tfrac{32}{3} (\partial_\t \phi \partial_- \phi)^2 +\tfrac{32}{3} (\partial_\t \phi) (\partial_- \phi) (\vec \nabla_T \phi)^2 -\tfrac{8}{3} (\vec \nabla_T \phi)^4.
\ee
Using the rule, this becomes
\be
\label{rgsjn}
\begin{split}
- \tfrac{8}{3} (\partial \phi)^4 \ \to \  & \
   -\tfrac{32}{3}\Bigl[   {1\over 2 \partial_-}  (\vec \nabla_T^2 \phi + \phi + \phi^2) \Bigr] (\partial_\t \phi) (\partial_- \phi)^2 \\
   & \  +\tfrac{32}{3} \Bigl[   {1\over 2 \partial_-}  (\vec \nabla_T^2 \phi + \phi + \phi^2) \Bigr]  (\partial_- \phi) (\vec \nabla_T \phi)^2 -\tfrac{8}{3} (\vec \nabla_T \phi)^4\,,
   \end{split}
\ee
after the field redefinition
\be
\delta \phi =   \tfrac{32}{3}  {1\over 2 \partial_-} \bigl[ (\partial_\t \phi) (\partial_- \phi)^2 \bigr] - \tfrac{32}{3}  {1\over 2 \partial_-} \bigl[(\partial_- \phi) (\vec \nabla_T \phi)^2\bigr]\,.
\ee
A second redefinition is needed for the first term in~(\ref{rgsjn}).  Indeed, we now get
\be
\label{rgsjnm}
\begin{split}
- \tfrac{8}{3} (\partial \phi)^4 \ \to \  & \
   -\tfrac{32}{3}\Bigl[   {1\over 2 \partial_-}  (\vec \nabla_T^2 \phi + \phi + \phi^2) \Bigr]^2  (\partial_- \phi)^2 \\
   & \  +\tfrac{32}{3} \Bigl[   {1\over 2 \partial_-}  (\vec \nabla_T^2 \phi + \phi + \phi^2) \Bigr]  (\partial_- \phi) (\vec \nabla_T \phi)^2 -\tfrac{8}{3} (\vec \nabla_T \phi)^4\,,
   \end{split}
\ee
after the field redefinition
\be
\delta' \phi =  \tfrac{32}{3} {1\over 2\partial_-}  \Biggl(
\Bigl[   {1\over 2 \partial_-}  (\vec \nabla_T^2 \phi + \phi + \phi^2) \Bigr] (\partial_- \phi)^2     \Biggr) \,.
\ee
The interaction has now been stripped of the offending light-cone time derivatives.

Combining the above results with the rest of $\tilde L_6$, we find
\begin{equation}
\begin{aligned}
\tilde L_6
&\simeq
-\big(\tfrac{3}{2}\phi^3 + \tfrac{178}{9} \phi^4 + \tfrac{472}{9} \phi^5 + \tfrac{112}{3} \phi^6 \big)
-\tfrac{8}{3} ((\vec \nabla_T \phi)^2)^2
\\ &\hspace{0.3cm}
+\tfrac{32}{3}  (\partial_- \phi) (\vec \nabla_T \phi)^2  \tfrac{1}{2\partial_-} \big(\vec \nabla_T^2 \phi + \phi + \phi^2 \big)
-\tfrac{32}{3} \big(\partial_- \phi\big)^2 \big[\tfrac{1}{2\partial_-} \big(\vec \nabla_T^2 \phi + \phi + \phi^2 \big)\big]^2.
\end{aligned}
\end{equation}
Including every field redefinition performed at this order to $\d_6 \phi$, the total field redefinition needed to reach this term is
\be
\begin{split}
	\d_6 \phi &= -\tfrac{1}{6} \partial^4 \phi^2 -\phi \partial^4 \phi-\tfrac{1}{3} (\partial^2 \phi)
	^2 + 18\phi^2 \partial^2 \phi + \tfrac{8}{3} \phi \partial^2 \phi
	\\&\quad\
	+\tfrac{70}{3} \phi (\partial \phi)^2 + \tfrac{7}{3} (\partial \phi)^2
	- \tfrac{85}{3} \phi^4 -\tfrac{151}{9} \phi^3 - \tfrac{3}{2} \phi^2
	\\&\quad\
	+ \tfrac{32}{3} \tfrac{1}{2\partial_-} \big[(\partial_\t \phi) (\partial_- \phi)^2 \big]
	- \tfrac{32}{3}\tfrac{1}{2\partial_-} \big[(\partial_- \phi) (\vec \nabla_T \phi)^2 \big]
	+\tfrac{32}{3} \tfrac{1}{2\partial_-} \big[(\partial_- \phi)^2 \tfrac{1}{2\partial_-} (\vec \nabla_T^2 \phi + \phi + \phi^2)\big].
\end{split}
\ee

In conclusion, we find the field-redefined Lagrangian is given by
\begin{equation}
\begin{aligned}
\tilde L
&\simeq \bigg[ \tfrac{1}{2} \, \phi \partial^2 \phi
+ \tfrac{1}{2} \, \phi^2
+ \tfrac{1}{3} \, \phi^3 \bigg]
- \xi^2 \bigg[\phi^3 + \phi^4 \bigg]
+ \xi^4 \bigg[\tfrac{3}{2} \phi^3 + \tfrac{19}{3} \phi^4 + \tfrac{16}{3} \phi^5\bigg]
\\ &\
+ \xi^6 \bigg[
-\big(\tfrac{3}{2}\phi^3 + \tfrac{178}{9} \phi^4 + \tfrac{472}{9} \phi^5 + \tfrac{112}{3} \phi^6 \big)
-\tfrac{8}{3} ((\vec \nabla_T \phi)^2)^2
\\ &\
+\tfrac{32}{3}  (\partial_- \phi) (\vec \nabla_T \phi)^2  \tfrac{1}{2\partial_-} \big(\vec \nabla_T^2 \phi + \phi + \phi^2 \big)
-\tfrac{32}{3} \big(\partial_- \phi\big)^2 \big[\tfrac{1}{2\partial_-} \big(\vec \nabla_T^2 \phi + \phi + \phi^2 \big)\big]^2
\bigg]
+ \mathcal{O} (\xi^8),
\end{aligned}
\end{equation}
after performing the following field redefinition  
\be
\phi \to \phi + \d \phi = \phi - \xi^2 \phi^2 + \xi^4 \Bigl[ -\tfrac{1}{2} \, \partial^2 \phi^2 - \phi \partial^2 \phi  +\tfrac{3}{2} \phi^2 + \tfrac{13}{3} \phi^3 \Bigr]
	+ \delta_6 \phi  + \mathcal{O} (\xi^8),
	\ee
with $\delta_6\phi$ given above. As desired,  the only $\tau$ derivative is in the kinetic term; the interactions became
 highly nonlocal in the $x^-$ direction. Notice that having commuting and off-diagonal derivatives $\partial_\pm$ was crucial to be able to run this argument, which doesn't have analog in the covariant approach we considered in the previous subsections.

It is quite simple to argue that one can always eliminate light-cone time derivatives.
Consider a higher derivative $\partial_\t^m \phi$ ($m>1$) acted by the derivatives $\vec \nabla_T$ and $\partial_-$ (possibly including the inverse factors), and multiplied by products of $\phi, \vec \nabla_T \phi, \partial_- \phi$,  and $\partial_\t \phi $.  We write such
a term as follows:
\be
T = [W (\vec\nabla, \partial_-) \, \partial_\t^m \phi] \,  Z[ \phi,  \partial_\t \phi, \vec\nabla , \partial_-] .
\ee
Up to a sign, we can integrate by parts all the spatial derivatives in $W$ and all but one of the
$\tau$ derivatives, finding
\be
T = \pm  (\partial_\t \phi)  \,  \partial_\t^{m-1} W (\vec\nabla, \partial_-) Z[ \phi,  \partial_\t \phi, \vec\nabla , \partial_-] .
\ee
The replacement discussed in the rule~(\ref{lc-rule}) now gives
\be
\begin{split}
T = & \ \pm  \Bigl[   {1\over 2 \partial_-}  (\vec \nabla_T^2 \phi + \phi + \phi^2) \Bigr]
 \,  \partial_\t^{m-1} W (\vec\nabla, \partial_-) Z[ \phi,  \partial_\t \phi, \vec\nabla , \partial_-] \\
 = &  \  W (\vec\nabla, \partial_-)  \Bigl[   {1\over 2 \partial_-} \partial_\t^{m-1} (\vec \nabla_T^2 \phi + \phi + \phi^2) \Bigr]
 \,  Z[ \phi,  \partial_\t \phi, \vec\nabla , \partial_-].
\end{split}
\ee
We integrated by parts back in the second equality above.  We see that we have reduced by one unit the number of $\tau$ derivatives.  Doing this recursively we can eliminate them all.

\section{Causality from superluminality} \label{causfromsup}

In this section, we discuss causality from the point of view of dispersion relations and superluminality.
The general approach consists in linearizing the equations of motion around an on-shell background.
To study this linear equation one considers a plane wave and computes the refractive index, defined as the ratio between the norm of the
spatial momentum
and the frequency, from which the phase and group velocities can be extracted~\cite{Leander:1996:RelationWavefrontSpeed, Aharonov:1969:SuperluminalBehaviorCausality,Shore:2003:CausalitySuperluminalLight, Adams:2006:CausalityAnalyticityIR, Shore:2007:SuperluminalityUVCompletion}.
As we review below, however,
these two velocities are not necessarily physical, and a proper assessment of superluminality asks whether the wavefront velocity (the infinite-frequency limit of the phase velocity) is larger than the speed of light.
Another way to understand this claim is to look at the effective light-cone of the wave equation~\cite{Aharonov:1969:SuperluminalBehaviorCausality, Shore:2003:CausalitySuperluminalLight, Shore:2007:SuperluminalityUVCompletion, Babichev:2008:KEssenceSuperluminalPropagation}.
Studying the propagation of a wave in the WKB approximation, one finds again that the relevant speed is the wavefront velocity.

The Lagrangian of the redefined theory is of the k-inflation type~\cite{ArmendarizPicon:1999:KInflation}, up to $\mathcal O(\xi^6)$ or up to the non-covariant terms at higher orders.
Causality of these theories have been investigated in~\cite{Garriga:1999:PerturbationsKinflation, Babichev:2008:KEssenceSuperluminalPropagation} and no problem has been found.
Hence, this provides a strong hint that the redefined theory is also perfectly causal.
We will not discuss here this approach and refer the reader to the original literature~\cite{ArmendarizPicon:1999:KInflation, Garriga:1999:PerturbationsKinflation, Babichev:2008:KEssenceSuperluminalPropagation} for more details.

\subsection{Velocities, refractive index, and effective light-cone}

Several velocities can be introduced when describing the propagation of a wave,
making it important to determine which ones are relevant for causality.
In this subsection, we briefly review the definitions of the most common velocities and refer to the literature for more details~\cite{Leander:1996:RelationWavefrontSpeed,Shore:2003:CausalitySuperluminalLight,Shore:2007:SuperluminalityUVCompletion, Babichev:2008:KEssenceSuperluminalPropagation,deRham:2020:CausalityCurvedSpacetimes}.

Using $c=1$ for the speed of light, we consider the $d$-dimensional momentum $k^\mu = (\omega, \vec k)$ associated to the wave
\be
e^{i (\vec k \cdot  \vec x  - \omega t )} \,.
\ee
We take $\omega$ to be real and positive but $\vec k$ can be complex.
The dispersion relation of a given theory is obtained by considering the linearized equation of motion in momentum space and the wave above.
The dispersion relation provides a relation between $\omega$ and $\vec k$.
This relation determines  $\vec k \cdot \vec k$ in terms of $\omega$.
Taking the square root, we have the function
\be
k(\omega)= \sqrt{\vec k \cdot \vec k}\,,     \ \ k(\omega) \in \mathbb{C},
\ee
which is complex in general.
The function $k(\omega)$ is related to the refractive index $n(\omega)$ as
follows:
\be
n(\omega)  =  {k(\omega) \over \omega} \,  \,.
\ee
This is in accord to the familiar relation $n = c/v$, with $c$ the speed of light and
$v$ the phase velocity $v_{\text{p}}= \omega/k(\omega)$.
The refractive index can be complex, indicating attenuation in the direction
of wave propagation if
Im$(n) > 0$ and gain if Im$(n) < 0$.
The phase velocity $v_{\text{p}}(\omega)$ and the group velocity $v_{\text{g}}(\omega)$ are defined by:
\begin{equation}
	v_{\text{p}}(\omega)
		\equiv \frac{1}{n(\omega)},
	\qquad
	v_{\text{g}}(\omega)
		\equiv \left( \frac{d k(\omega)}{d \omega} \right)^{-1}
		= \left(
			n(\omega)
			+ \omega \, \frac{d n(\omega)}{d \omega}
			\right)^{-1}.
\end{equation}
Finally, the \emph{wavefront} velocity
 is defined as the infinite-frequency limit of the phase velocity, that is argued to coincide with the infinite-frequency limit of the group
velocity~\cite{Leander:1996:RelationWavefrontSpeed}:
\begin{equation}
	\label{eq:wavefront}
	v_{\text{wf}}
		\equiv \lim_{\omega\to \infty} v_{\text{p}} (\omega)
		 		   = \lim_{\omega \to \infty} v_{\text{g}}(\omega)\,.
\end{equation}

It is sometimes stated that causality requires $v_{\text{p}}(\omega) \le 1$ or $v_{\text{g}}(\omega) \le 1$, but this is not correct.
Indeed, the phase velocity is not physical because it describes the propagation of a single frequency.
The group velocity is often
 more physical because it describes the propagation of a wave packet made of the superposition of several frequencies and, in this case, equals the speed of energy propagation.
There are cases, however, where this velocity is not physical: in particular, superluminal group velocities have been measured experimentally~\cite{groupvel}.
The wavefront velocity measures the effective propagation of a disturbance in an empty medium, and can be seen to be the relevant velocity from the theory of PDEs.

For later comparison, let's discuss the case of a free massive scalar
field~\cite{Babichev:2008:KEssenceSuperluminalPropagation}.
The dispersion relation reads
\begin{equation}
\label{eoifd}
	k^2 = - \omega^2 + k^2(\omega)
		= - m^2\,.
	\end{equation}
	Taking differentials we note that the group velocity equals
	the index of refraction and therefore it is the inverse of
	the phase velocity:
\be
- \omega d\omega + k (\omega) d k(\omega) = 0 \quad \implies \quad
v_{\rm g} = {d\omega \over dk(\omega)} = {k(\omega)\over \omega} = n(\omega)= {1\over v_{\rm p}}  \,.
\ee
Moreover, factoring $\omega^2$ in (\ref{eoifd}) we get
\be
\omega^2 \big( - 1 + n^2(\omega) \big)
= - m^2\,,
\ee
from which we see that
\begin{equation}
	n(\omega) = \sqrt{1 - {m^2 \over \omega^2}} \,,
	\qquad
	v_{\rm p}(\omega) = \frac{1}{\sqrt{1 - {m^2 \over \omega^2}}} \,.
\end{equation}
Consequently, for both $m^2 >0$ and $m^2 <0$ we find, on account of \eqref{eq:wavefront}:
\begin{equation}
	v_{\text{wf}} = \lim_{\omega\to \infty} v_{\rm p}(\omega)  = 1.
\end{equation}
Hence, signals in {\em both}  theories are causal.
We also see that:
\begin{itemize}
	\item $m^2 > 0$: \
		$v_{\text{g}}(\omega) \le 1 \le v_{\text{p}}(\omega)$, \
		for \ $  m< \omega < \infty$.

		\item $m^2 < 0$: \
		$v_{\text{p}}(\omega) \le 1 \le v_{\text{g}}(\omega)$, \
		for  \  $ 0 < \omega < \infty$.

\end{itemize}
The tachyonic theory has group velocity larger than $c=1$, but this
signals no acausality, it is a sign of instability.

\subsection{Nonlocal theory dispersion}

We consider the Lagrangian \eqref{clbbsnsdlinterm}:
\begin{equation}
	L= - \frac{1}{2} \, \phi (- \partial^2 + \epsilon) \phi
			+ \frac{1}{3} \, \big( e^{\xi^2 \partial^2} \phi \big)^3,
\end{equation}
where $m^2 =\epsilon = 1$ for the massive theory,  and $m^2 =\epsilon = -1$ for
 the tachyonic theory.
The linearized equations of motion of this theory (and the p-adic string)
were considered  in~\cite{Gomis:2003xv,Barnaby:2007ve}, and the following
discussion can be viewed as an elaboration of their analysis, geared towards
the questions of superluminality and focused on the nonlocality dependence.

Expanding the field around a background $\phi_0(x)$ which solves the equation of motion
\begin{equation}
	\phi(x)
		= \phi_0(x) + \psi(x),
\end{equation}
we find that the Lagrangian $L$ above becomes $L_\psi$ with:
\begin{equation}
	L_\psi  \equiv - \frac{1}{2} \, \psi \big( - \partial^2 + \epsilon) \psi
			+ (e^{\xi^2 \partial^2} \phi_0)
				\big( e^{\xi^2 \partial^2} \psi \big)^2
			+ \frac{1}{3} \, \big( e^{\xi^2 \partial^2} \psi \big)^3.
\end{equation}
The linearized equation of motion reads:
\begin{equation}
	0 = (- \partial^2 + \epsilon) \psi
			- 2 \, e^{\xi^2 \partial^2}
				\big( e^{\xi^2 \partial^2} \phi_0 \, e^{\xi^2 \partial^2} \psi \big).
\end{equation}
The constant solutions to the equations of motion are $\phi_0 = 0, \epsilon$.  The solution with $\phi_0=0$ is too simple:  the associated
field equation for $\psi$ is just that of a free scalar.   For the tachyonic theory
$(\epsilon=-1)$ this
is the unstable vacuum and for the massive theory $(\epsilon=1)$ this is the stable vacuum.

 Our focus here will be on the nontrivial solution $\phi_0 = \epsilon$.  For the tachyonic
theory  this represents the stable tachyon
vacuum, for the massive theory
 this represents the unstable vacuum.  Taking $\phi_0=\epsilon$,  the dispersion relation becomes:
\begin{equation}
	0
		= k^2 + \epsilon
			- 2 \epsilon \, e^{- 2 \xi^2 k^2} \quad \implies \quad
			k^2 = -\epsilon + 2\epsilon\,   e^{-2\xi^2 k^2} \,.
\end{equation}
Following~\cite{Gomis:2003xv,Barnaby:2007ve},
we can solve for $k^2$ using the Lambert $W$ function, also called
the product log function, and
defined as the solution for $w$ in the equation $z = w \, e^w$:
\begin{equation}
	z = w \, e^w \quad \Longrightarrow \quad
	w = W(z).
\end{equation}
Indeed, one can quickly show that
\be
x = a + b e^{cx} \,  \ \ \hbox{is solved for $x$ by:} \ \ \ x = a - \tfrac{1}{c} W  (-bc e^{ac})\,.
\ee
Using this result, we find
\be
k^2 = -\epsilon  + {1\over 2\xi^2}  W ( 4 \epsilon \xi^2 e^{2 \epsilon \xi^2}) \,.
\ee
This defines an effective mass through $k^2 = - m_{\text{eff}}^2$, so that we have:
\begin{equation}
	\label{eq:refractive-index}
	n^2(\omega) = 1 - \frac{m_{\text{eff}}^2}{\omega^2},
	\qquad
	m_{\text{eff}}^2 := \epsilon
			- \frac{1}{2 \xi^2} \,
	W\Big( 4 \epsilon \xi^2 e^{2 \epsilon \xi^2} \Big) \,.
\end{equation}
We have cast the solution in the form of a free massive scalar. This time, however, the effective mass of the scalar is a nontrivial function of the nonlocality
parameter.\footnote{
	One may also wonder if keeping or removing the mass from the exponential following~\eqref{clbb} changes the result.  It does not:  keeping the mass modifies the background solution in such a way that the mass cancels from the exponential in the argument of $W$.}

Before continuing, observe that the Lambert $W$ function is multi-valued and has an infinite number of branches $W_n$ with $n \in \mathbb Z$.
Above, we implicitly take the principal branch, i.e. $W \equiv W_0$.
Nonetheless, the dispersion relation admits an {\em infinite} number of solutions for the refractive index, with $W$ replaced by $W_n$ in \eqref{eq:refractive-index}.
The only real solutions are $W_0(x)$ for $x \ge -1/e$, and $W_{-1}(x)$ for $x \in [-1/e, 0)$.
Moreover, all $W_n$ with $n \neq 0$ has a branch point at the origin such that one does not recover the local theory for $\xi^2 \to 0$.
These solutions are non-perturbative in $\xi$, and one may wonder if they
are at all relevant to the redefined theory, as discussed in the introduction.
From an effective field theory perspective obtained as a perturbative
series in $\xi$, it makes sense to
ignore these solutions by requiring that the limit $\xi^2 \to 0$ is well-defined.
But from the viewpoint of causality of the original nonlocal theory, they are
certainly relevant. In the following, we focus mostly on the solution with $W \equiv W_0$ and comment on the other solutions at the end.

Some useful properties of the Lambert function $W \equiv W_0$ are as follows:
\begin{equation}
	\begin{gathered}
		W(0)= 0, \qquad  W(-1/e) = - 1,
		\qquad W(-\pi/2) = \tfrac{i \pi}{2},
		\qquad
		\lim_{x \to -\infty} W(x)
			= \infty + i \pi,
		\\
			W(x) \ge 0 \quad \text{for $x \ge 0$},
		\qquad
		W(x) \le 0 \quad
			\text{for $x \in [-1/e, 0]$},
		\\
		\Re W(x) \in [-1, \infty]
			\quad
			\text{for $x < -1/e$},
		\qquad
		\Im W(x) \in [0, \pi]
			\quad
			\text{for $x < -1/e$} \\
			\lim_{x\to \infty}  W (x) = \log x - \log \log x + {\cal O} (1) \,, \quad
			W(x) = x - x^2 + {\cal O} (x^3) .
	\end{gathered}
\end{equation}

\begin{figure}[!t]
	\centering
	\includegraphics[width=0.475\linewidth]{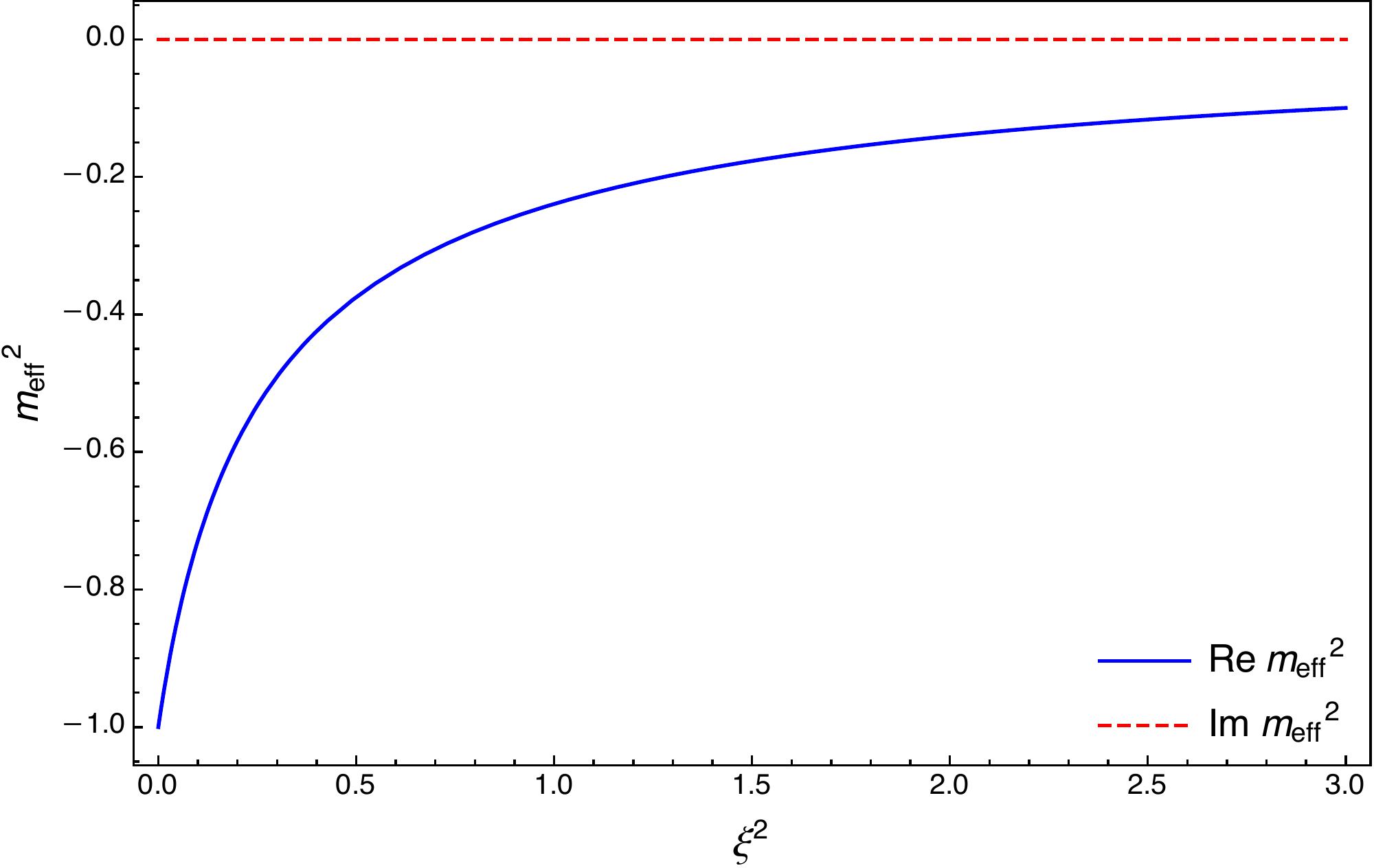}
	\includegraphics[width=0.475\linewidth]{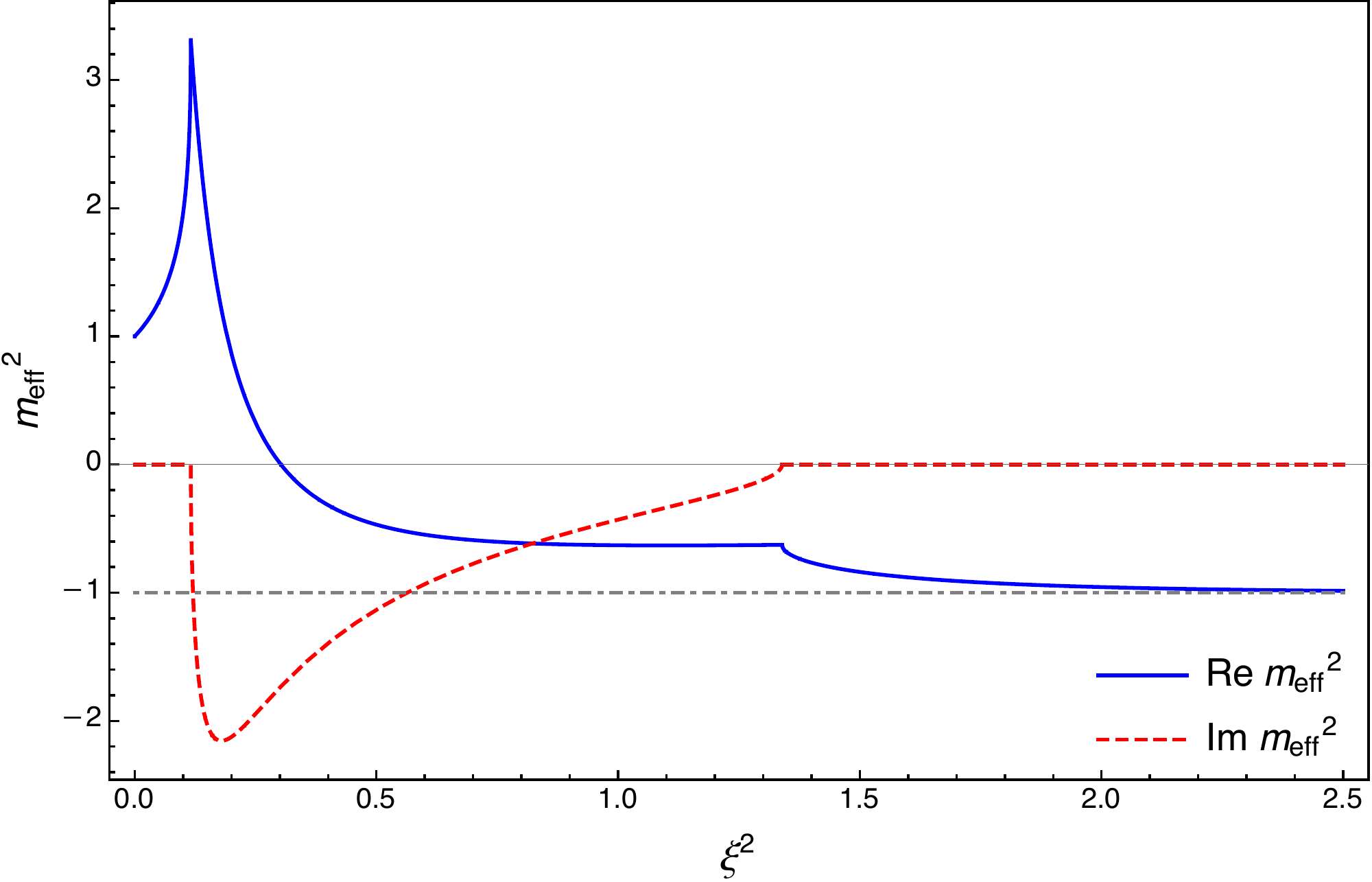}
	\caption{Effective mass-squared as a function of $\xi^2$ on the principal
	branch of the Lambert function.
	Left:  massive theory.  Since this represents the unstable vacuum, the mass-squared goes to $-1$ as $\xi\to 0$.  The mass-squared is real.
	 Right:  tachyonic theory. Since this represents
	the stable vacuum, the mass-squared goes to $+1$ as  $\xi\to 0$.  Here, the mass-squared acquires a non-zero imaginary part for a range of values of $\xi$.  }
	\label{fig:meff-neg}
\end{figure}

As expected, we recover the local result for $m_{\text{eff}}^2$ when $\xi^2 \to 0$.  Indeed, using the
expansion of $W$ for small argument we find
\begin{equation}
	\lim_{\xi^2 \to 0} m_{\text{eff}}^2
		= - \epsilon\,.
\end{equation}
This is the mass-squared at the $\phi_0$ vacuum, opposite to the one at the
$\phi=0$ vacuum.
The effective mass-squared is also finite as $\xi^2 \to \infty$, as
can be checked both for $\epsilon$ equal plus or minus one.
In the massive theory it goes to zero and in the tachyonic theory it goes to
$-1$.  Since the effective mass is always finite, the wavefront velocity is not affected by the nonlocality and we still find:
\begin{equation}
	v_{\text{wf}} = 1.
\end{equation}
The effective mass-squared $m_{\text{eff}}^2$ is plotted in figure~\ref{fig:meff-neg}.
The argument of $W$ is positive for the massive theory ($\epsilon = 1$)
and negative for the tachyonic theory ($\epsilon = -1$).
In the massive case, $W$ is always real such that there is no attenuation.
As $\xi^2$ goes from zero to infinity, the effective mass-squared
goes from $-1$ to zero, and nonlocality does not seem to have any specific influence.    Note that for the tachyonic theory the mass-squared is complex over a finite range
of values of $\xi^2$, as will be explained below.  The real part of the mass-squared goes from $1$ to $-1$ as the
nonlocality goes from zero to infinity.

\begin{figure}[!t]
	\centering
	\includegraphics[width=0.7\linewidth]{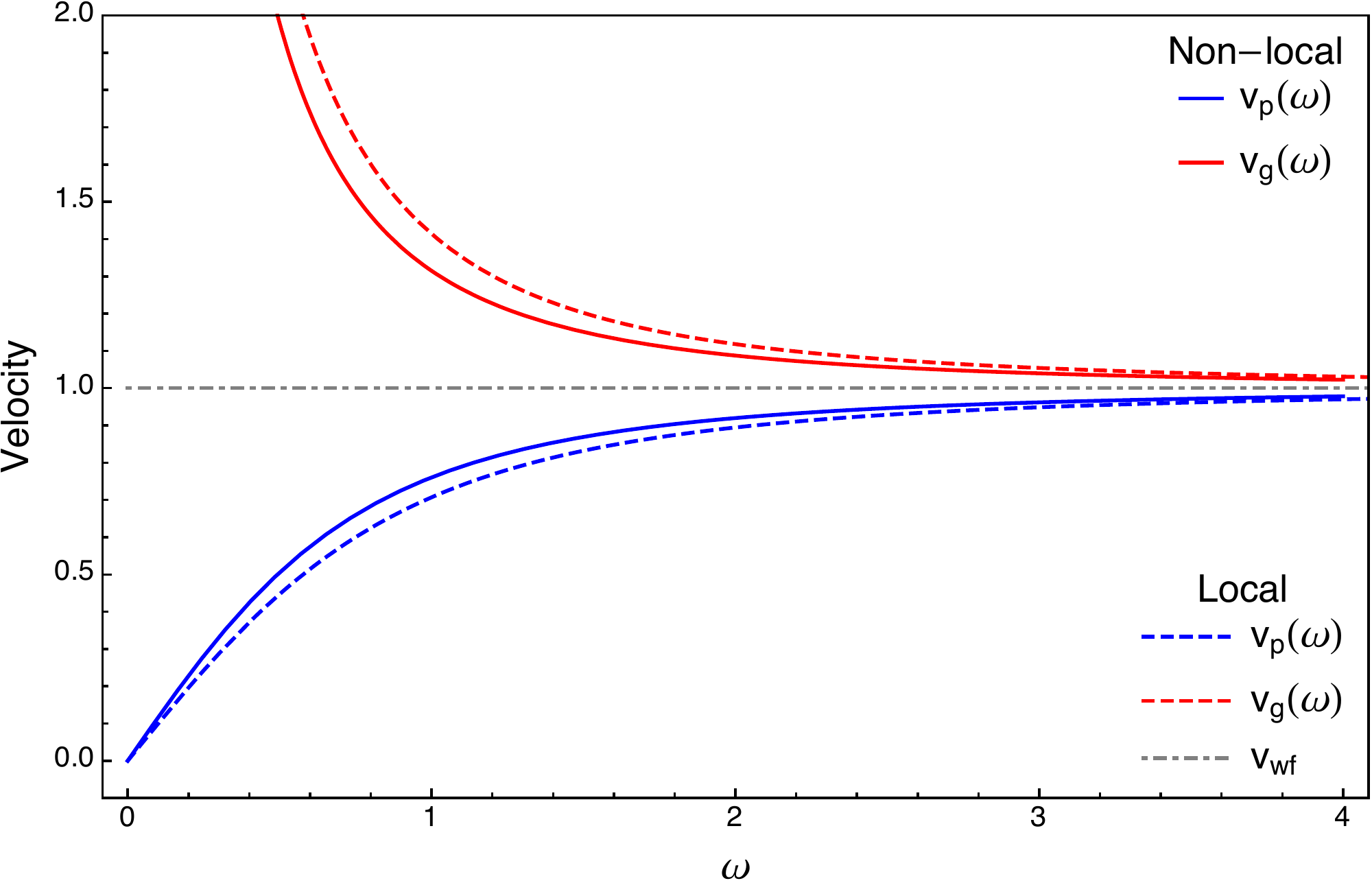}
	\caption{Phase and group velocities for a scalar with $m^2 > 0$ as a function of $\omega$.  The continuous lines correspond to the case $\xi^2 = 0.1$ (nonlocal).  The dashed lines correspond to $\xi=0$ (local).}
	\label{fig:velocities-pos}
\end{figure}

As noted before, we have $v_{\text{g}}(\omega) = n(\omega) = v_{\text{p}}(\omega)^{-1}$. The velocities for the massive theory, with $\xi^2 = 0.1$ for illustration,  are given are given as a function of frequency in Figure~\ref{fig:velocities-pos}. The figure also shows these quantities for the $\xi=0$ theory.

\begin{figure}[!t]
	\centering
	\includegraphics[width=0.7\linewidth]{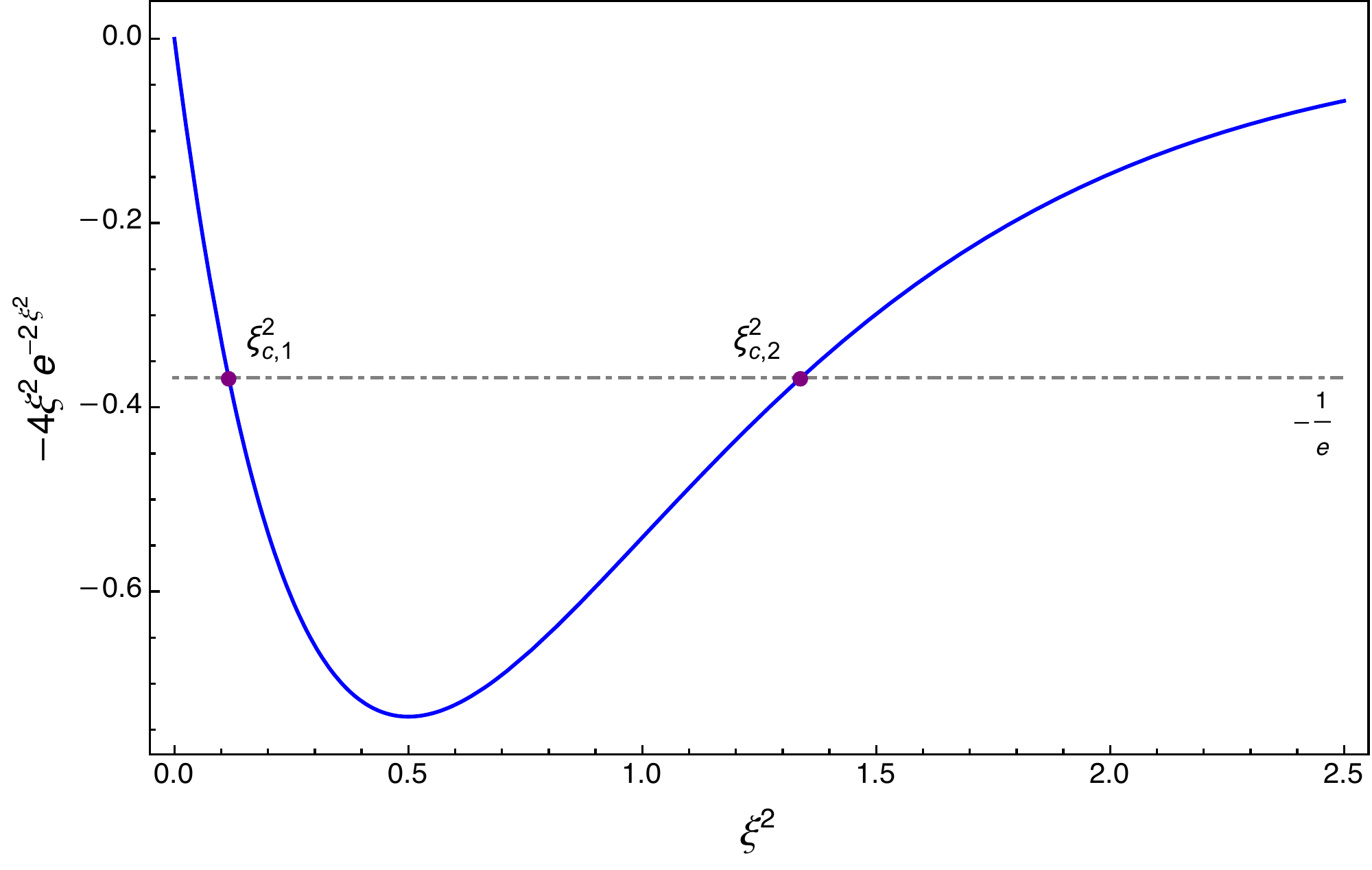}
	\caption{Behavior of the argument of $W$ in \eqref{eq:refractive-index} for
	the tachyonic theory.}
	\label{fig:meff-neg-Warg}
\end{figure}

The complex effective mass-squared of the tachyonic theory
arises because the argument of $W= W_0$ is negative and
 $W_0(z)$ has a branch point at $z = - 1/e$, and a branch cut
 extending along the real axis all the way to minus infinity.
The nonlocality parameter $\xi_c$ at such branch point must satisfy the equation
\begin{equation}
	-4  \xi_c^2 e^{-2\xi_c^2}
		= - \frac{1}{e}  \,.
\end{equation}
This equation admits two real solutions:
\begin{equation}
	\xi_{c,1}^2
		:= -\tfrac{1}{2} \, W\left( - \tfrac{1}{2 e} \right)
		\approx 0.116,
	\qquad
	\xi_{c,2}^2
		:= -\tfrac{1}{2} \, W_{-1}\left( - \tfrac{1}{2 e} \right)
		\approx 1.339.
\end{equation}
The behavior of the argument is plotted in Figure~\ref{fig:meff-neg-Warg}.
For $\xi^2$ on the interval
$(\xi^2_{c,1}, \xi^2_{c,2})$,  the argument of $W_0$ is less than $-1/e$ and
$W_0$ is complex, with positive imaginary part  (this corresponds to going
above the branch cut).  This results in a negative
imaginary part for the effective mass-squared.
This  can be interpreted as attenuation in propagation, because as $\Im W(x) \in [0, \pi]$, we find that the imaginary part of the effective mass satisfies:
\begin{equation}
	- \frac{\pi}{2 \xi^2}
		\le \Im m_{\text{eff}}^2
		\le 0\,,
\end{equation}
such that
\begin{equation}
	\Im n(\omega)^2
		\sim \frac{1}{2 \xi^2 \omega^2} > 0.
\end{equation}
If we use the principal branch of the square root, any sign for
 Im$\, n^2 (\omega)$ gives $\Im n(\omega) > 0$.

In local QFT, the imaginary part of the refractive index must be positive~\cite{Toll:1956cya,Shore:2007:SuperluminalityUVCompletion,deRham:2020:CausalityCurvedSpacetimes} as a consequence of causality of the retarded Green function, analyticity and unitarity.
While analyticity properties for nonlocal theories are only partially understood~\cite{DeLacroix:2018arq,Bhattacharya:2020gar,Bhattacharya:2021riu} (see~\cite{Hollowood:2007:CausalityMicroCausalityCurved} for a case where the usual argument fails), it is reassuring to see that we can consistenly
work with $\Im n(\omega) > 0$ here.

\begin{figure}[!t]
	\centering
	\includegraphics[width=0.475\linewidth]{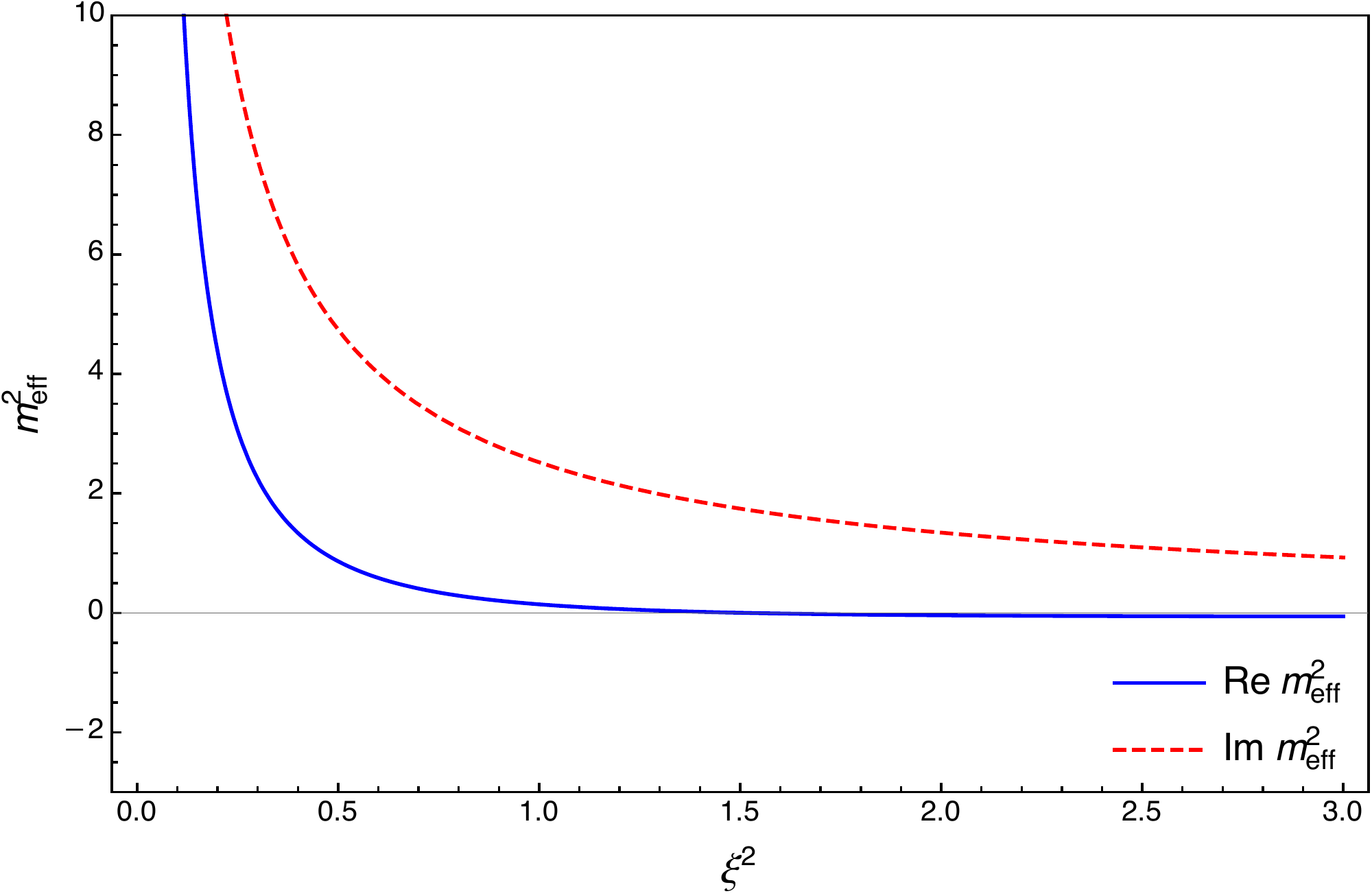}
	\includegraphics[width=0.475\linewidth]{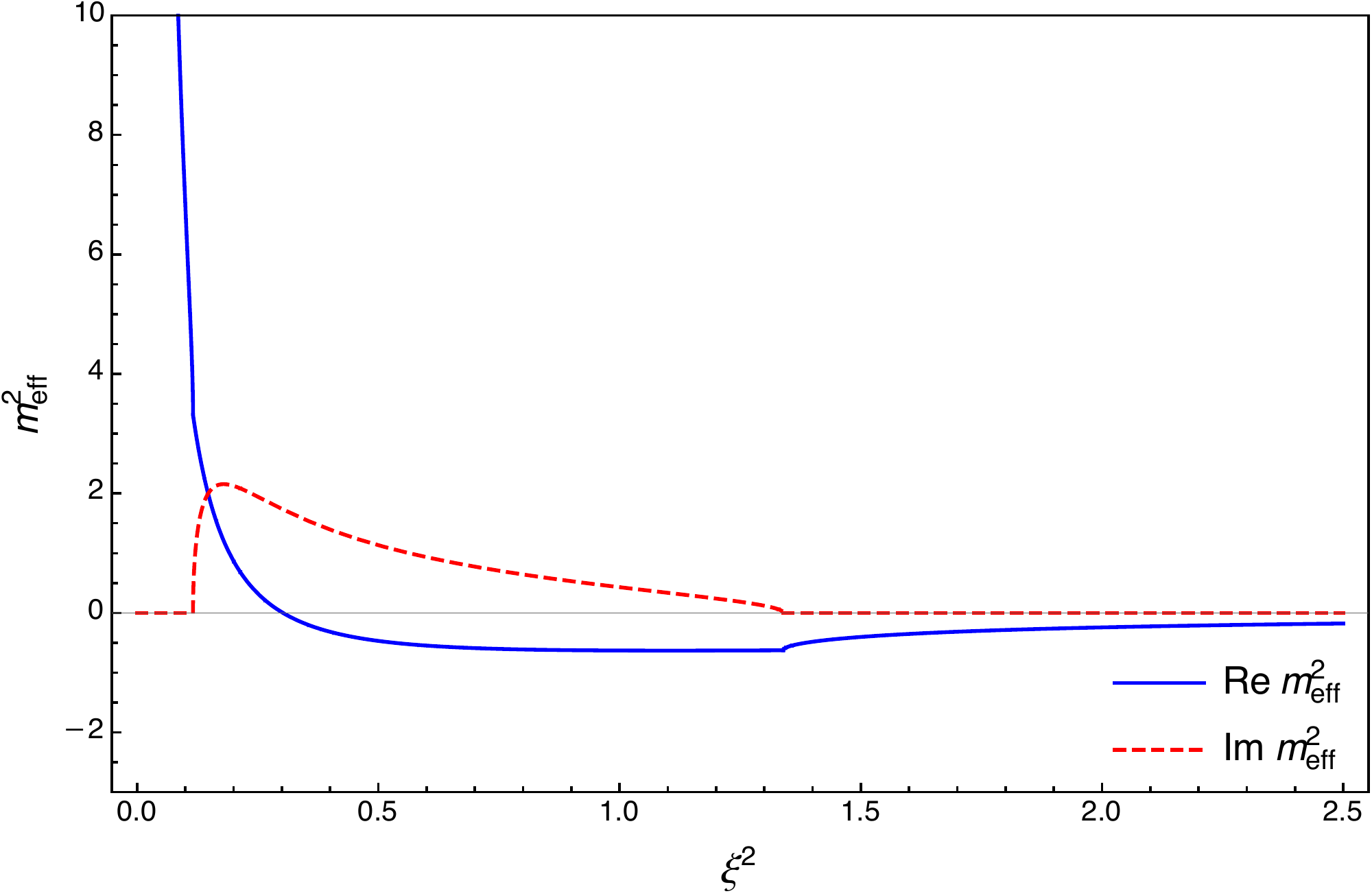}
	\caption{Effective mass-squared as a function of $\xi^2$
	using the  $W_{-1}$ branch of the Lambert function.  Left:  massive theory. Right: tachyonic theory.
	In both cases the limit $\xi^2 \to 0$ does not give the local theory; the mass-squared diverges.}
	\label{fig:meff-pos}
\end{figure}

Let us comment briefly on the other solutions with $W_n$ and $n \neq 0$.
In this case, the solutions are generically complex. The effective mass \eqref{eq:refractive-index}, however, is finite for all $n$ and $\xi^2 > 0$, implying that the wavefront velocity is still $v_{\rm wf} = 1$ and there is no superluminal propagation.
Moreover, given arbitrary finite $\Im n^2$ of any sign, we can always take the branch such that $\Im n > 0$. This means that any equation of the form $k^2 = - m_{\text{eff}}^2$ is consistent with $\Im n > 0$.  In Figure~\ref{fig:meff-pos} we show the $\xi^2$ dependence of the effective mass-squared for a solution that uses
the branch $W_{-1}$.

\section{Remarks and open questions}

We have analyzed nonlocal field theories that appear in string field theory using
an effective field theory perspective.  We used a derivative expansion of the Lagrangian, and at each order used field redefinitions to adjust the theory
and to put it into some canonical
form.  The results appear to be technically novel and show a set of subtleties and
possibilities.  We have seen that full locality can be achieved for the purely time-dependent situation and an initial value problem can be formulated for general configurations at the cost of losing manifest Lorentz covariance.  Moreover, our analysis suggests, but does not demonstrate, that the theory in question is causal.

There are a number of open questions that could be investigated:

\begin{enumerate}

\item  It would be significant progress if a closed form expression could
be found for the potential $\tilde V (\phi; \xi^2)$ relevant to the purely time-dependent theory.  This could require a clever choice for the quasi-symmetry discussed in
section~\ref{nonuniscapot} that is used to simplify the potential.  The availability of such a closed
form would help settle if the redefinition of the theory is valid non-perturbatively.

\item In a similar vein, it would be useful if some simpler forms or exact subsectors
of the redefined Lagrangians were available for the case of fully spacetime dependent configurations.
Here, a comparison of the S-matrix elements of the nonlocal and the redefined
theory could help---as it allowed us to fix the cubic term in the potential of the redefined theory nonperturbatively.

\item  We have argued that the rolling solution in the redefined canonical model
poses a puzzle, as it appears to represent the decay and subsequent
re-creation of the unstable spacefilling D-brane.  This is in sharp
contrast with the SFT intuition and evidence that the final state of the tachyon
is not the same as the original state.   It would be important to confirm that the
wild oscillations of the tachyon model map under the redefinition to a smooth
rolling solution.  At stake is whether the tachyon model captures {\em any}
of the physics of the string theory.  We need to learn how to compute
pressure in nonlocal field theory, and test this against the string field theory
results.    The pressure in rolling solutions does
not seem to go to zero,  as expected for tachyon matter. Instead, calculations in
p-adic theories and the model discussed here (see~\cite{Yang:2002nm}) exhibit
a pressure with ever growing oscillations.  It is conceivable that improvements of the
stress tensor could change this conclusion.

\item A full investigation of the causality of the original
nonlocal (quantum) theory is warranted.
A path-integral approach to the question could use a condition
derived by Bogoliubov~\cite{bogoliubov:tp} and discussed extensively by Tomboulis~\cite{Tomboulis:2017rvd} as well as 't Hooft and Veltman~\cite{tHooft:1973wag}.

\item  An even deeper question asks if our removal of time
nonlocalities of the theory,
done here perturbatively, is valid beyond perturbation theory and thus
{\em defines} an equivalent, manifestly causal
formulation of the theory.   This claim requires two necessary but perhaps
not sufficient conditions.  One, the redefined theory must be well defined---that is,
physically unique.  Second,  the original nonlocal theory is found to be
completely causal, once this analysis can be done reliably.  Any progress
on this question would help rule out causality troubles in string theory.

\end{enumerate}

\section*{Acknowledgments}

We thank Joaquim Gomis, Daniel Harlow, Olaf Hohm, and Iain Stewart for stimulating discussions.
We are very grateful to Ted Erler, Yuji Okawa, and Ashoke Sen for valuable comments and suggestions concerning the material in this paper.

This material is based upon work supported by the U.S. Department of Energy, Office of Science, Office of High Energy Physics of U.S. Department of Energy under grant Contract Number  DE-SC0012567. This project has received funding from the European Union's Horizon 2020 research and innovation program under the Marie Sklodowska-Curie grant agreement No 891169.


\begin{thebibliography}{99}



\bibitem{Pius:2016jsl}
R.~Pius and A.~Sen,
``Cutkosky rules for superstring field theory,''
JHEP \textbf{10}, 024 (2016)
[erratum: JHEP \textbf{09}, 122 (2018)]
doi:10.1007/JHEP10(2016)024
[arXiv:1604.01783 [hep-th]].

\bibitem{deLacroix:2017lif}
C.~de Lacroix, H.~Erbin, S.~P.~Kashyap, A.~Sen and M.~Verma,
``Closed Superstring Field Theory and its Applications,''
Int. J. Mod. Phys. A \textbf{32}, no.28n29, 1730021 (2017)
doi:10.1142/S0217751X17300216
[arXiv:1703.06410 [hep-th]].

\bibitem{Susskind:1994sm}
L.~Susskind and J.~Uglum,
``Black hole entropy in canonical quantum gravity and superstring theory,''
Phys. Rev. D \textbf{50}, 2700-2711 (1994)
doi:10.1103/PhysRevD.50.2700
[arXiv:hep-th/9401070 [hep-th]];
D.~A.~Lowe, L.~Susskind and J.~Uglum,
``Information spreading in interacting string field theory,''
Phys. Lett. B \textbf{327}, 226-233 (1994)
doi:10.1016/0370-2693(94)90722-6
[arXiv:hep-th/9402136 [hep-th]];
D.~A.~Lowe, J.~Polchinski, L.~Susskind, L.~Thorlacius and J.~Uglum,
``Black hole complementarity versus locality,''
Phys. Rev. D \textbf{52}, 6997-7010 (1995)
doi:10.1103/PhysRevD.52.6997
[arXiv:hep-th/9506138 [hep-th]].

\bibitem{Lowe:1994ah}
D.~A.~Lowe and A.~Strominger,
``Strings near a Rindler or black hole horizon,''
Phys. Rev. D \textbf{51}, 1793-1799 (1995)
doi:10.1103/PhysRevD.51.1793
[arXiv:hep-th/9410215 [hep-th]].

\bibitem{Polchinski:1995ta}
J.~Polchinski,
``String theory and black hole complementarity,''
[arXiv:hep-th/9507094 [hep-th]].


\bibitem{Hata:1995di}
H.~Hata, H.~Oda and S.~Yahikozawa,
``String field theory in Rindler space-time and string thermalization,''
Prog. Theor. Phys. \textbf{96}, 985-1020 (1996)
doi:10.1143/PTP.96.985
[arXiv:hep-th/9512206 [hep-th]].

\bibitem{Giddings:2006vu}
S.~B.~Giddings,
``Locality in quantum gravity and string theory,''
Phys. Rev. D \textbf{74}, 106006 (2006)
doi:10.1103/PhysRevD.74.106006
[arXiv:hep-th/0604072 [hep-th]];
S.~B.~Giddings, D.~J.~Gross and A.~Maharana,
``Gravitational effects in ultra high-energy string scattering,''
Phys. Rev. D \textbf{77}, 046001 (2008)
doi:10.1103/PhysRevD.77.046001
[arXiv:0705.1816 [hep-th]];
S.~B.~Giddings,
``Nonlocality versus complementarity: A Conservative approach to the information problem,''
Class. Quant. Grav. \textbf{28}, 025002 (2011)
doi:10.1088/0264-9381/28/2/025002
[arXiv:0911.3395 [hep-th]];
S.~B.~Giddings,
``Nonviolent nonlocality,''
Phys. Rev. D \textbf{88}, 064023 (2013)
doi:10.1103/PhysRevD.88.064023
[arXiv:1211.7070 [hep-th]].

\bibitem{Lowe:2014vfa}
D.~A.~Lowe and L.~Thorlacius,
``Black hole complementarity: The inside view,''
Phys. Lett. B \textbf{737}, 320-324 (2014)
doi:10.1016/j.physletb.2014.08.062
[arXiv:1402.4545 [hep-th]].

\bibitem{Dodelson:2017hyu}
M.~Dodelson and E.~Silverstein,
``Long-Range Nonlocality in Six-Point String Scattering: simulation of black hole infallers,''
Phys. Rev. D \textbf{96}, no.6, 066009 (2017)
doi:10.1103/PhysRevD.96.066009
[arXiv:1703.10147 [hep-th]];
M.~Dodelson and E.~Silverstein,``String-theoretic breakdown of effective field theory near black hole horizons,''
Phys. Rev. D \textbf{96}, no.6, 066010 (2017)
doi:10.1103/PhysRevD.96.066010
[arXiv:1504.05536 [hep-th]];
A.~Mousatov and E.~Silverstein,
``Recovering Infalling Information via String Spreading,''
[arXiv:2002.12377 [hep-th]].

\bibitem{Naseer:2020lwr}
U.~Naseer,
``Entanglement Entropy in Closed String Theory,''
[arXiv:2002.12148 [hep-th]].


\bibitem{llosa}
J.~Llosa and J.~Vives,
``Hamiltonian formalism for nonlocal Lagrangians,''
J. Math. Phys. 35, 2856 (1994)
doi:10.1063/1.530492;
J.~Gomis, K.~Kamimura and J.~Llosa,
``Hamiltonian formalism for space-time noncommutative theories,''
Phys. Rev. D \textbf{63}, 045003 (2001)
doi:10.1103/PhysRevD.63.045003
[arXiv:hep-th/0006235 [hep-th]].

\bibitem{Gomis:2003xv}
J.~Gomis, K.~Kamimura and T.~Ramirez,
``Physical degrees of freedom of non-local theories,''
Nucl. Phys. B \textbf{696}, 263-291 (2004)
doi:10.1016/j.nuclphysb.2004.06.046
[arXiv:hep-th/0311184 [hep-th]].

\bibitem{Tomboulis:2015gfa}
E.~T.~Tomboulis,
``Nonlocal and quasilocal field theories,''
Phys. Rev. D \textbf{92}, no.12, 125037 (2015)
doi:10.1103/PhysRevD.92.125037
[arXiv:1507.00981 [hep-th]].


\bibitem{Eliezer:1989cr}
D.~A.~Eliezer and R.~P.~Woodard,
``The Problem of Nonlocality in String Theory,''
Nucl. Phys. B \textbf{325}, 389 (1989)
doi:10.1016/0550-3213(89)90461-6

\bibitem{Erler:2004hv}
T.~Erler and D.~J.~Gross,
``Locality, causality, and an initial value formulation for open string field theory,''
[arXiv:hep-th/0406199 [hep-th]].

\bibitem{Erler:2020beb}
T.~Erler and H.~Matsunaga,
``Mapping between Witten and Lightcone String Field Theories,''
[arXiv:2012.09521 [hep-th]].

\bibitem{Criado:2018sdb}
J.~C.~Criado and M.~P\'erez-Victoria,
``Field redefinitions in effective theories at higher orders,''
JHEP \textbf{03}, 038 (2019)
doi:10.1007/JHEP03(2019)038
[arXiv:1811.09413 [hep-ph]].

\bibitem{Hohm:2019jgu}
O.~Hohm and B.~Zwiebach,
``Duality invariant cosmology to all orders in $\alpha$',''
Phys. Rev. D \textbf{100}, no.12, 126011 (2019)
doi:10.1103/PhysRevD.100.126011
[arXiv:1905.06963 [hep-th]].



\bibitem{Moeller:2002vx}
N.~Moeller and B.~Zwiebach,
``Dynamics with infinitely many time derivatives and rolling tachyons,''
JHEP \textbf{10}, 034 (2002)
doi:10.1088/1126-6708/2002/10/034
[arXiv:hep-th/0207107 [hep-th]].



\bibitem{Moeller:2003gg}
N.~Moeller and M.~Schnabl,
``Tachyon condensation in open closed p adic string theory,''
JHEP \textbf{01}, 011 (2004)
doi:10.1088/1126-6708/2004/01/011
[arXiv:hep-th/0304213 [hep-th]].

\bibitem{Fujita:2003ex}
M.~Fujita and H.~Hata,
``Time dependent solution in cubic string field theory,''
JHEP \textbf{05}, 043 (2003)
doi:10.1088/1126-6708/2003/05/043
[arXiv:hep-th/0304163 [hep-th]].


\bibitem{Yang:2002nm}
H.~t.~Yang,
``Stress tensors in p-adic string theory and truncated OSFT,''
JHEP \textbf{11}, 007 (2002)
doi:10.1088/1126-6708/2002/11/007
[arXiv:hep-th/0209197 [hep-th]].


\bibitem{Minahan:2001pd}
J.~A.~Minahan,
``Quantum corrections in p-adic string theory,''
[arXiv:hep-th/0105312 [hep-th]].



\bibitem{Barnaby:2007ve}
N.~Barnaby and N.~Kamran,
``Dynamics with infinitely many derivatives: The Initial value problem,''
JHEP \textbf{02}, 008 (2008)
doi:10.1088/1126-6708/2008/02/008
[arXiv:0709.3968[hep-th]].

\bibitem{Zwiebach:2000dk}
B.~Zwiebach,
``A Solvable toy model for tachyon condensation in string field theory,''
JHEP \textbf{09}, 028 (2000)
doi:10.1088/1126-6708/2000/09/028
[arXiv:hep-th/0008227 [hep-th]].

\bibitem{Sen:2002nu}
A.~Sen,
``Rolling tachyon,''
JHEP \textbf{04}, 048 (2002)
doi:10.1088/1126-6708/2002/04/048
[arXiv:hep-th/0203211 [hep-th]].
\bibitem{Sen:2002in}
A.~Sen,
``Tachyon matter,''
JHEP \textbf{07}, 065 (2002)
doi:10.1088/1126-6708/2002/07/065
[arXiv:hep-th/0203265 [hep-th]].



\bibitem{Schnabl:2007az}
M.~Schnabl,
``Comments on marginal deformations in open string field theory,''
Phys. Lett. B \textbf{654}, 194-199 (2007)
doi:10.1016/j.physletb.2007.08.023
[arXiv:hep-th/0701248 [hep-th]].


\bibitem{Kiermaier:2007ba}
M.~Kiermaier, Y.~Okawa, L.~Rastelli and B.~Zwiebach,
``Analytic solutions for marginal deformations in open string field theory,''
JHEP \textbf{01}, 028 (2008)
doi:10.1088/1126-6708/2008/01/028
[arXiv:hep-th/0701249 [hep-th]].


\bibitem{Kiermaier:2010cf}
M.~Kiermaier, Y.~Okawa and P.~Soler,
``Solutions from boundary condition changing operators in open string field theory,''
JHEP \textbf{03}, 122 (2011)
doi:10.1007/JHEP03(2011)122
[arXiv:1009.6185 [hep-th]].

\bibitem{Ellwood:2008jh}
I.~Ellwood,
``The Closed string tadpole in open string field theory,''
JHEP \textbf{08}, 063 (2008)
doi:10.1088/1126-6708/2008/08/063
[arXiv:0804.1131 [hep-th]].

\bibitem{Kishimoto:2008zj}
I.~Kishimoto,
``Comments on gauge invariant overlaps for marginal solutions in open string field theory,''
Prog. Theor. Phys. \textbf{120}, 875-886 (2008)
doi:10.1143/PTP.120.875
[arXiv:0808.0355 [hep-th]].


\bibitem{Kiermaier:2008qu}
M.~Kiermaier, Y.~Okawa and B.~Zwiebach,
``The boundary state from open string fields,''
[arXiv:0810.1737 [hep-th]].


\bibitem{Kudrna:2012re}
M.~Kudrna, C.~Maccaferri and M.~Schnabl,
``Boundary State from Ellwood Invariants,''
JHEP \textbf{07}, 033 (2013)
doi:10.1007/JHEP07(2013)033
[arXiv:1207.4785 [hep-th]].

\bibitem{Ellwood:2007xr}
I.~Ellwood,
``Rolling to the tachyon vacuum in string field theory,''
JHEP \textbf{12}, 028 (2007)
doi:10.1088/1126-6708/2007/12/028
[arXiv:0705.0013 [hep-th]].




\bibitem{Hellerman:2008wp}
S.~Hellerman and M.~Schnabl,
``Light-like tachyon condensation in Open String Field Theory,''
JHEP \textbf{04}, 005 (2013)
doi:10.1007/JHEP04(2013)005
[arXiv:0803.1184 [hep-th]].


\bibitem{Erler:2019xof}
T.~Erler, T.~Masuda and M.~Schnabl,
``Rolling near the tachyon vacuum,''
JHEP \textbf{04}, 104 (2020)
doi:10.1007/JHEP04(2020)104
[arXiv:1902.11103 [hep-th]].


\bibitem{Gutperle:2002ai}
M.~Gutperle and A.~Strominger,
``Space - like branes,''
JHEP \textbf{04}, 018 (2002)
doi:10.1088/1126-6708/2002/04/018
[arXiv:hep-th/0202210 [hep-th]].

\bibitem{bogoliubov:tp}
N. N. Bogoliubov and D. V. Shirkov, Fortschr. der Physik, 3 (1955) 439; N. N.
Bogoliubov, B. Medvedev and M. Polivanov, ibid, 6 (1958) 169; N. N. Bogoliubov and D. V. Shirkov, Introduction to the theory of quantized fields, Wiley-Interscience (1959).

\bibitem{Tomboulis:2017rvd}
E.~T.~Tomboulis,
``Causality and Unitarity via the Tree-Loop Duality Relation,''
JHEP \textbf{05}, 148 (2017)
doi:10.1007/JHEP05(2017)148
[arXiv:1701.07052 [hep-th]].


\bibitem{Adams:2006:CausalityAnalyticityIR}
A.~Adams, N.~Arkani-Hamed, S.~Dubovsky, A.~Nicolis, and R.~Rattazzi,
“Causality, Analyticity and an IR Obstruction to UV Completion,”
JHEP, vol. 2006, no. 10, pp. 014–014 (2006)
doi: 10.1088/1126-6708/2006/10/014
[hep-th/0602178]

\bibitem{Aharonov:1969:SuperluminalBehaviorCausality}
Y.~Aharonov, A.~Komar, and L.~Susskind,
“Superluminal Behavior, Causality, and Instability,”
Phys. Rev., vol. 182, no. 5, pp. 1400–1403 (1969),
doi: 10.1103/PhysRev.182.1400.

\bibitem{Leander:1996:RelationWavefrontSpeed}
J.~L.~Leander,
“On the relation between the wavefront speed and the group velocity concept,”
The Journal of the Acoustical Society of America, vol. 100, no. 6, pp. 3503–3507, Dec. 1996,
doi:10.1121/1.417249.

\bibitem{Shore:2003:CausalitySuperluminalLight}
G.~M.~Shore,
“Causality and Superluminal Light,”
presented at the Time and Matter, Venice, Italy, Feb. 2003
[arxiv:gr-qc/0302116].

\bibitem{Shore:2007:SuperluminalityUVCompletion}
G.~M.~Shore,
“Superluminality and UV Completion,”
Nuclear Physics B, vol. 778, no. 3, pp. 219–258, Sep. 2007,
doi:10.1016/j.nuclphysb.2007.03.034
[arxiv:hep-th/0701185].

\bibitem{Sen:1999nx}
A.~Sen and B.~Zwiebach,
``Tachyon condensation in string field theory,''
JHEP \textbf{03}, 002 (2000)
doi:10.1088/1126-6708/2000/03/002
[arXiv:hep-th/9912249 [hep-th]].

\bibitem{Cho:2019anu}
M.~Cho,
``Open-closed Hyperbolic String Vertices,''
JHEP \textbf{05}, 046 (2020)
doi:10.1007/JHEP05(2020)046
[arXiv:1912.00030 [hep-th]].

\bibitem{Firat:2021ukc}
A.~H.~F\i{}rat,
``Hyperbolic three-string vertex,''
doi:10.1007/JHEP08(2021)035
[arXiv:2102.03936 [hep-th]].

\bibitem{Brekke:1987ptq}
L.~Brekke, P.~G.~O.~Freund, M.~Olson and E.~Witten,
``Nonarchimedean String Dynamics,''
Nucl. Phys. B \textbf{302}, 365-402 (1988)
doi:10.1016/0550-3213(88)90207-6

\bibitem{Halverson:2020trp}
J.~Halverson, A.~Maiti and K.~Stoner,
``Neural Networks and Quantum Field Theory,''
Mach. Learn. Sci. Tech. \textbf{2}, no.3, 035002 (2021)
doi:10.1088/2632-2153/abeca3
[arXiv:2008.08601 [cs.LG]].

\bibitem{Erbin:2021kqf}
H.~Erbin, V.~Lahoche and D.~O.~Samary,
``Nonperturbative renormalization for the neural network-QFT correspondence,''
[arXiv:2108.01403 [hep-th]].

\bibitem{deRham:2020:CausalityCurvedSpacetimes}
C.~de Rham and A.~J.~Tolley,
“Causality in Curved Spacetimes: The Speed of Light \& Gravity,”
Phys. Rev. D, vol. 102, no. 8, p. 084048, Oct. 2020,
doi:10.1103/PhysRevD.102.084048
[arxiv:2007.01847].

\bibitem{ArmendarizPicon:1999:KInflation}
C.~Armendariz-Picon, T.~Damour, and V.~Mukhanov,
“k-Inflation,”
Physics Letters B, vol. 458, no. 2–3, pp. 209–218, Jul. 1999,
doi: 10.1016/S0370-2693(99)00603-6
[hep-th/9904075].

\bibitem{Garriga:1999:PerturbationsKinflation}
J.~Garriga and V.~Mukhanov,
“Perturbations in k-inflation,”
Physics Letters B, vol. 458, no. 2–3, pp. 219–225 (1999),
doi: 10.1016/S0370-2693(99)00602-4
[hep-th/9904176].

\bibitem{Babichev:2008:KEssenceSuperluminalPropagation}
E.~Babichev, V.~Mukhanov, and A.~Vikman,
“k-Essence, superluminal propagation, causality and emergent geometry,”
J. High Energy Phys., vol. 2008, no. 02, pp. 101–101, Feb. 2008,
doi:10.1088/1126-6708/2008/02/101
[arxiv:0708.0561].

\bibitem{groupvel}
N.~Brunner, V.~Scarani, M.~Wegmüller, M.~Legré and N.~Gisin,
“Direct Measurement of Superluminal Group Velocity and Signal Velocity in an Optical Fiber,”
Phys. Rev. Lett., vol. 93, issue 20, Nov. 2004,
doi:10.1103/PhysRevLett.93.203902
[arXiv:quant-ph/0407155].

\bibitem{Hollowood:2007:CausalityMicroCausalityCurved}
T.~J.~Hollowood and G.~M.~Shore,
“Causality and Micro-Causality in Curved Spacetime,”
Physics Letters B, vol. 655, no. 1–2, pp. 67–74, Oct. 2007,
doi: 10.1016/j.physletb.2007.08.073
[0707.2302].


\bibitem{Toll:1956cya}
J.~S.~Toll,
``Causality and the Dispersion Relation: Logical Foundations,''
Phys. Rev. \textbf{104}, 1760-1770 (1956)
doi:10.1103/PhysRev.104.1760

\bibitem{DeLacroix:2018arq}
C.~De Lacroix, H.~Erbin and A.~Sen,
``Analyticity and Crossing Symmetry of Superstring Loop Amplitudes,''
JHEP \textbf{05}, 139 (2019)
doi:10.1007/JHEP05(2019)139
[arXiv:1810.07197 [hep-th]].

\bibitem{Bhattacharya:2020gar}
R.~Bhattacharya and R.~Mahanta,
``Analyticity of off-shell Green\textquoteright{}s functions in superstring field theory,''
JHEP \textbf{01}, 010 (2021)
doi:10.1007/JHEP01(2021)010
[arXiv:2009.03375 [hep-th]].

\bibitem{Bhattacharya:2021riu}
R.~Bhattacharya and R.~Mahanta,
``Analyticity Domain for Off-shell Five-point Superstring Loop Amplitudes,''
[arXiv:2110.13215 [hep-th]].

\bibitem{tHooft:1973wag}
G.~'t Hooft and M.~J.~G.~Veltman,
``DIAGRAMMAR,''
NATO Sci. Ser. B \textbf{4}, 177-322 (1974)
doi:10.1007/978-1-4684-2826-1\_5



\end{thebibliography}
\end{document}